\documentclass[11pt]{amsart}

\usepackage{times}

\usepackage{cite}   
\usepackage{eucal}  
\usepackage{epsfig} 

\DeclareMathOperator{\dist}{dist}
\DeclareMathOperator{\spec}{spec}
\DeclareMathOperator{\supp}{supp}
\DeclareMathOperator{\tmod}{mod}
\DeclareMathOperator{\diam}{diam}

\theoremstyle{plain}
\newtheorem{prop}{Proposition}[section]

\newtheorem{theorem}{Theorem}[section]

\theoremstyle{remark}
\newtheorem*{remark}{Remark}
\newtheorem*{remarks}{Remarks}

\theoremstyle{definition}
\newtheorem{defin}{Definition}[section]

\numberwithin{equation}{section}
\numberwithin{figure}{section}

\newcommand{\NNN}{\mathbb{N}}
\newcommand{\ZZZ}{\mathbb{Z}}
\newcommand{\RRR}{\mathbb{R}}

\newcommand{\TTT}{\mathbb{T}}

\newcommand{\MMM}{\mathcal{M}}

\newcommand{\Z}{{\mathbb Z}}
\newcommand{\R}{{\mathbb R}}

\newcommand{\cv}[1]{\mathcal{\uppercase{#1}}}

\begin{document}

\title[]{The Spectral Asymptotics
of the Two-Dimensional Schr\"odinger operator
with a Strong Magnetic Field}

\author[]{Jochen Br\"uning}
\address{Institute of Mathematics, Humboldt University
at Berlin, Rudower Chaussee~25, Berlin 12489 Germany}
\email{bruening@mathematik.hu-berlin.de}

\author[]{Serguei Yu. Dobrokhotov}
\address{Institute for Problems in Mechanics,
Russian Academy of Sciences, Vernadskogo~prosp.~101, Moscow 117526 Russia}
\email{dobr@ipmnet.ru}

\author[]{Konstantin V. Pankrashkin}
\address{Institute of Mathematics, Humboldt University
at Berlin, Rudower Chaussee~25, Berlin 12489 Germany, and
Institute for Problems in Mechanics,
Russian Academy of Sciences, Vernadskogo~prosp.~101, Moscow 117526 Russia}
\email{const@mathematik.hu-berlin.de}

\thanks{A slightly revised version
in published in \emph{Russian Jourmal of Mathematical Physics 2002, Volume~9;}
Part~I (Sections~1--6): no.~1, pp.~14--43,
Part~II (Section~7--Appendix~A): no.~4, pp.~400--416.}

\begin{abstract}
We consider the~spectral problem for the two-dimensional Schr\"odinger operator
for a charged particle in strong uniform magnetic and periodic electric fields.
The related classical problem is analyzed first by means of the
Krylov-Bogoljubov-Alfven and Neishtadt averaging methods. It allows us
to show ``almost integrability'' of the the original two-dimensional
classical Hamilton system,  and  to
reduce it  to a~one-dimensional one on the phase space which is a~two-dimensional
torus. Using the topological methods for integrable Hamiltonian system
and elementary facts from the Morse theory, we give a~general
classification of the classical motion. According this classification the
classical motion is separated into different regimes with different
topological characteristics (like rotation numbers and Maslov indices).
Using these regimes, the semiclassical approximation, the Bohr-Sommerfeld rule
and the correspondence principle, we give a~general asymptotic description
of the (band) spectrum of the original Schr\"odinger operator and, in
particular, estimation for the number of subbands in each Landau band.
From this point of view the regimes, are the classical preimages
of ``spectral series'' of the Schr\"odinger operator.  We also
discuss the relationship between this spectrum and the spectrum
of one-dimensional difference operators.
\end{abstract}

\maketitle

Both classical and quantum problems describing the motions of particles under
the influence of a uniform magnetic and a periodic electric fields have very
curious properties even in two dimensions. This has caused a large number of
publications; which we mention here only some of them which
are most relevant to our considerations \cite{BSu,LL,LP,No1,No2,NV,NV2,Z1,Z2,Z3,
BV,Gey,AZ,AS,ASS,DN1,DN2,DSS,Ly,Ha,
HS1,HS2,HS3,KS,MN,TKNN,W1,W2,shubin1}.

If the magnetic field is strong enough, then a large
parameter appears in both  classical and quantum
mechanics. Hence it is possible to use the averaging
methods \cite{A2,AF,BM,BP,BZ,Li,MS,Ne,NY} and semiclassical approximation
\cite{BB,CdV,Gu,Hel,M1,M2,MF}. Even though this circle of problems is well
studied, we propose here some apparently new formulas and interpretations.

\section{Formulation of the problem and brief description of the results}\label{s1}

\subsection{The two-dimensional magnetic Schr\"odinger operator in a periodic
electric field. Assumptions and parameters}\label{ss1.1}

We want to describe certain asymptotic spectral properties of the
Schr\"odinger operator
\begin{equation}
{\widehat H}\Psi:=\Big[\frac12\big(-ih\frac{\partial}{\partial x_1}+x_2\big)^2+
\frac12\big(-ih\frac{\partial}{\partial x_2}\big)^2+
\varepsilon v(x_1,x_2)\Big]\Psi
                                 \label{1.1}
\end{equation}
in $L^2(\RRR^2)$, which is essentially self-adjoint on $C^\infty_0(\RRR^2)$,
as $h,\varepsilon\to0$.
We assume that the potential $v(x_1,x_2)$ is real analytic in $\RRR^2$
and periodic with
respect to the lattice $\Gamma$ generated by two linearly independent vectors
$a_1=(a_{11},a_{12})\equiv (2\pi,0)$,
$a_2=(a_{21},a_{22})$, i.~e. we have $v(x+a_1)= v(x+a_2)=v(x)$.

Such a problem arises in the following physical situation.
Consider the motion of a particle with charge $-e$ and mass $m$
in the plane $\RRR^2_y$ in a uniform magnetic
and periodic electric field. If the magnetic field is perpendicular
to the $y$-plane and has strength $B>0$, then this motion
is described (in the Landau gauge) by the operator
\begin{equation*}
\widehat{H}_B=\frac{1}{2m}\bigg(\Big(-i\hbar\frac{\partial}{\partial y_1}+
\frac{eB}{c} y_2\Big)^2
-\hbar^2\frac{\partial^2}{\partial y_2^2}\bigg) + V(y_1,y_2),
\end{equation*}
where $V$ is the potential of the electric field and
$c$, $\hbar$ are physical constants.
Let the potential $V$ is periodic with respect to
the lattice spanned on two vectors
\begin{equation*}
l_1=(L_0,0),\quad l_2=(l_{21},l_{22}),\quad l_{22}\ne 0.
\end{equation*}
Introducing new variables $x= 2\pi y/L_0$
we reduce the spectral problem for $\widehat{H}_B$ to the form
\begin{equation}
                       \label{B-spectrum}
\frac{(e B L_0)^2}{4\pi^2 m c^2}\widehat{H}\Psi' = E'\Psi',
\end{equation}
where
\begin{gather*}
h=(2\pi)^2\Bigl(\frac{l_M}{L_0}\Bigr)^2,
\varepsilon=h\frac{|V|_{\text{max}}}{\hbar\omega_c},\\
\text{$\omega_c=\frac{|eB|}{cm}$~is the \emph{cyclotron frequency},\quad
$l_M=\sqrt{\frac{\hbar}{m\omega_c}}$ is the~\emph{magnetic length}
of the system,}\\
|V|_{\text{max}}=\max|V|, \qquad
v=\frac{1}{|V|_{\text{max}}}V\big(L_0 x/(2\pi)\big).
\end{gather*}
Therefore, the smallness of $h$ means that the \emph{characteristic
size} $L_0$ of the lattice is much greater than the magnetic length
$l_M$; then $\varepsilon$ is small if, for example,
the electric energy $|V|_{\text{max}}$ is comparable with
the magnetic energy $\hbar\omega_c$.

The smallness of $h$ indicates that the number $\eta:=a_{22}/h$
(which is the \emph{number of the magnetic flux quanta
through the elementary cell})
is large.

Such a situation can be realized e.~g. in periodic arrays of~quantum dots or
antidots or in super-lattices~\cite{BvH,EP,NPF,We2}).

It is well known that the spectral properties of the operator $\widehat H$
depend crucially on the parameter $\eta$.
If $\eta=N/M$ is rational, then the spectrum of $\widehat H$
has band structure (in this case, $\widehat H$ has the \emph{Kadison property},
see~\cite{Be,Gey}). For each
spectral value $E$ of $\widehat H$ it is then possible to construct a
basis of $M$ generalized eigenfunctions $\Psi^j(x,q)$,  $j=0,\dots,M-1$,
depending on two new parameters $q=(q_1,q_2)$, $q_1\in [0,1/M]$,
$q_2\in[0,1]$, with the following \emph{magneto-Bloch} properties
(see \cite{Be,No1,Gey}):
\begin{gather}
\Psi^j(x+a_1,q)=\Psi^{j+1}(x,q)e^{-2\pi i(q_1-j\eta)},\quad j=0,\dots,M-1,
                                \label{1.2}\\
\begin{gathered}
\Psi^j(x+a_2,q)=\Psi^{j+1}(x,q)e^{-i \eta (x_1+a_{21}/2)},\quad j=0,\dots,M-2,\\
\Psi^{M-1}(x+a_2,q)=\Psi^0(x,q)e^{-i\eta(x_1+a_{21}/2)-2\pi iq_2}.
\end{gathered}                  \label{1.3}
\end{gather}
Thus the spectral value $E$ becomes a
function of $(q_1,q_2)$ and both $E$ and $\Psi^j$ depend on the parameters
$h$ and $\varepsilon$ (and also on some others);
we omit this dependence to simplify the notation.
The structure of the spectrum of~$\widehat H$ becomes much more complicate
if $\eta$ is irrational; in particular, Cantor sets may arise~\cite{HS2}.

\subsection{The Correspondence Principle.  The goals
and the structure of the paper}

We want to exploit the small parameter $h$
to obtain asymptotic information about $\spec \widehat H$ by means
of semiclassical approximation as $h\to +0$.

The fact that the parameter~$1/B$ plays the same role for
the operator $\widehat H$ as the Planck
constant for the ordinary Schr\"odinger equation
was first pointed out in~\cite{BV,AZ}.
We emphasize that all our assumptions on the parameters $h$,
$\varepsilon$, and the vectors $a_1$, $a_2$ are
essential for our method. If, for example, $|a_j|\sim h$,
then $v= v(x_1/h,x_2/h)$, and instead of ``standard'' semiclassical methods
the Born-Oppenheimer (adiabatic) approximation has to be used in this
situation, and it leads to quite different results (see \cite{MN}).

Semiclassical asymptotics are used very widely in problems with
discrete spectrum (see e.g.~\cite{MF,Gu,La,CdV})
but they are not commonly used
in (multidimensional) problems with continuous spectrum. Hence one of the
goals of this paper is to point out the potential of semiclassical methods for
these problems. In particular, we want to understand what the conditions
\eqref{1.2} and~\eqref{1.3} mean for the semiclassical approximation.

It is a well known fact that usually there are no
universal asymptotic formulas for the spectrum even in quite simple
situations; one has to describe different parts of the spectrum
by different formulas. In the  discrete  case, the various  parts of the
spectrum  arising in this way are referred to as~\emph{spectral series}
in physics literature; we keep this notation for our
situation where the spectrum is continuous.
Let us recall that the semiclassical approximation realizes the
\emph{Correspondence Principle}: it allows us  to
describe asymptotic properties of the spectrum of the quantum mechanical
system via some objects related to the associated
classical Hamiltonian system.
Thus it is natural to ask which ``regimes'' of
the classical phase space should correspond to these spectral series;
this is the main motivation of this paper.

For the magnetic Schr\"odinger operator \eqref{1.1},
the classical Hamiltonian is
\begin{equation}
H(p,x,\varepsilon)=\frac12(p_1+x_2)^2+\frac12p^2_2+\varepsilon v(x_1,x_2).
                                  \label{1.4}
\end{equation}
For the operator $\widehat H$
we  want to show that at least in low-dimensional classically integrable
situations  these pre-images are certain subsets or ``regimes``,
to be denoted by  $\MMM_r$, in the phase space which
allow a convenient description in terms of certain graphs.

Of course, we now have to explain the relationship between the magnetic
Schr\"odin\-ger operator~\eqref{1.4} and
integrable systems, since the Hamiltonian
system associated with the latter one is, generally speaking, non-integrable.
The connection is brought about through the small parameter $\varepsilon$:
it turns out that  the  Hamiltonian system associated with~\eqref{1.4}
is ``almost integrable'', modulo corrections which are exponentially small
with respect to this parameter. This observation follows from the
averaging methods, which were applied in~\cite{AF,BM,BP,BZ,Li,MS}
to the analysis of the motion of classical
particles  subject to a strong uniform magnetic
and certain electric fields from different pints of view.

The averaging  (see~\ref{s3}) allow us to reduce the original
Hamiltonian system, with two degrees of freedom, to a system
with one degree of freedom i.e. an integrable system.
Actually, this reduction does not depend on the periodicity
of the  electric potential $v$. But in the two-dimensional periodic case
averaging leads to a Hamiltonian system with phase space
the two-dimensional torus $\TTT^2$,  and the ``reduced''
Hamiltonian turns out  to be a \emph{Morse function} on~$\TTT^2$.
This observation leads naturally to a complete
classification of the  classical motion  in terms of ``regimes''
(see sections~4 and~5).
The Bohr-Sommerfeld  quantization
rule defines some subsets of~$\RRR$ consisting of
points and intervals which form the desired  \emph{spectral series}. For
each  point from such a set we can construct  a collection  of asymptotic
eigenfunctions  (or quasimodes) of the operator
$\widehat H$, which are given as power series in the small parameter and
are localized in a neighborhood of certain  domains in the original
configuration space $\RRR^2$ (see section~\textup{6}).

The next question is to understand how the actual spectrum and the actual
(generalized) eigenfunctions are related to the constructed spectral series
and quasimodes. For instance, in th case of rational flux,
the constructed quasimodes do not satisfy the Bloch conditions~\eqref{1.2},
\eqref{1.3}, and the Bohr-Sommerfeld
rule gives a discrete subset in contrast to the band structure of
spec $\widehat H$ in this case.
A strict mathematical answer to this question is beyond the scope
of the ``power'' approximations used in this paper; we will discuss it only
heuristically (section~7). In particular, we obtain a heuristic
``Weil formula'' for the number of subbands in each Landau band
(section 8) and discuss a connection between our quasimodes and
the Harper-type difference equations (sections 6 and 8).

To motivate our considerations, we begin by
describing some well known results from the semiclassical analysis
of one-dimensional periodic Schr\"odinger operators with
a small parameter in front of the second derivative (section~\ref{s2}).
This example allows us already to illustrate the main features
of our approach: the geometric description of the spectrum by
means of Reeb graphs, the semiclassical structure of quasimodes
and spectral series in problems with continuous spectra, the
relationship between different asymptotic formulas, and the correctness
of certain heuristic considerations.

\subsection{Table of notation}
We will have to use a somewhat elaborate notation which we summarize
here for easy reference.

\begin{itemize}
\item $\varepsilon$ is a small classical parameter
in the classical problem,
\item $h$ is a small semiclassical parameter in the quantum problem,
\item
$K$ is an integer number describing the accuracy of the expansion
with respect to $\varepsilon$;
\item
$L$ is an integer number describing the accuracy of the expansion with
respect to $h$;
\item
$\eta$ is the number of the magnetic flux
quanta through the elementary cell
(which we denote by $\eta=N/M$ in the rational case);
\item
$a_1$ and $a_2$ are the generators of the lattice $\Gamma$;
\item
$d=(d_1,d_2)\in\ZZZ^2$ is the drift vector  of classical
trajectories, $d_1/d_2$ is the rotation number;
\item
$f=(f_1,f_2)\in\ZZZ^2$ is a vector
that is conjugate to $d$, i.~e. $d_1f_1+d_2f_2=1$;
\item
the over-line index $\widetilde{}$ (tilde) indicates a connection
with infinite motion;
\item
$r \in \NNN$ numbers the regimes, $\MMM_r$ (finite motion)
and $\widetilde{M}_r$ (finite motion), of the classical motion;
\item
$q=(q_1,q_2)$ is the vector of quasimomenta;
\item
$l=(l_1,l_2)\in \ZZZ^2$ is a multi-index indexing  closed (contractible)
curves on the two-dimensional torus belonging to the boundary regimes
and implied quasimodes;
\item
$k \in \ZZZ$ is the index of open curves or two-dimensional cylinders belonging
to the interior regimes and implied quasimodes;
\item
$\mu$ is the quantum number of the Landau level $\cv{I}^{(\mu)}_1$;
\item
$\nu$ is the (quantum) number of the ``slow drift'' action $\cv{I}^{(\nu)}_2$
(it appears in the boundary regimes only);
\item
$\delta$ characterizes the neighborhood of the singular manifolds of
the classical motions;
\item
$j\in \NNN$ numbers the magneto-Bloch eigenfunctions;
\item
$s$ is the number of the collection of the magneto-Bloch functions, satisfying
\eqref{1.2}, \eqref{1.3};
\item
$n^\pm $ is the index of a band in the interior regimes.
\end{itemize}

\subsection{Averaging, almost integrability,
and classification of the classical motion (sections 3--5)}

\begin{figure}\centering
\begin{minipage}{100mm}
\noindent\epsfig{file=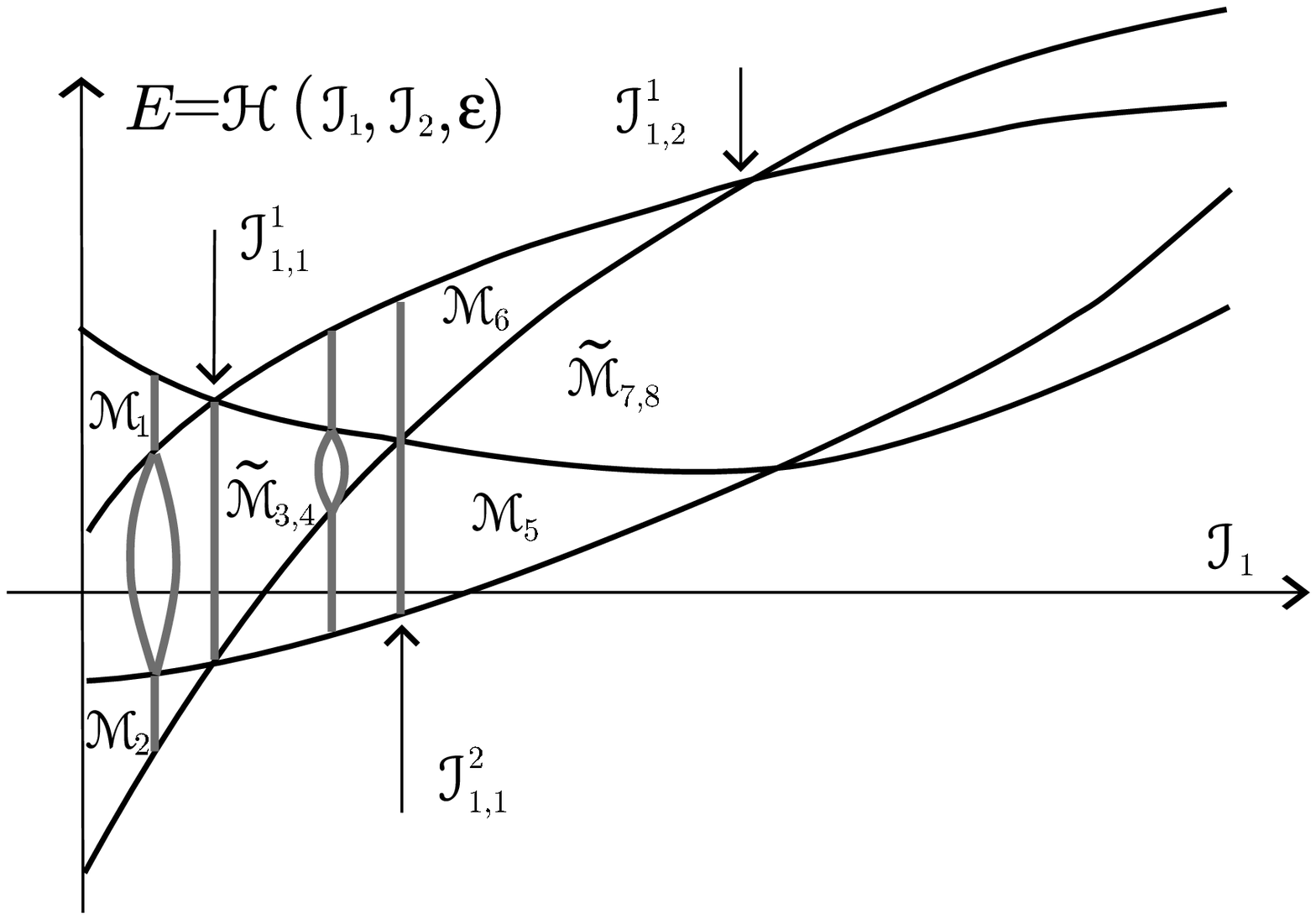,width=95mm}
\caption{Global classification of the classical motion\label{f1.1}}
\end{minipage}
\begin{minipage}{60mm}\centering
\noindent\epsfig{file=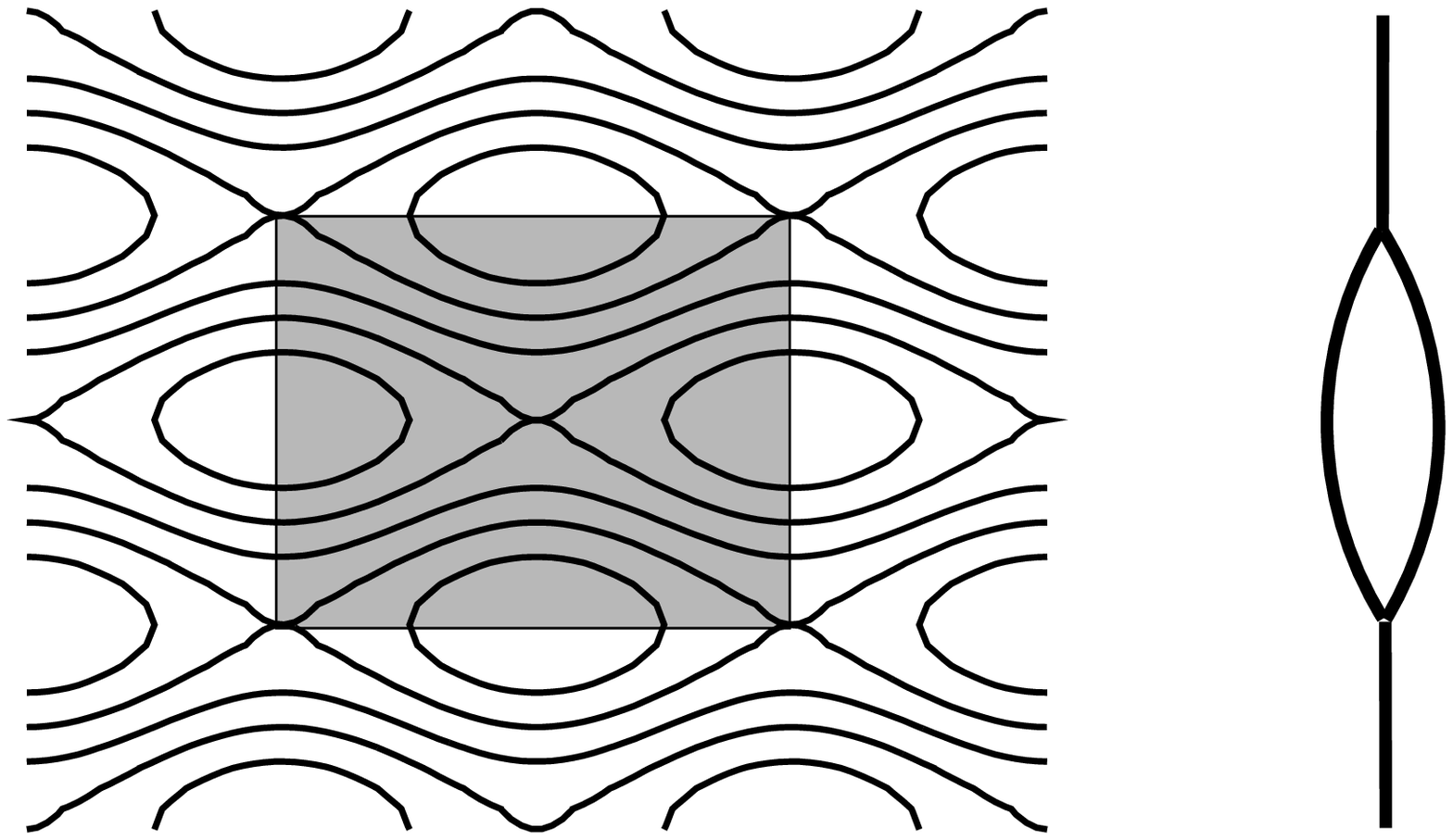,width=55mm}
\noindent\caption{Level curves and the Reeb graph\label{f1.2}}
\end{minipage}
\end{figure}

The averaging process gives us an averaged Hamiltonian,
$\cv{H}$, such that
in new ``corrected'' symplectic
coordinates (with generalized momenta $\cv{I}_1$, $\cv{y}_1$  and generalized
coordinates $\Phi$, $\cv{y}_2$) we can write
\begin{gather}
H=\cv{H}(\cv{I}_1,\cv{y},\varepsilon)+O(e^{-C/\varepsilon}),
                                                 \label{1.5}\\
\cv{H}=\Bar{H} +O(\varepsilon^2),\quad
\Bar{H}(\cv{I},\cv{y},\varepsilon)=
\cv{I}_1+\varepsilon J_0(\sqrt{-2\cv{I}_1\Delta_{\cv{y}}})v(\cv{y}),\notag
\end{gather}
Here $J_0(z)$ is the Bessel function of order zero and
$\Delta_{\cv{y}}=\partial^2/\partial\cv{y}_1^2+\partial^2/\partial \cv{y}_2^2$.
Our main example in this paper is connected with the potential
\begin{equation}
v(x)=A \cos x_1+B \cos (\beta x_2),
                                                 \label{1.6}
\end{equation}
where $A$, $B$, and $\beta$ are positive constants. Then we have
\begin{equation}
\Bar{H}(\cv{I}_1,\cv{y},\varepsilon)=
\cv{I}_1+\varepsilon\bigl(A J_0(\sqrt {2\cv{I}_1})\cos \cv{y}_1+
B {J}_0(\beta \sqrt {2\cv{I}_1})\cos (\beta \cv{y}_2)\bigr).
                                                 \label{1.7}
\end{equation}
Now for almost all $\cv{I}_1$, the Hamiltonian
$\Bar H$ (or $\cv{H}$) may be considered as
a Morse function on the two-torus $\TTT^2 =\RRR^2/(a_1,a_2)$.
Using the topological theory of Hamiltonian systems~\cite{BF,Fom},
for each fixed $\cv{I}_1$ we may separate the motion defined
by the averaged Hamiltonian into different topological regimes,
which are conveniently described by means of its Reeb graph.
After a change of the action variable $\cv{I}_1$ we obtain
the~\emph{regimes} as the sets of points in phase space which
correspond to topologically similar edges
of the Reeb graph. Then classical motions through points from a fixed regime
are topologically similar.  It is convenient to present the regimes
on the half-plane $\{(\cv{I}_1, E)\in \RRR^2; \cv{I}_1\ge 0 \}$
where $E$ is the classical energy of the averaged system.
We give the complete description of the regimes in section~3;
the picture for example~\eqref{1.6} is given in Fig.~\ref{f1.1}.

The motion defined by the
averaged Hamiltonian $\Bar{H}$ takes place in the domain
\begin{equation}
\begin{gathered}
\Sigma_0 =\left\{(\cv{I}_1, E)\in \RRR^2; \cv{I}_1\ge 0,\quad
|E-\cv{I}_1|\le  \varepsilon (A |J_0(\sqrt {2\cv{I}_1})|+
B|J_0(\beta \sqrt {2\cv{I}_1})|)\right\}.
\end{gathered}
                                            \label{1.8}
\end{equation}
This domain is the projection of the
the actual motion surface, $\Sigma$;
any its cutting by the plane $\cv{I}_1=\text{const}$
is then homeomorphic to the Reeb graph
of the Morse function  $\Bar{H}$ (see Fig.~\ref{f1.1}).

Also, $\Sigma$ decomposes
into \emph{regimes} along the curves
\begin{equation*}
E=\cv{I}_1\pm \varepsilon\big|A|J_0(\sqrt {2\cv{I}_1})|\pm
B|J_0(\beta\sqrt {2 \cv{I}_1})|\big|
\end{equation*}
which, in this example, form the common boundaries of the boundary
and interior regimes. We distinguish between the regimes
$\MMM_r$ corresponding to finite classical motion and
$\widetilde{\MMM}_r$ corresponding to infinite classical motion.

Also, it is natural to distinguish between \emph{regular}
and \emph{singular} boundaries of the regimes,
according to whether they are external or internal.
The internal boundaries may have intersection points which are their
\emph{singularities}.

With each regime, one can associate topological and analytical
characteristics.
These are
the \emph{drift vector},
the \emph{Maslov index},
the \emph{action variables},
and
the \emph{form of the Hamiltonian in the action variables}.

In fact, to each inner point of a regime there corresponds a family of
closed trajectories on $\TTT^2$, hence a family of closed (for boundary
regimes) or  open trajectories (for inner regimes) on
the covering $\RRR^2_{\cv{y}}$. To these corresponds in turn
a family of Lagrangian (or Liouville) tori, for boundary regimes,
and Lagrangian (or Liouville) cylinders, for interior regimes, in the original
phase space  $\RRR^4_{p,x}$. To the Lagrangian (or Liouville) tori or cylinders
(and hence to the regime under consideration) we may associate
(\emph{a}) the  vector $d=(d_1,d_2)$ of the
drift in the original configuration
space $\RRR^2_x$, or equivalently
the rotation number $d_1/d_2$
of the related closed trajectory on the
torus,
and (\emph{b})
the Maslov indices of the related Lagrangian (or Liouville) tori or
cylinders.

The rotation number of a boundary regime is equal to $0/0$,
there is no drift, and there is no preferred direction. The Maslov
indices of natural cycles on a Liouville torus is equal to~$2$.

On the other hand, the rotation number of an inner regime is not trivial,
there exists a~preferred  direction, but each cylinder has only one cycle,
and hence only one Maslov index, which again is equal to $2$.

Also, in each regime one can introduce a second action variable $\cv{I}_2$
and find
(\emph{c}) the analytic representation of the Hamiltonian in action variables
$\Bar{H}=\Bar{H}^r(\cv{I}_1,\cv{I}_2,\varepsilon)$,
which \emph{depends on the regime}.
The drift vector and the function  $\Bar{H}^r(\cv{I}_1,\cv{I}_2,\varepsilon)$
changes discontinuously when one passes from one regime to another.
The correction $\cv{H}-\Bar{H}$ does not change
neither this rough description of the classical motion nor the
general asymptotic  description of the spectrum, even though a complete
analysis of the effected changes may be of importance in certain physical
problems.
In this paper we do not analyze  the  classical motion in the
neighborhood of singular boundaries or the  behavior of the corresponding part
of the spectrum. Thus we introduce some small number $\delta$,
and remove certain $\delta$-neighborhood
of the singular boundary from all regimes $\MMM_r$ and $\widetilde{\MMM}_r$.
These new sets we also refer to as regimes;
we denote them  by ${\MMM}_{r,\delta}$ and $\widetilde{\MMM}_{r,\delta}$,
respectively.

\subsection{The global asymptotic structure of the spectrum (section~6)}
\begin{figure}\centering
\epsfig{file=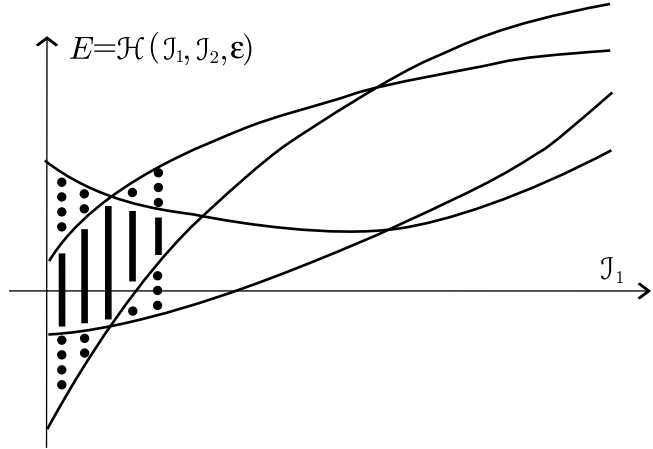,width=90mm}
\caption{Global structure of the spectrum\label{f1.3}}
\end{figure}

The Bohr-Sommerfeld quantization of the regimes
${\MMM}_{r,\delta}$ and $\widetilde{\MMM}_{r,\delta}$  results in
quantized regimes  on the ``Reeb surface``, which after
the projection onto the energy axis $E$ defines
the first approximation in the asymptotics of the spectrum
of the original operator. The quantization conditions are different
for boundary and interior regimes. In both cases, we can
quantize the variable $\cv{I}_1$ thus defining the so-called
\emph{Landau level}
\begin{equation}
\cv{I}_1^{(\mu)} =(\frac12+\mu)h.
                                      \label{1.9}
\end{equation}
For boundary regimes, we have in addition a quantization of $\cv{I}_2$, given by
\begin{equation}
\cv{I}_2^{(\nu)} =(\frac12+\nu)h,
                                      \label{1.10}
\end{equation}
Here, $\mu$ and $\nu$ are integers with
$(\cv{I}_1^{(\mu)},\cv{I}_2^{(\nu)})\in  \MMM_{r,\delta}$.
However, $\cv{I}_2$ is not quantized in interior regimes.
Now consider the numbers
$\Bar{H}^r(\cv{I}_1^{(\mu)},\cv{I}_2^{(\nu)},\varepsilon)$,
$(\cv{I}_1^{(\mu)},\cv{I}_2^{(\nu)})\in  \MMM_{r,\delta}$
for boundary regimes, and the functions
$\Bar{H}_r(\cv{I}_1^{(\mu)},\cv{I}_2,\varepsilon)$,
$(\cv{I}_1^{(\mu)},\cv{I}_2)\in  \widetilde{\MMM}_{r,\delta}$,
for interior regimes.
We then derive the quantized regimes or the~\emph{spectral series} on the
surface $\Sigma$ and their projections onto the domain $\Sigma_0$ in the
plane $(E,\cv{I}_1)$).
These sets consist of points (for boundary regimes)
and intervals  (for interior regimes);
for example~\eqref{1.6}, the result is sketched in Fig.~\ref{f1.3}.

Projecting  this set onto the $E$-axis we obtain the set which
describes the first order asymptotics of the spectrum of
the operator $\widehat H$.
Indeed, we have the following result.

\begin{prop}
For each $r$ and suitable $(\mu,\nu)$ or $(\mu,\cv{I}_2)$
related to  $\MMM_{r,\delta}$ or $\widetilde{\MMM}_{r,\delta}$
and arbitrary $K$, $L\in \NNN$, there exist numbers
\begin{equation*}
E^{\mu,\nu}_r=\Bar{H}^r(\cv{I}_1^{(\mu)},\cv{I}_2^{(\nu)},\varepsilon)
+O(h^2+\varepsilon^2)
\end{equation*}
for the boundary regimes and functions
\begin{equation*}
E^{\mu}_r(\cv{I}_2)=\Bar{H}^r(\cv{I}_1^{(\mu)},\cv{I}_2,
\varepsilon) +O(h^2+\varepsilon^2)
\end{equation*}
for the interior regimes,
such that the distance between
them and the spectrum of the operator $\widehat H$ is
$O({\varepsilon}^K+h^L)$.
\end{prop}

We have already mentioned that semiclassical methods will also allow us
also to construct asymptotic
(generalized) eigenfunctions (or quasimodes) for the operator $\widehat H$.
Actually,  each number in the sets just described leads to the construction
of infinitely many quasimodes, with support localized in a neighborhood of the
image of the  invariant Liouville tori or cylinders in the
configuration plane $\RRR^2_x$ (see Fig.~\ref{f1.4}).

Of course, this ``degeneration'' in the construction of quasimodes stems from
the fact that $\widehat H$ has continuous spectrum. We emphasize that the
described construction does not depend  on the rationality of the flux $\eta$,
i.e. we do not feel any rationality effects. Our results concerning the
spectrum of the operator $\widehat H$ and its quasimodes cannot be improved using
semiclassical approximations in powers of the parameters, not taking in account
tunneling. However, the  description of the spectrum on the plane
$E$, $\cv{I}_1$ by the quantized regimes carries more information about
the original  operator than the description of the spectrum on the  energy
axis: for example, it separates the spectrum according to the different Landau
bands, numbered by the index  $\mu$, and allows us to estimate
their  width. In the special case \eqref{1.6}, this width is
(see section~6)
\begin{equation*}
2\varepsilon \Big(A \big|J_0(\sqrt{2\cv{I}_1^{\mu}})\big|+
B\big|J_0(\beta \sqrt {2\cv{I}_1^{(\mu)}})\big|+O(h)\Big).
\end{equation*}

\subsection{The case of rational flux (sections~7--8)}

We now turn to the connection between the constructed
set and the spectrum of $\widehat H$.
If $\varepsilon$ is smaller than $h$, then the
asymptotic Landau bands do not intersect.
So in this case the constructed
semiclassical ``asymptotic'' spectrum consists of intervals and points
on the axis $E$. As in the case with rational flux  the
spectrum of the operator $\widehat H$ has  band structure,
it means that in this situation
discrete points define something like ``traces'' of the (exponentially)
small bands, and on the other hand there can be (exponentially)
small gaps in the intervals in inner regimes, which one cannot catch by means
of ``power'' semiclassical  asymptotics. Moreover, there exists probably
their fuzziness on the surface $\Sigma$ (and the plane $\cv{I}_1,E$) in the
direction $\cv{I}_1$.

To clarify this situation (in heuristic
level) one can  look at these quasimodes from
the point of view of magneto-Bloch conditions~\eqref{1.2} and~\eqref{1.3}.
It is clear that the described
quasimodes do not satisfy these conditions,
but one can use them as a base for constructing the functions
satisfying~\eqref{1.2}--\eqref{1.3}.
The corresponding pure algebraic procedure
(it does not depend on concrete form of the potential $v$) gives
the following results.
First, it defines certain  points on the intervals from the inner
regimes describing the ``traces'' of gaps on them.
Secondly, it takes off infinite degeneration in such a sence,
that for each Bohr-Sommerfeld point
$\cv{H}_r(\cv{I}^{(\mu)}_1,\cv{I}^{(\nu)}_2,\varepsilon)$ and quasimomentum $q$,
from the Bloch conditions we obtain $M$ collection of
linear independent (``Bloch'') quasimodes
(It is interesting that the structure of these ``Bloch''
quasimodes related to boundary regimes does not depend on the
choice of the coordinates $x_1$, $x_2$. It is not the case for quasimodes
related  to inner regimes: they have the simplest form if
the Bloch conditions in coordinates  $x_1$, $x_2$ agree with
the drift vector (rotation number, which is a topological invariant)
in such a way, that the latter one is $(1,0)$.)
On the other hand, the typical degeneration gives
the multiplicity $M$, which means that indeed
the Bohr-Sommerfeld points corresponds to $M$ exponentially
small subbands, separated by exponentially small gaps, satisfying to Bloch
conditions~\eqref{1.2}--\eqref{1.3}.
(Recall that $M$ is the denominator of the flux $\eta$.)

\begin{figure}\centering
\begin{minipage}{75mm}\centering
\noindent\epsfig{file=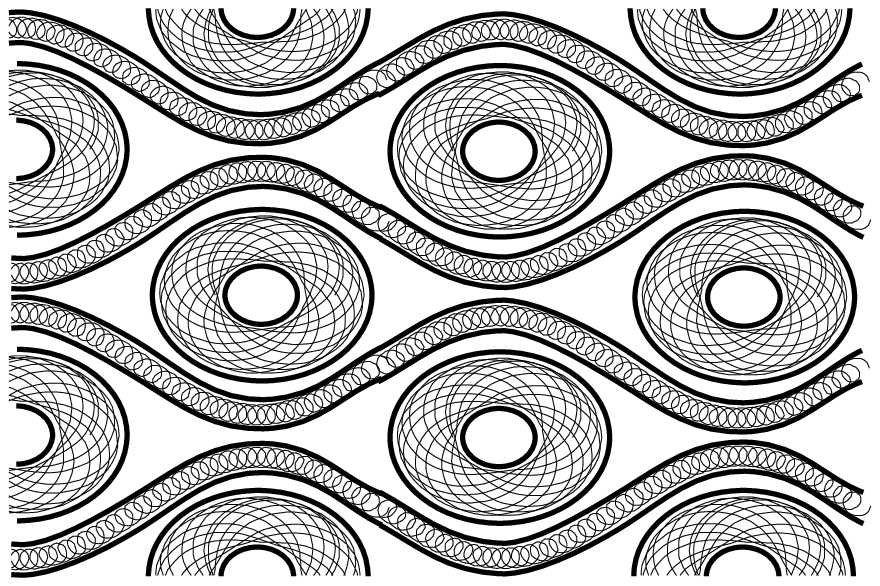,width=60mm, height=50mm}
\caption{Projections of trajectories\label{f1.4}}
\end{minipage}
\begin{minipage}{70mm}\centering
\noindent\epsfig{file=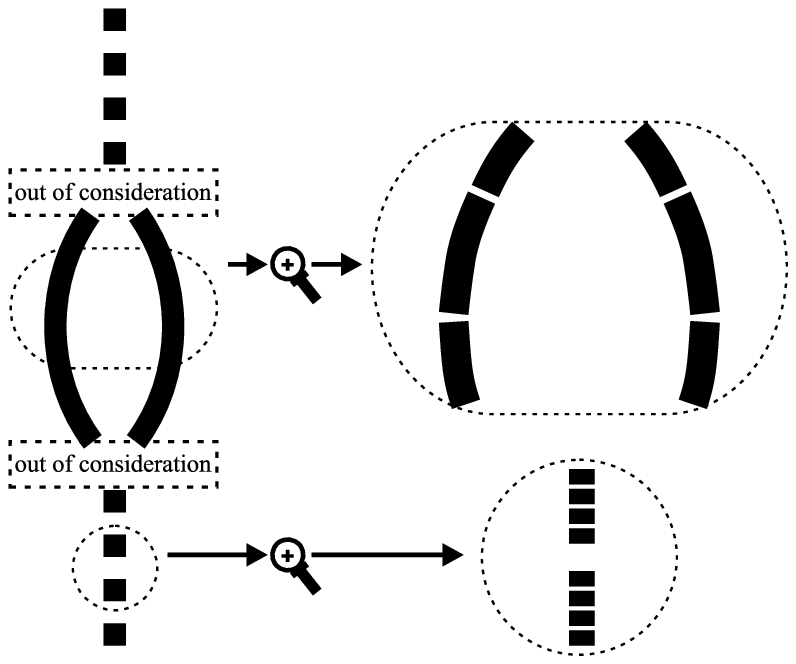,width=60mm, height=50mm}
\noindent\caption{Separation of subbands\label{f1.5}}
\end{minipage}
\end{figure}

So if one takes a magnifying glass (i.e. construct
a more precise approximation) and look at the Reeb graph
corresponding to a certain fixed Landau level,
and its (exponentially)  small neighborhood, the following
picture appears (see Fig.~\ref{f1.5}).

These consideration gives the heuristic
``geometrical Weyl''  estimates for number $\mathcal N$  of subbands for
the fixed Landau  level. The idea is that first one has to count them on
the edges of the Reeb
graph, and then to project the result to the energy axis. Final formula
for  $\mu$-th Landau level in example \eqref{1.6} is
$\mathcal N(\cv{I}_1^{(\mu)})\approx N$,
where $N$ is the nominator of flux $\eta$.

The construction  of ``Bloch'' quasimodes gives also in the first
approximation the simple dependence on quasimomenta or the dispersion relations.

At last we obtain difference Harper-like equations, if
quantize the averaged Hamiltonian $\Bar{H}$ in naive way. It implies the
correspondence between constructed ``Bloch'' quasimodes and quasimodes of the
difference equations. We discuss this correspondence in sections~6 and 8.

\subsection*{Acknowledgments}
In carrying out this work we had useful discussions with S.~Albeverio,
J.~Avron, E.~D.~Belokolos, V.~S.~Buslaev, P.~Exner, V.~A.~Geyler,
M.~V.~Karasev, E.~Korotyaev, V.~A.~Margulis, A.~I.~Neishtadt,
L.~A.~Pastur, M.~A.~Poteryakhin, A.~I.~Shafarevich, P.~Yuditskii, and J.~Zak. 
To all them we express a gratitude.

The work is supported by the collaborative research
project of the German Research Society
(Deutsche Forschungsgemeinschaft) no. 436 RUS 113/572
and by the grant INTAS-00-257.
K.~V.~P. is also thankful for financial support
to the Graduate College ``Geometry and Nonlinear Analysis''
(DFG GRK~46) of Humboldt University at Berlin.

\section{Example: ``graph'' semiclassical analysis of the spectral
Sturm-Liouville problem on the circle}\label{s2}

\subsection{The periodic  Sturm-Liouville problem}\label{ss2.1}
\begin{figure}\centering
\begin{minipage}{65mm}\centering
\epsfig{file=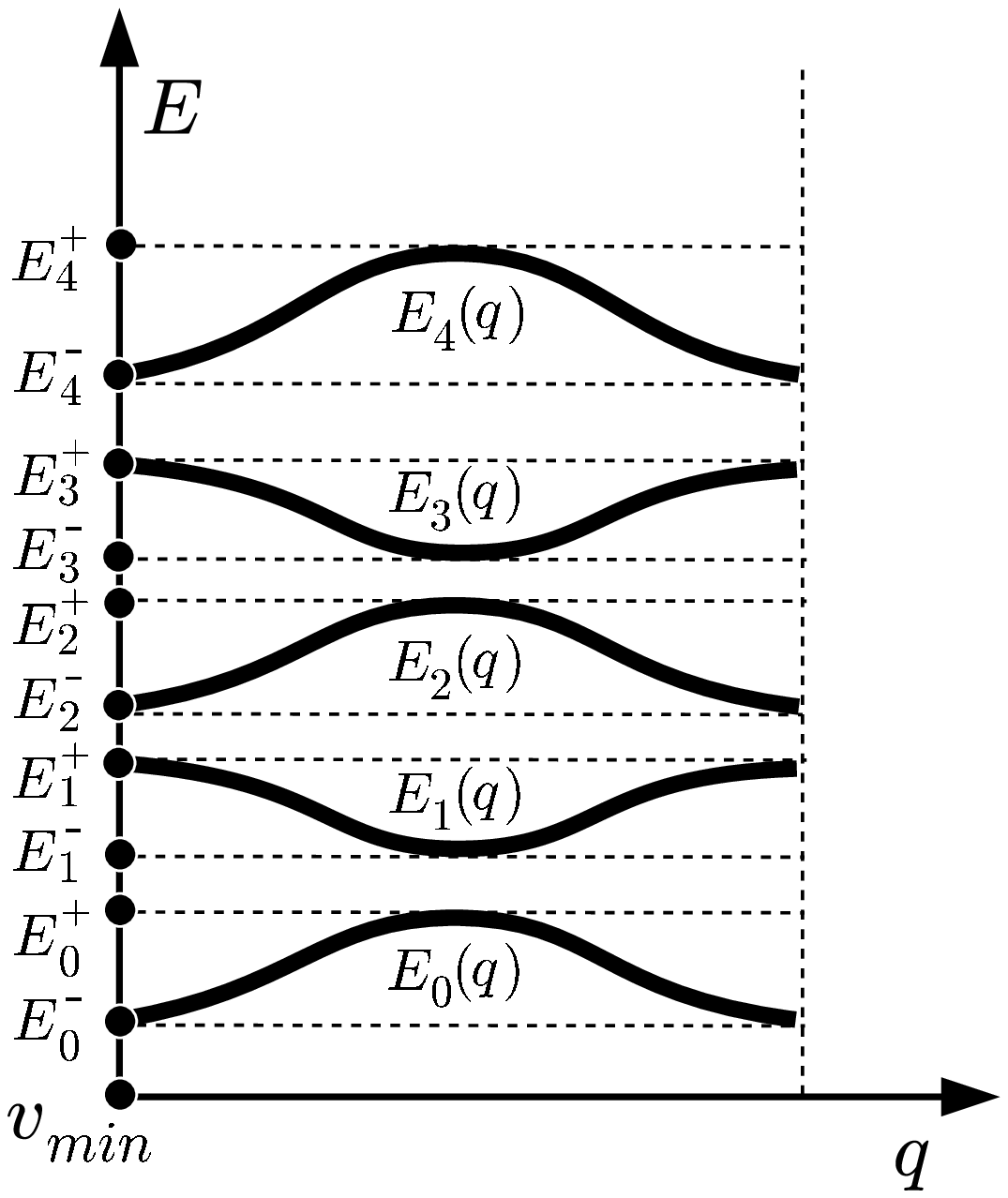,height=70mm}
\caption{Typical dispersion relations\label{f2.1}}
\end{minipage}
\begin{minipage}{80mm}\centering
\epsfig{file=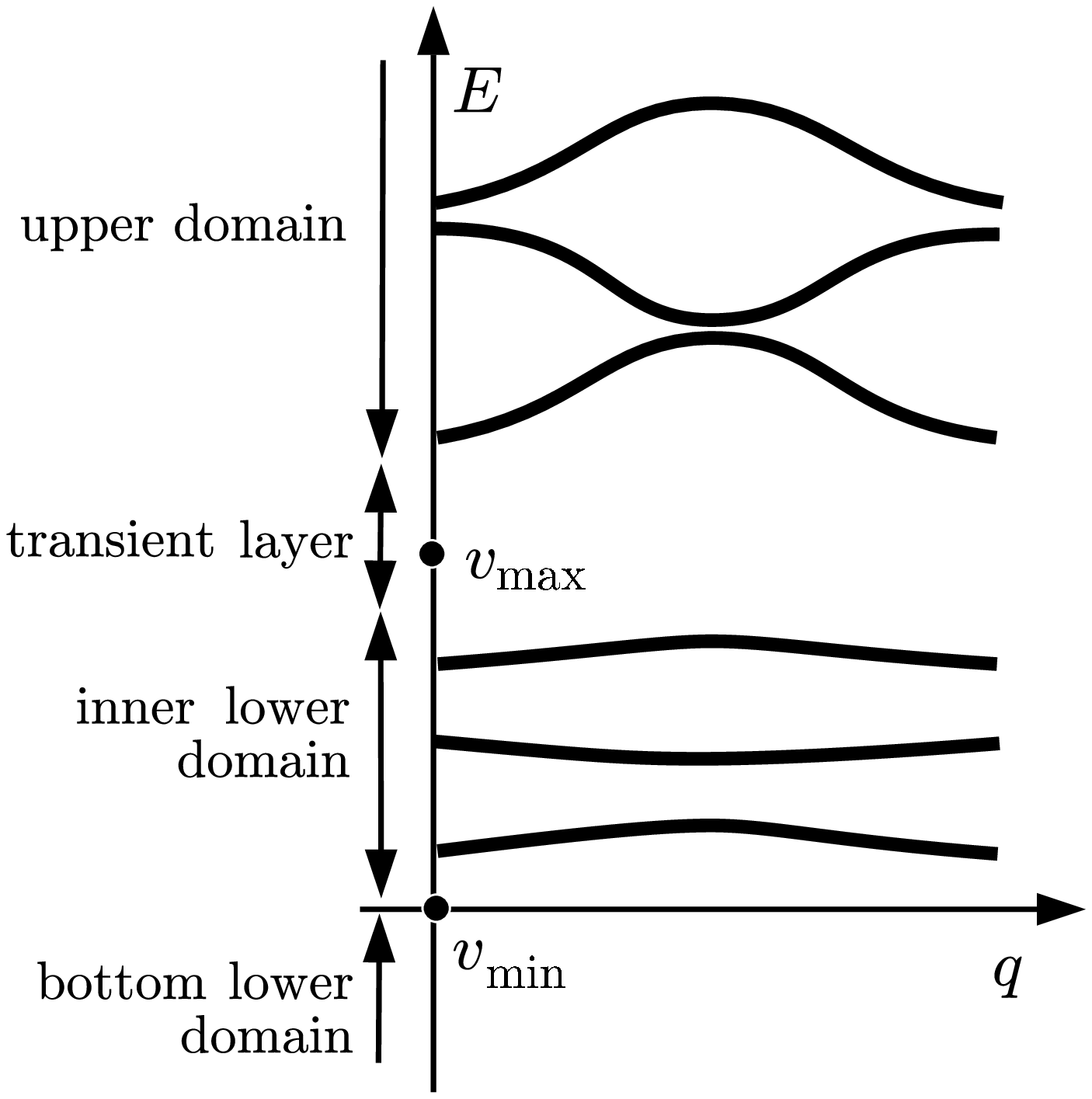,height=70mm}
\caption{Asymptotics of the dispersion relations}
\label{f2.2}
\end{minipage}
\end{figure}

To explain what kind of results we obtain for the two-dimensional
Schr\"odinger operator $\widehat H$, let us consider the spectral problem
in $L^2(\RRR_x)$ with a~small parameter $h>0$,
\begin{equation}
\widehat{L}\Psi(x) = -h^2\frac {d^2\Psi(x)}{dx^2}+v(x)\Psi(x) =E\Psi(x),
                        \label{2.1}
\end{equation}
where $v(x)$ is a smooth  $2\pi-$periodic function. The structure of
the spectrum of $\widehat L$ is well known, but the presence
of the~semiclassical parameter $h$ introduces certain additional aspects and allows
in particular to construct certain explicit semiclassical
asymptotic formulas for the spectrum.
According to the theory developed by
Floquet, Krein, Gelfand (we refer to the original works~\cite{Kr,Gel}
and to the reviews \cite{Marc,Fe,KMS,RS,Sk}) the
spectrum of \eqref{2.1} is continuous and, along
the energy axis, separated into
bands $\Delta _{\nu}=[E_{\nu}^-,E_{\nu}^+]$
and gaps $(E_{\nu}^+,E_{\nu+1}^-)$, $\nu=1,2,\cdots $,
$E^-_{\nu}<E^+_{\nu} \le E^-_{\nu+1}$, $E^-_0 > \min v$.
(Some gaps may be closed, such that $E_{\nu}^+=E_{\nu+1}^-$.)
Each point $E\in (E_{\nu}^-,E_{\nu}^+)$ has multiplicity two,
and \eqref{2.1} has two linear independent Floquet (or Bloch) solutions.
It is convenient to parameterize the points in each band by the {\it
quasi momentum} $q$, $0\le q \le 1$
(under the assumption that one separates the gaps in cases when $E_{\nu}^+=
E_{\nu+1}^-$),
and to write the \emph{dispersion relations} as
\begin{equation}
E=E_{\nu}(q).
                            \label{2.2}
\end{equation}
To each point $q\in [0,1]$ corresponds a~Bloch function,
i.~e. a solution $\Psi_ {\nu}(x,q)$ of \eqref{2.1} with
$E=E_\nu(q)$   satisfying the  Bloch condition
\begin{equation}
\Psi_{\nu}(x+2\pi,q)=e ^{2\pi iq}\Psi_{\nu}(x,q).
                              \label{2.3}
\end{equation}
Of course, the functions $E_{\nu}$ and $\Psi_{\nu}(x,q)$
depend on $h$, but to simplify the notation
\emph{we omit this dependence.}
The points $q=0$ and $q=1/2$, ($q=1$ is identified with $q=0$) correspond
to the ends of the bands
and give periodic and anti-periodic solutions of \eqref{2.1},
respectively. If $E_{\nu}^+<E_{\nu+1}^-$
then for $E=E_{\nu}^+$ and $E=E_{\nu+1}^-$,
\eqref{2.1} has only one solution, the Bloch solutions
associated to quasimomenta $q$ and $1-q$ are complex conjugate.
Clearly, $\Psi_{\nu}(x,q)\notin L_2(\RRR^1_x)$.
Typical dispersion relations are illustrated in Fig.~\ref{f2.1}.

There are no explicit analytic formulas expressing the dispersion
relations in terms of the potential $v$ even for the simplest
potentials (except
in the case of finitely many gaps, see \cite[Chapter~2]{DMN},
\cite[\S1.5]{Marc}).
Semiclassical asymptotic
formulas are available only for large $E$.
{\it We consider now the situation
when $h\to+0$ in \eqref{2.1}}. This problem was studied
in many papers and monographs,
cf. e.~g.~\cite{LP,Fe,DKM,DoSh,KMS,Ol,WK1,WK2,Hel},
see also~\cite{Mar} and references there-in,
but with a somewhat different point of view.
We recall here several results in a form suitable
for us, keeping in mind their multidimensional generalization.

\subsection{The asymptotics of the spectrum}\label{ss2.2}
To simplify the discussion we now assume that $v$ is analytic
and has only one non-degenerate minimum point $x_{\text{min}}$  on $S^1_x$;
we may and will assume $v_{\text{min}}:=v(x_{\text{min}})=0$.
Then there also exists only one
global maximum point $x_{\text{max}}$ of $v$;
we put $v_{\text{max}}=v(x_{\text{max}})$.
Under this assumption the dispersion picture is divided
in four domains. The bands situated under
$v_{\text{max}}$ become exponentially small with respect to $h$,
and the corresponding dispersion curves
are almost horizontal  segments $E_{\nu}=\text{const}\mod O(h^{\infty})$
(see Fig.~\ref{f2.2})
with distance $O(h)$ between them; this means that the number
of bands increases as  $h$ tends to zero, and $\nu$ is allowed to be large.
Let $\delta$ denote a small number independent of $h$.

\begin{figure}
\centering
\epsfig{file=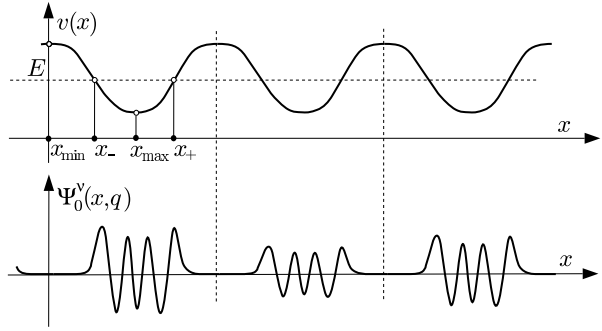,height=50mm}
\caption{Structure of Bloch functions\label{f2.3}}
\end{figure}

\begin{prop}                          \label{p2.1}
\textup{(a)}
For $E_{\nu}(q)< v_{\text{max}}-\delta$ we have
\begin{equation}
E_{\nu}(q)=E_{1,\nu}+o(h),
                                 \label{2.4}
\end{equation}
where $E_{1,\nu}$ is defined by the  {\it Bohr-Sommerfeld rule}
\begin{equation}
\frac{1}{\pi}\int_{x_-}^{x_+}\sqrt {E_{1,\nu}-v(x)}dx= h(\frac12+\nu ),
                                 \label{2.5}
\end{equation}
with $x_{\pm}(E_{1,\nu})$ solutions of the equation $v(x)=E_{1,\nu}$
\textup{(}see Fig.~\textup{\ref{f2.3}}\textup{)}.

If $\nu h$ is small enough then $E_{1,\nu}$
is also small and one can simplify~\eqref{2.5}
using a~Taylor expansion; this leads
to the ``harmonic'' oscillator approximation
for $E_{1,\nu}$,
\begin{equation}
E_{1,\nu}=h(\frac{1}{2}+\nu)\omega_0+O(h^2),\quad \omega_0=
\sqrt{2v^{\prime\prime}(x_{\text{min}})}.
                               \label{2.6}
\end{equation}

\textup{(b)}
Let $v_{\text{min}}+\delta< E_{1,\nu}< v_{\text{max}}-\delta$, then
\begin{equation}
E_{\nu}(q)-E^-_{\nu}=
\frac {\omega ( E_{1,\nu})h}{\pi}((-1)^{\nu+1}
\cos (2\pi q)+1)e^{-\rho/h}\bigl(1+O(h)\bigr).
                              \label{2.7}
\end{equation}
Here  $\omega (E)=2\pi (\int _{x_-}^{x_+}\frac{1}{\sqrt{E-v(x)}}dx)^{-1}$
is the frequency \textup{(}see subsection~\textup{\ref{ss2.4}}\textup{)},
and $\rho= \int _{x_+}^{x_-}\sqrt {v(x)-E_{1,\nu}}dx$ is
the Agmon distance~\textup{\cite{Ag}}.

If $\nu h$ is small enough then
\begin{equation}
\begin{gathered}
E_{\nu}(q)-E^-_{\nu}=
\frac {2^{4\nu+5/2}{\omega_0}^{\nu+3/2}h^{1/2-\nu}}{\nu ! \sqrt {\pi}}\times\\
{\exp} {\bigl(\frac{1}{4}(2\nu+1)\int_{x_{\text{min}}}^{x_{\text{min}}+2\pi}
(\frac {\omega_0}{\sqrt{v(x)}}-\frac{1}{\sin{(\frac{x-x_{\text{min}}}{2})}})dx \bigl )}
\times\\((-1)^{\nu+1} \cos (2\pi q)+1)e^{-\rho/h} \bigl(1+O(h^{1/2})\bigr),
\end{gathered}
                              \label{2.8}
\end{equation}
with Agmon distance $\rho=\int _{x_{\text{max}}}^{x_{\text{max}}+2\pi}\sqrt {v(x)-E_{1,\nu}}dx$.
\end{prop}

For the~proof we refer to~\cite{WK2},
see also review of~results in~\cite{Mar}.

\begin{remarks}~

\textup{(1)}
If $\nu\sim 1/h$ then $o(h)$ in \eqref{2.4} can be replaced
by $O(h^2)$; the estimate $o(h)$ appears during the passage from ``small'' to
large ``large'' $\nu$, see \cite{KMS,Mar}.

\textup{(2)}
In these asymptotic formulas the potential $v$ appears
only through the frequency and the Agmon distance:
the dependence on the quasimomentum $q$ is the same for different
potentials.

\textup{(3)}
If we \emph{formally} take the~limit $\nu h\to 0$ in~\eqref{2.7}
for  ${\nu h}<<1$, we obtain~\eqref{2.8}, but these arguments
are not rigorous, so~\eqref{2.8} has to be proved by other means.
In such a~situation we say that formula~\eqref{2.7}
\emph{allows a~formal limit}.

\textup{(4)}
The subtraction of
$1/\sin{(\frac{x-x_{\text{min}}}{2})}$ from the integrand in \eqref{2.8}
is just one possible type of regularization.

\end{remarks}

In the upper domain, on the other hand, the gaps above $v_{\text{max}}$
become exponentially small, the bands have length
$O(h)$ and almost cover  the spectral half axis.

\begin{prop}                       \label{p2.2}
For $E_{\nu}(q)> v_{\text{max}}+\delta$ we have
$E_{\nu}^+ -E_{\nu+1}^-=O_\nu(h^{\infty})$, and
the ends of the gaps are defined by a~Bohr-Sommerfeld
quantization rule in the~form
\begin{equation}
E_{\nu}^+ = \mathcal{E}_{\nu}+O_\nu(h^2),\quad
\frac{1}{{2\pi }}\int _0^{2\pi }\sqrt {\mathcal{E}_{\nu}-v(x)}dx=\frac{h\nu}{2}.
                                  \label{2.9}
\end{equation}
We also have the following asymptotic formulas  for the dispersion
relation in the $\nu$-th band \textup{(}Fig~.\textup{\ref{f2.2}}\textup{):}
\begin{equation}
E_{\nu}(q)= E_{2,\nu}(q)+O_\nu(h^2),\quad
\frac{1}{{2\pi }}\int _0^{2\pi }\sqrt {E_{2,\nu}(q)-v(x)}dx=I_{\nu}(q,h),
                                  \label{2.10}
\end{equation}
where for even $\nu=2s$
\begin{subequations}\label{2.11}
\begin{equation}
I_{\nu}(q,h)=\begin{cases}
h(\frac{\nu}{2}+q), & 0< q< \frac12,\\
h(\frac{\nu}{2}+1-q), & \frac12 < q < 1,
\end{cases}
                                 \label{2.11a}
\end{equation}
and for odd $\nu=2 s+1$
\begin{equation}
I_\nu(q,h)=\begin{cases}
h(\frac{\nu+1}{2}-q), & 0< q<\frac12,\\
h(\frac {\nu-1}{2}+q), & \frac12< q< 1.
\end{cases}
                                \label{2.11b}
\end{equation}
\end{subequations}
\end{prop}

For the proof of~\eqref{2.9}, see~\cite[Appendix B]{DoSh},
or~\cite[\S7]{Fe}. Formulas~\eqref{2.11a} and~\eqref{2.11b}
in a~slightly different form are contained in~\cite{CUMS}
or~\cite[Part~II]{KMS}.

\begin{remark}
The description of the splitting $E_{\nu+1}^--E_{\nu}^+$
between the  ends of the
gaps is not so simple in this case as in~\eqref{2.7}. In the physics
literature, this splitting is associated with the so-called
``over barrier'' reflection.
Under additional assumptions~\cite{WK1,WK2,Fe,DoSh},
the splitting has the more explicit form:
\begin{equation}
E^-_{\nu+1}-E_{\nu}^+=
\frac {\omega ( E_{1,\nu})h}{\pi}e^{-\widetilde \rho/h}\bigl(1+O(h)\bigr).
                              \label{2.12}
\end{equation}
The definition of the Agmon distance $\widetilde{\rho}$,
however, is now quite different.
\end{remark}

Of course, there is a~transient layer
in a neighborhood of $v_{\text{max}}$ where the band length is
comparable with the gap length;
formulas \eqref{2.5}, \eqref{2.7}, \eqref{2.10}, and \eqref{2.11} are not
valid there. We do not consider this situation
(see, nevertheless,~\cite{Bl,Mar,WK1}).

\subsection{Quasimodes and Bloch solutions}\label{ss2.3}
The behavior of the Bloch solutions differs sharply in
the lower and upper  domains.
Moreover, it is natural  to  isolate
a certain neighborhood of the bottom of the lower domain,
because the behavior of
the corresponding eigenfunctions is also different there.
The asymptotic formulas depend, of course, on the accuracy
of the approximation: they must be
more complicated e.~g. in case  of the subtle
dispersion relations~\eqref{2.8}, \eqref{2.9}.

Recall the following definitions (see e.~g.
\cite{MF,M1,M2,HS1,HS2,HS3,La}).
\begin{defin}                           \label{d2.1}
Let  $L>1$ be a~real number.

\begin{itemize}
\item A~pair $(\Psi^L,E^L)$ is called a \emph{formal asymptotic
solution} or \emph{quasimode of order~$L$}, relative to some function
space $\mathcal{F}$, if
\begin{equation}
\|(\widehat{L}-E^L)\Psi^L\|_{\mathcal{F}}=O(h^L).
                                \label{2.13}
\end{equation}
We can use, for example, $\mathcal{F}=C(\RRR)$,
or $\mathcal{F}=L^2(\RRR_x)$.
\item Let $\Psi$ be a solution of the~equation
$(\widehat{L}-E)\Psi=0$.
The function  $\Psi^L $ is called an~\emph{asymptotic part of order $L$} of~$\Psi$
if  $\|\Psi-\Psi^L\|_{\mathcal{F}}=O(h^L)$ as $h\to0$.

\item Let $\|\Psi^L\|\ge c>0$ as $h\to 0$.
The function $\Psi^0$ is called \emph{a~leading term
of the~quasimode} $\Psi^L$ if
$\|\Psi^L-\Psi^0\|_{\mathcal{F}}=o(1)$ as $h\to 0$.
\end{itemize}
\end{defin}

\begin{remarks}~

\textup{(1)} Note that in the~definition of~quasimode,
the Bloch condition \eqref{2.3} is not required.

\textup{(2)} Definition~\eqref{d2.1} describes so-called
``power'' or ``additive'' asymptotics; these notions
are used in contrast to ``multiplicative'' asymptotics, which
we will define later.

\textup{(3)} An~asymptotic part contains
more information about the true solution of~\eqref{2.1} than
a quasimode, even though both concepts can coincide
in specific examples. Nevertheless, one can derive information
about the spectrum of~$\widehat{L}$ from quasimodes.
The following proposition is essentially well known.
\end{remarks}

\begin{prop}                \label{p2.3}
Let $\Psi^L$ be a~smooth function
and $E^L\in\RRR$ with the property

\begin{itemize}
\item[(a)]
$(\Psi^L,E^L)$ is a~quasimode of~$\widehat{L}$ of order~$L$
in $L^2(\RRR)$,
and $\|\Psi\|_{L^2(\RRR)}\ge c> 0$ as~$h\to 0$\textup{;}
\item[\textit{or}] ~~
\item[(b)]
$(\Psi^L,E^L)$ is a~quasimode of~$\widehat{L}$ of~order~$L$
in~$L^2[-\pi,\pi]$, $\Psi^L$~satisfies~\eqref{2.3},
and $\|\Psi^L\|_{L_2[-\pi,\pi]}\ge c>0$
as~$h\to 0$.
\end{itemize}
Then the distance between $E^L$ and the~spectrum
of~$\widehat{L}$ is $O(h^L)$.
\end{prop}

\begin{proof}
The proof is well known for the case (a), see, for example
\cite[Lemma 1.3]{M1} or \cite[Lemma 13.1]{MF}.
Let us give the proof for the case (b),
which is a~simple generalization and probably also known.

For any function $\varphi$ satisfying~\eqref{2.3} we have
$\|\varphi\|_{L^2[-m\pi,m\pi]}=\sqrt{m}\|\varphi\|_{L^2[-\pi,\pi]}$,
so this holds, in~particular, for $\Psi^L$, for~$(\Psi^L)'$,
and for the~discrepancy $f:=(\widehat{L}-E^L)\Psi^L$.
Let us choose a~smooth cut~off function $e$ with
$0\le e\le 1$, $e(x)=1$ for $x\in(-\pi,\pi)$,
and $e(x)=0$ for~$x\notin(-2\pi,2\pi)$,
and let $|e(x)|+|e'(x)|+|e''(x)|\le c_1$ for $x\in\RRR$.
For $m\in\NNN$ we put $e_m(x):=e(x/m)$.
Since $\widehat{L}$ is self-adjoint, we have
\begin{equation}
    \label{est-0}
\|e_m \Psi^L\|_{L^2(\RRR)} \dist(\spec\widehat{L},E^L)
\le\|(\widehat{L}-E^L)(e_m\Psi^L)\|_{L^2(\RRR)},
\end{equation}
and the~left-hand side satisfies
\begin{equation}
       \label{est-1}
\|e_m\Psi^L\|_{L^2(\RRR)}\ge c\sqrt{m},
\end{equation}
by assumption.
To~estimate the right-hand side we use
\begin{equation*}
(\widehat{L}-E^L)(e_m\Psi^L)=e_m f -h^2\frac{d^2 e_m}{dx^2}\Psi^L
-2h^2 \frac{de_m}{dx}\frac{d\Psi^L}{dx}.
\end{equation*}
Then
\begin{equation}
             \label{est-2}
\|(\widehat{L}-E^L)(e_m\Psi^L)\|_{L^2(\RRR)} \le
\sqrt{2m}\|f\|_{L^2[-\pi,\pi]}+
\frac{2\sqrt{2}h^2 c_1}{m\sqrt{m}}\|\Psi^L\|_{L^2[-\pi,\pi]}+
\frac{2\sqrt{2}h^2 c_1}{\sqrt{m}}\left\|\frac{d\Psi^L}{dx}\right\|_{L^2[-\pi,\pi]}.
\end{equation}

Combining \eqref{est-0}, \eqref{est-1}, and~\eqref{est-2}
we obtain in the~limit $m\to+\infty$ the~desired estimate
\begin{equation*}
\dist(\spec\widehat{L},E^L)\le\frac{\sqrt{2}}{c}
\|(\widehat{L}-E^L)\Psi^L\|_{L^2[-\pi,\pi]}=O(h^L).
\end{equation*}
\end{proof}

To~describe the asymptotics of~Bloch functions  let us start from the
simplest level of~complexity related to~\eqref{2.6}.
One obtains the following picture: Bloch functions associated
to~the lowest bands are localized
in $O(\sqrt{h})$-neighborhoods of~the minimum points
$x_{\text{min}}+2\pi l$, $l\in\ZZZ$,
of~the potential $v$, where they coincide to first order
with the~eigenfunctions of a~harmonic oscillator.
More precisely, in some $O(\sqrt{h})$-neighborhood
of~$x_{\text{min}}$ one has the following formula for the leading term
in the asymptotics of~\emph{all} Bloch solutions:
\begin{equation}
\psi^{\nu}_0(x)=C^{\nu}\exp{(-\frac{\omega_0(x-x_{\text{min}})^2}{4h})}
H_{\nu}(\frac{\sqrt{\omega_0}(x-x_{\text{min}})}{\sqrt{2h}}),
                          \label{2.14}
\end{equation}
where $C^{\nu}$ is a~normalizing constant
and $H_{\nu}$ denotes the~$\nu$-th Hermite polynomial,
whereas the Bloch functions are $O(h^\infty)$ in all other
points of the segment $[x_{\text{max}},x_{\text{max}}+2\pi]$.
This together with the~Bloch condition
completely defines a~leading term in~suitable neighborhoods
of all other minimum points $x_{\text{min}}+2\pi l$, $l\in\ZZZ$,
by the~formula
\begin{equation}
\Psi^{\nu}_0(x,q) =\sum_{l\in\mathbb{z}}
e^{2\pi i q l}\psi^{\nu }_0(x-2\pi l).
                            \label{2.15}
\end{equation}
More precisely, for any Bloch function $\Psi^\nu$ there exists
a~function $\psi^\nu$ such that
\begin{equation}
\Psi^{\nu}(x,q) =\sum_{l=-\infty}^{\infty}
e^{2\pi i ql}\psi^\nu(x-2\pi l);
                            \label{2.16}
\end{equation}
this is the so-called Gelfand representation
(see~\cite{Gel,Sk}, \cite[XIII.16]{RS}.
Let us record the fact that
$\psi^\nu=\psi^\nu_0+O(\sqrt{h})$.

Using the~terminology introduced above, one can prove that
\eqref{2.14} gives  asymptotics of~certain quasimodes
of~order $L$ for~\eqref{2.1},
and the~functions \eqref{2.15} are the~leading terms of~asymptotics
of the~Bloch solutions.
In~a~way, \eqref{2.15} presents the~asymptotics of~modes via quasimodes,
and the~approximation \eqref{2.14} and~\eqref{2.15} allow us
to~derive~\eqref{2.6}. Note that \eqref{2.15} gives
more information about the~Bloch solutions than~\eqref{2.15}
but no better spectral information than~\eqref{2.6}.

The Bloch solutions corresponding to~the~higher
bands are localized in a~neighborhood of~the~segments
$[x_-+2 \pi l,x_+ + 2 \pi l]$, $l\in\ZZZ$,
where $x_{\pm}$ are solutions of~the~equation $v(x)=E_{1,\nu}$
introduced above; they can be represented in the~form~\eqref{2.15}.
This~means precisely that in~inner points of~the~interval $(x_-,x_+)$
a~leading term of~all Bloch solutions is~given by
\begin{equation}
\psi_0^{\nu }(x):= \frac{C^{\nu}(h)}{\bigl(E_{1,\nu}-v(x)\bigr)^{1/4}}
\big (\cos (\frac {1}{h}
\int_{x_-}^{x}
\sqrt {E_{1,\nu}-v(x)}dx+\frac{\pi}{4})
+O(h)\big),
                                 \label{2.17}
\end{equation}
with $C^{\nu}$ a~normalizing constant.

In a~neighborhood of the~turning points
$x_-$ and $x_+$, the~functions $\psi_0^{\nu }(x,h)$ are given
in~terms of~Airy functions and have large amplitudes;
but they are still $O(h^{\infty})$ outside
certain neighborhoods of the segment $[x_-,x_+]$.
Thus it follows from~\eqref{2.15} again that
there exist gaps in the~asymptotic support
of the~Bloch solutions (see below for the definition).
A~global uniform ``power'' asymptotic of~$\psi_0^{\nu}$
can be  given in terms of~Maslov's canonical
operator (we will return to~this representation later).
Using quasimodes as~before, one derives
the~spectral information given in~\eqref{2.4} and \eqref{2.5}.

It~is convenient to~use some terminology taken from
the~theory of~short-wavelength approximation in~optics
Let us consider a~certain asymptotic solution $\Psi(x,h)$.
The closure of the~domain
where $\lim_{h\to 0}\Psi(x,h)\ne0$ is called
its \emph{asymptotic support} or~\emph{light region}.
The~domain where $\Psi(x)=O(h^{\infty})$ as~$h\to 0$
is called the \emph{shadow region}.
In~some neighborhood of~the boundary
(this neighborhood is~small together with~$h$)
$\Psi(x,h)$ has order $h^L$;
sometimes this neighborhood is~called  the \emph{penumbra}
(this definition, of~course, is not rigorous).
So for~\eqref{2.14} (or~\eqref{2.15}) the~light
region is the union of the~minimum points~$x_{\text{min}}+2\pi l$, $l\in\ZZZ$,
all other points belong to the shadow region, and
the~penumbra is some neighborhood of~$\{x_{\text{min}}+2\pi l$, $l\in\ZZZ\}$.
The~light region for the asymptotic solutions related
to the~higher bands consists of the union of the segments
$[x_-+2\pi l,x_{+}+2\pi l]$, $l\in\ZZZ$,
all other points belong to the shadow region, and
the~penumbra is the~union of~certain neighborhoods
of the~turning points $x_-+2\pi l$, $x_++2\pi l$, $l\in\ZZZ$.
In~quantum mechanics, the~shadow region
is~also sometimes called the~\emph{under-barrier region.}

Now let discuss the representation~\eqref{2.15} in greater detail.
\begin{prop}                \label{p2.4}

We fix $\varepsilon>0$ and denote by $e$ some smooth cut off function
with $e(x)=1$ for $x\in(b-\varepsilon,b+2\pi+\varepsilon)$ and
$e(x)=0$ for $x\notin(b-2\varepsilon,b+2\pi+2\varepsilon)$
\textup{(}the number $b$ is defined later\textup{)}.

\textup{(a)} Let $\nu\in\NNN$ be a~fixed number,
then the~Bloch function $\Psi^{\nu}$ associated
with the~$\nu$-th band has the~form \eqref{2.16},
where $\psi^\nu(x)$ coincides up to~$O(\sqrt{h})$
with the~function~\eqref{2.14} in a~certain neighborhood of~$x_{\text{min}}$.
Outside this neighborhood, in the~interval $(b-2\varepsilon, b+2\pi+2 \varepsilon)$
we have
\begin{equation}
\begin{gathered}
\psi ^\nu(x)= e(x)
C^{\nu}\sqrt {\frac{2^{5\nu+1}\omega_0^{\nu+1}}{h^{\nu}}}\exp{\big(-\frac{1}{h}
\Phi(x)\big)} \sqrt{\frac {\tan{(\frac{x-x_{\text{min}}}{4})}}{\xi(x)}}\cdot \\
{\exp} {\big(\frac{1}{4}(2\nu+1)\int_{x_{\text{min}}}^x
(\frac {\omega_0}{\xi(x)}-\frac{1}{\sin{(\frac{x-x_{\text{min}}}{2})}})dx \big )}
(\tan{(\frac{x-x_{\text{min}}}{4})})^{\nu}
 \bigl(1+O(h)\bigr).
\end{gathered}
                                \label{2.18}
\end{equation}
Here $\Phi(x):=|\int_{x_{\text{min}}}^x \sqrt{v(x)}dx|$,
$\xi(x)=\Phi'(x)=\sqrt{v(x)}$ if $x\ge x_{\text{min}}$ and
$\xi(x)=\Phi'(x)= -\sqrt{v(x)}$ otherwise.
The~point $b$ is defined as the unique solution of the~equation
$\Phi(b)=\Phi(b+2\pi)$, i.~e.
$\int_{b}^{x_{\text{min}}} \sqrt{v(x)}dx =
\int_{x_{\text{min}}}^{b+2\pi}\sqrt{v(x)}dx$.

\textup{(b)}
Let $c_1\le h\nu\le c_2$ for some $c_1$, $c_2>0$, $\nu\in\NNN$.
Then the~Bloch function
$\Psi^{\nu}$ associated with the~$\nu$th band has the~form~\eqref{2.16},
where $\psi^\nu(x)$
coincides up to~$o(h)$ with~\eqref{2.17} in the~interior
of the~interval $(x_-,x_+)$. Moreover,
\begin{equation}
\psi ^\nu(x)=\begin{cases}
C^\nu e(x)
\dfrac{1}{2(v(x)-E_{1,\nu})^{1/4}}\exp{\left(
\frac{1}{h}\int_{x}^{x_+}\sqrt{v(x)-E_{1,\nu}}\,dx
\right)}\bigl(1+O(h)\bigr), & x>x_++\varepsilon\\[\medskipamount]
C^\nu e(x)
\dfrac{(-1)^\nu}{2(v(x)-E_{1,\nu})^{1/4}}
\exp{\left(
\frac{1}{h}\int_{x_-}^x\sqrt{v(x)-E_{1,\nu}}\,dx
\right)}\bigl(1+O(h)\bigr), & x<x_--\varepsilon,\\
\end{cases}
                                \label{2.19}
\end{equation}
and $b$ is defined by the~equation
$\int_{b}^{x_-}\sqrt{v(x)-E_{1,\nu}}\,dx=
\int_{x_+}^{b+2\pi}\sqrt{v(x)-E_{1,\nu}}\,dx$.

The~normalizing constants $C^{\nu}$ in~\eqref{2.18} and~\eqref{2.19}
are the~same as in~\eqref{2.14} and~\eqref{2.17}.

Both asymptotics~\eqref{2.18} and~\eqref{2.19} admit
differentiation with respect to~$x$,
i.e. in these formulas $\bigl(O(h)\bigr)'_x=O(h)$.
\end{prop}

For the~proof, we refer to~\cite[\S\S{} IV.1, IV.4]{Fe}.

\begin{remarks}~\par

\textup{(1)}
Formula \eqref{2.18} does not admit a~formal limit
for $x\to x_{\text{min}}$; in particular, \eqref{2.18}
contains only the highest degree term
$\bigl(\sqrt {\omega_0/2}(x-x_{\text{min}})\bigr)^{\nu}$
of the~corresponding Hermite polynomial, but
the other terms also play a~role in a~neighborhood
of the minimal points. Of course, it is possible
to~extend the~construction accordingly, but, as we will see below,
it is not necessary for obtaining the dispersion relation~\eqref{2.8}.
The presence of the $\tan$-like term in~\eqref{2.18} is
only one possible way of regularization; another
way of regularization has been used in~\cite{Hel}.

\textup{(2)}
One has different
constructions for the~Bloch functions corresponding
to~the~bottom lower and the~inner lower domains.
In the~first case, they are defined by a~real-valued phase
and decay exponentially with $h$,
while in the~second case one needs complex phases,
i.~e., in this case the Bloch functions have
both oscillating and~exponentially decaying parts.
This phenomenon reflects deep properties
of the asymptotics given by~Maslov's canonical operator;
this distinction appears more clearly in multidimensional problems.

It is necessary to~emphasize that it is not complicated
to obtain the~asymptotic formulas for the~spectrum
in this one-dimensional situation,
but it is difficult to~prove the formulas for the true asymptotics
of the~Bloch functions.
The standard way of doing this in the one-dimensional case
is based on WKB methods for ordinary differential equations
and matching
solutions in the complex plane, see \cite{WK1,WK2,Fe,Ol,Sl}.
These methods are applicable in both bottom and~inner parts
of~the~lower domain,
and allow to obtain also the~corresponding
dispersion relations. But up to now there
is no \emph{rigorous} generalization of~this method
to multidimensional problems.
On the~other hand, there are some methods (see e.g.
\cite{M1,M2,Ag,DKM,Hel,HS1,HS2,HS3,Si1,Si2}) which
are applicable also to~multidimensional spectral
problems (like tunneling problems or problems with purely imaginary phase),
but they work in the~bottom parts of~the spectrum only.
One may call all these methods ``semiclassical approximations''
(although not in sense of~\cite{LL}), because they use
certain objects from~classical mechanics.

\textup{(3)}
Usually, Bloch functions corresponding to the~same band and
to the quasimomenta $q$ and $1-q$ are normalized in such a~way
that their Wronskian is equal to~$2i$. This leads
to normalizing constants
in~\eqref{2.14} and~\eqref{2.17} which are exponentially
large in~$h$. Otherwise, the~behavior
of the~Bloch functions is quite strange:
in~each segment $[2\pi l,2\pi(l+1)]$ some their linear combination is~$O(h^\infty)$.
\emph{The~appearance of~large normalizing constants destroys this
effect.}

Clearly, the~exponential smallness
of the~lower bands makes it difficult to~calculate
the~spectrum and the~Bloch functions numerically.

\textup{(4)}
If for some solution $\Psi$ of~\eqref{2.1} we have
a~representation $\Psi=\bigl(1+O(h)\bigr)\Psi_0$ which can be
differentiated in~$x$, then the~function
$\Psi_0$ is sometimes called a~\emph{multiplicative asymptotic}
of~$\Psi$; both formulas \eqref{2.18}, \eqref{2.19} together
with~\eqref{2.16} provide examples.
In~contrast to~additive asymptotics, multiplicative
asymptotics make sense also in the~shadow region.
Multiplicative asymptotics are~sometimes also called
\emph{exponential} or~\emph{tunneling} asymptotics, because
knowing them allows to~construct
asymptotics of~the~spectrum with an~error~$O(e^{-C/h})$ which is
necessary to deal with tunneling effects.

\end{remarks}

Let us show now that from the~formulas \eqref{2.18}
and~\eqref{2.19} it is easy
to~derive the~dispersion relations~\eqref{2.7} and~\eqref{2.9}.
Of course, this can be done using matching
solutions in the~complex plane and the one-dimensional WKB method~--
as mentioned above, --
but we give here a~derivation based on the simple integral formula
suggested by I.~M.~Lifshits (see~\cite[\S VI.55, Problem 3]{LP}).

Recall that if $(\Psi_1, E_1)$ and $(\Psi_2, E_2)$ are
solutions of~\eqref{2.1} and
$\int_{\alpha}^{\beta}\Psi_1 \Psi_2 dx \ne 0$, then
\begin{equation}
E_1-E_2=h^2 \frac {\big({\Psi}_1{\Psi'}_ 2-
{\Psi}_2 {\Psi'}_1\big)\Big|^{\beta}_{\alpha}}
 {\int_{\alpha}^{\beta}\Psi_1 \Psi_2 dx}.
                                \label{2.20}
\end{equation}

Let us choose in~\eqref{2.20}
$\Psi_{i}(x)= \Psi^\nu(x,q_{i})$,
$E_{i}=E_\nu(q_{i})$, $i=1,2$, $\alpha=b$, $\beta=b+2\pi$.
Note that both $\Psi_1$ and~$\Psi_2$ are defined by~\eqref{2.16},
and thus we have
\begin{equation*}
\big({\Psi}_1{\Psi'}_ 2-{\Psi}_2 {\Psi'} _1\big)|^{b+2\pi}_{b}=
2\big(
(\psi^\nu)'(b+2\pi)\psi^\nu(b)-\psi^\nu(b+2\pi)(\psi^\nu)'(b)\big)
\bigl(\cos(2\pi q_1)-\cos(2\pi q_1)\bigr),
\end{equation*}
because the~supports of~the functions $\psi^\nu(\cdot-2\pi l)$
do not intersect.
In the~inner lower domain, for the~denominator of~\eqref{2.20} we have
\begin{equation}
                   \label{2.20-1}
\begin{gathered}
\int_{b}^{b+2\pi} \Psi_1(x) \Psi_2(x)\,dx=
\sum_{l_1,l_2=-1,0,1}e^{2\pi i (l_1 q_1+l_2 q_2)}
\int_{b}^{b+2\pi}\psi^\nu(x-2\pi l_1)\psi^\nu(x-2\pi l_2)\,dx\\
{}=\int_{b}^{b+2\pi} \bigl(\psi^\nu(x)\bigr)^2 dx+O(h^\infty)=
\int_{b}^{b+2\pi} \bigl(\psi^\nu_0(x)\bigr)^2 dx + o(h),
\end{gathered}
\end{equation}
and, therefore,
\begin{equation}
\begin{gathered}
 E_\nu(q_1)-E_\nu(q_2)=2h^2\frac{\bigl(\psi^\nu\bigr)'(b+2\pi)\psi^\nu(b)-
\psi^\nu(b+2\pi)\bigl(\psi^\nu\bigr)'(b)}
{\int_{b}^{b+2\pi}
(\psi^\nu_{0})^2 dx}\cdot \\
\bigl(\cos(2\pi q_1)-\cos(2\pi q_2)\bigr)\bigl(1+o(h)\bigr).
\end{gathered}
                             \label{2.21}
\end{equation}
For the~ground lower domain we have~\eqref{2.20-1} and~\eqref{2.21}
with $o(h)$ replaced by~$O(\sqrt{h})$.

To~prove \eqref{2.7} and~\eqref{2.8} one has only to~substitute
\eqref{2.18}, \eqref{2.19} into~\eqref{2.21}.

We observe again that the~main term of the dispersion
relation is independent of~$v$; the~potential appears only
in its coefficients
in terms of~higher order in~$h$.
To~determine the~nominator in~\eqref{2.21},
one should use a~multiplicative asymptotics,
while the denominator is defined by power asymptotics.

The~Bloch functions for~all quasimomenta
in the bands from the upper
domain are bounded as $h\to 0$ and  oscillate everywhere on $\RRR_x$,
there are no gaps in their asymptotic
support; they can be expressed by means of simple formulas outside
neighborhoods of~size~$O(h^\infty)$
of the~points $q=0,\frac12$:
\begin{equation}
\Psi (x,q)=C^{\nu,\pm}(q)\left(\frac{\exp\left(\pm \frac {i}{h}
\int_a^x\sqrt {E_{2,\nu}(q,h)-v(x)}dx\right)}
{\bigl(E_{2,\nu}(q,h)-v(x)\bigr)^{1/4}}+O(h)\right).
                                    \label{2.22}
\end{equation}

Here  $C^{\nu,\pm}(q)$ and $a$ are normalizing constants, $E_{2,\nu}(q)$ is defined by
\eqref{2.10} and~\eqref{2.11}, and one has to~take signs + and
-- according to $q\in(\frac12,1)$ and  $q\in (0,\frac12)$,
respectively.

The~formula~\eqref{2.22} does not give asymptotics
of the~solutions of~\eqref{2.1} in the points
$q=0, \frac12$, i.~e. in the ends of the bands. 
Due to~resonances and~tunnel
effects between these points, the~asymptotics of the true eigenfunctions
(the~periodic and~antiperiodic solutions) are given by~the~even
and~odd combinations of~the functions~\eqref{2.22} provided
that the~constants $C^\nu$ and $a$~in~\eqref{2.22}
are chosen appropriately, see~\cite{DoSh} for~details.
In~some cases, the~constants $C^{\nu,\pm}$ can be
expressed through each other; to do this one can
normalize the~corresponding Bloch function $\Psi^\nu(x,q)$
by the~condition $\Psi^\nu(0,q)=1$ (see~\cite{DMN}).

So~for these eigenvalues, \eqref{2.22} defines
quasimodes but not asymptotics of the modes.

The~Bloch functions related to~the~transient domain are close
to the latter ones, but in a~neighborhood of the critical points
$x_{\text{max}}+2 \pi l$, $l\in\ZZZ$, one can express them by means
of~certain special functions; e.~g. if $x_{\text{max}}$ is a non-degenerate
critical point of~$v$, then these are
the~Weber (parabolic cylinder) functions.

The~results we have mentioned so far are obtained
by a~variety of~techniques but with different levels
of~complexity. Thus it is considerably more difficult
to~derive the~dispersion relations~\eqref{2.7} and~\eqref{2.8}
with exponentially small bands~---
using ``multiplicative asymptotics'' corresponding to~tunnel
effects~--- than~\eqref{2.4} and~\eqref{2.6}.
The~analysis of~the~transient domain --- which we have not
explained here --- becomes even more complicated.

Some of~the~methods mentioned above have been extended
to~problems in~higher dimensions but not in~a~systematic way.
For such an~approach,  from the~general
philosophy of~quantum mechanics we should expect a~correspondence between
certain characteristic parts of~the~spectrum of~$\widehat{L}$ (so-called
\emph{spectral series}) and~certain
characteristic geometric objects in~the~phase space of~the~classical
motion. In~our one-dimensional example, the classical motion
is~integrable, such that inspiration gained here can be expected
to~extend at~least to~the~generic integrable case, and that is
what we want to~explain.

In~the~case at~hand, the~spectrum of~$\widehat{L}$ may be decomposed
into four domains having similar asymptotic behavior as~detailed above;
these are spectral series.
We are now going to~show that the~presence of~different types of~asymptotics
naturally corresponds to a~decomposition of~the~phase space
into ``regimes'' which each allow a~simultaneous treatment
of~the~flow. This decomposition, in~turn, is characterized by~a~single
graph which, in~this example, coincides with the~Reeb graph
of~the~corresponding classical Hamiltonian.

\subsection{The graph of the~classical motion}\label{ss2.4}
We now want to~construct classical preimages of the spectral series
described above. To do so, we
give a~suitable classification of~the~classical motion and
establish a~relationship with the ``quantum motion'' defined by~\eqref{2.1}.
Thus we have  to~consider the~corresponding classical problem
defined  by the~one-dimensional Hamiltonian
\begin{equation}
H(p,x):=p^2+v(x).
                                    \label{2.23}
\end{equation}
The related Hamiltonian system
\begin{equation}
\left.\begin{aligned}
\Dot p & =-v'(x)\\
\Dot x & =2p
\end{aligned}\right\}
\quad\Longleftrightarrow \quad \Ddot x=-2v'(x),
                                    \label{2.24}
\end{equation}
can be considered from two points of~view:
(1) as a~system with phase space $\RRR_{p,x}^2$;
(2) as a~system with phase space the cylinder
$Q^2_{p,x}:=\RRR_p\times S^1_x$, such that
$\RRR^2_{p,x}$ is the~universal covering of $Q^2_{p,x}$.

Then we can distinguish the~following types of the~trajectories:
\begin{enumerate}
\item[a)]
closed trajectories on $\RRR^2_{p,x}$ which correspond to~closed contractible
trajectories on $Q^2_{p,x}$ (on these we have $v_{\text{min}}< H< v_{\text{max}}$);
\item[b)]
open trajectories on $\RRR^2_{p,x}$ which correspond to~closed  but
not contractible trajectories on $Q^2_{p,x}$ (on these we have $H> v_{\text{max}}$);
\item[c)]
the~stable minimum points $(0,x_{\text{min}}+2\pi l)$
on $\RRR^2_{p,x}$ or on $Q^2_{p,x}$;
\item[d)]
the saddle points $(0,x_{\text{max}}+2\pi l)$ on $\RRR^2_{p,x}$ or on $Q^2_{p,x}$
and the singular manifolds (separatrices) on $\RRR^2_{p,x}$ or $Q^2_{p,x}$, which
belong to the ``singular'' energy level $v_{\text{max}}$.
\end{enumerate}

\begin{figure}
\centering
\epsfig{file=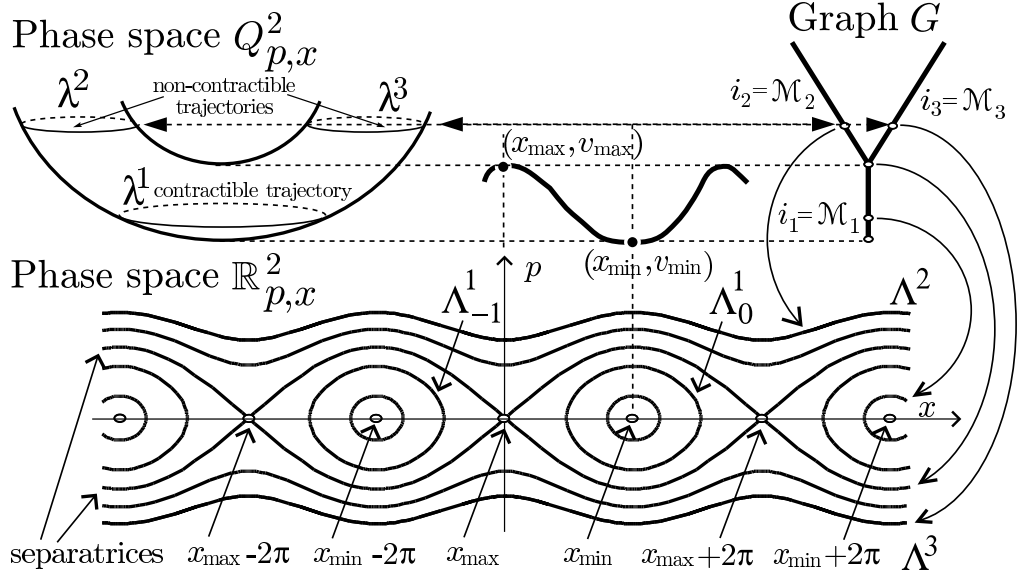,height=85mm}
\caption{Reeb graph and trajectories of Hamiltonian system\label{f2.4}}
\end{figure}

This correspondence can be easily illustrated
if one imagines that the trajectories of the Hamiltonian
system are level curves of the height function of the deformed
cylinder, see~Fig.~\ref{f2.4}.

Both phase pictures decompose qualitatively into
the~stationary point(s), the~separatrix, and the~three
connected components of~the~complement of~their union;
obviously, two of the~three components are equivalent
under the~map $p \mapsto -p$. Relating this to~the~energy
function, we see that the stationary point(s) correspond(s)
to~$v_{\text{min}}$ while the~separatrix corresponds to~$v_{\text{max}}$.
Hence the~Reeb graph~$G$~\cite{BF} of~$H$ describes the~situation nicely,
cf.~Fig.~\ref{f2.4}. The Reeb graph is constructed as follows:
each connected component of the level set of $H$ corresponds
to a~point of this graph, and connectivity in this set is
introduced in a~natural way.
In our case, the Reeb graph has four vertices, $v_{\text{min}}$,
$v_{\text{max}}$, and $\infty_{2/3}$, and edges $i_1:=(v_{\text{min}},
v_{\text{max}})$,
$i_{2/3}:=(v_{\text{max}},\infty_{2/3})$. Moreover, each edge may be
identified with an~interval of~the~energy axis.
We observe next that for each edge of~the~graph we have to~introduce
a~different action variable which we denote by~$I^r$, where $r=1,2,3$
numbers the~corresponding edge. For the~edge $i_1$ we have
\begin{equation}
I^1(H)=\frac{1}{{2\pi }}\oint_{\Lambda^1_l} pdx=
\frac{1}{\pi }\int _{x_-}^{x_+}\sqrt {H-v(x)}dx,\quad H<v_{\text{max}}.
                              \label{2.25}
\end{equation}
For $i_{2/3}$ we have:
\begin{equation}
I^{2/3}(H)=\frac{1}{2\pi}\int _0^{2\pi }\sqrt {H-v(x)}dx,\quad H> v_{\text{max}}.
                              \label{2.26}
\end{equation}
Introduce the action variable in the saddle points:
\begin{align*}
I^{1+}(v_{\text{max}}) & =\lim_{H\to v_{\text{max}}-0} I^1(H)=
\frac{1}{{\pi }}\int _0^{2\pi }\sqrt {v_{\text{max}}-v(x)}dx,\\
I^{2-/3-}(v_{\text{max}}) & =\lim_{H\to v_{\text{max}}+0} I^{2/3}(H)=
\frac{1}{2\pi}\int _0^{2\pi }\sqrt {v_{\text{max}}-v(x)}dx.
\end{align*}
Obviously, $\lim_{H\to v_{\text{max}}+0} I^1(H)=0$,
and $\lim_{H\to+\infty} I^{2/3}(H)=+\infty$,
so $I^1\in [0,I^{1+}(v_{\text{max}})]$ and
$I^{2/3}\in [I^{2+/3+},+\infty)$.
One has the~``Kirchoff law''
$I^{1+}(v_{\text{max}})=I^{2-}(v_{\text{max}})+I^{3-}(v_{\text{max}})$,
such that $I^{1+}(v_{\text{max}})=2I^{2-/3-}(v_{\text{max}})$.

Since the~functions $I^r$ are continuous
and strictly increasing, we can invert them to~find
the~dependence of~$H$ on $I$ for each edge,
\begin{equation}
H=\mathcal{H}^r(I),
                            \label{2.27}
\end{equation}
where $\mathcal{H}^2(I)=\mathcal{H}^3(I)$.
Next we will also parameterize the~trajectories by~the action
variables, separately for each edge~$i_r$, $r=1,2,3$.

We have obtained three open subsets in the~phase space~$\RRR^2$
corresponding to~the~edges $i_r$ to be denoted by~$\MMM_r$,
$r=1,2,3$; these are the~regimes mentioned above.

In~$\MMM_1$, we have only closed trajectories grouped
together by~their images in~$Q^2$. These curves can be parameterized
by~the~action variable as follows:
\begin{equation*}
\Lambda^1_l(I)=\bigl(p_{1l}(I,t),x_{1l}(I,t)\bigr),
\quad
I\in(0,I^{1+}),\, t\in[0,T(I)],\, l\in\ZZZ,
\end{equation*}
where $\bigl(p_{1l}(I,t),x_{1l}(I,t)\bigr)$ is the~solution of~\eqref{2.24}
satisfying
\begin{gather*}
p_{1l}(I,0)=0,\quad
x_{10}(I,0)=x_-\bigl(\mathcal{H}^1(I)\bigr),\\
\intertext{and}
x_{1l}(I,0)=x_{10}(I,0)+2\pi l.
\end{gather*}
Thus all the trajectories $\Lambda^1_l$ are uniquely determined
and periodic with period
\begin{equation*}
T(I)=\int_{x_-\bigl(\mathcal{H}^1(I)\bigr)}^{x_+\bigl(\mathcal{H}^1(I)\bigr)}
\frac{1}{\sqrt{\mathcal{H}^1(I)-v(x)}}\,dx;
\end{equation*}
all $\Lambda^1_l$ cover the same trajectory $\lambda^1$ on $Q^2$.

In $\MMM_2$ (and likewise in $\MMM_3$) we obtain
quite similarly families of trajectories
\begin{equation*}
\Lambda^2(I)=\bigl(p_2(I,t),x_2(I,t)\bigr),\quad
I\in (I^{2-},\infty),\,t\in\RRR,
\end{equation*}
where $\bigl(p_2(I,t),x_2(I,t)\bigr)$ is the solution of~\eqref{2.24}
satisfying
\begin{equation*}
p_{2/3}(I,0)=\pm\sqrt{\mathcal{H}^{2/3}(I)-v_{\text{max}}},\quad x_{2/3}(I,0)=x_{\text{max}}.
\end{equation*}
$\Lambda^{2/3}$ covers a unique trajectory, $\lambda^{2/3}$,
on $Q^2$ and enjoys the~periodicity property
\begin{gather*}
p_{2/3}\bigl(I,t+T(I)\bigr)=p_{2/3}\bigl(I,t\bigr),\quad
x_{2/3}\bigl(I,t+T(I)\bigr)=x_{2/3}\bigl(I,t\bigr)\pm 2\pi,\\
\intertext{where now}
T(I)=\frac{1}{2}\int_{-\pi}^{\pi}\frac{1}{\sqrt{\mathcal{H}^{2/3}(I)-v(x)}}\,dx.
\end{gather*}

All the trajectories $\Lambda^1_l$, $\Lambda^{2/3}$ are
one-dimensional Lagrangian
manifolds. The~closed curves $\Lambda^1_l$ have a~Maslov index
equal to~$2$; the~curves
$\Lambda^{2/3}$ are open and their Maslov index is not defined.

\emph{Finally, we see that the phase space is separated into regimes corresponding
to edges of the Reeb graph for~$H$. Trajectories from the same regime
have similar topological characteristics. Singular
manifolds form the~boundaries of the regimes.}

Of course, for a~general
potential~$v$ (even required to be a~Morse function) the~Reeb graph
can become very complicated. It is impossible to give a ``generic''
description of the Reeb graph because there exists no ``generic''
potential. But obviously the procedure described above is applicable
in any case.

\subsection{The relationship between the graph
and spectral asymptotics}\label{ss2.5}
Now it  is quite easy to see  that our regimes
are suitable objects for semiclassical quantization or, 
more precisely, that they explain the spectral series of
the Sturm-Liouville problem~\eqref{2.1} as described
in~\ref{ss2.2} above, corresponding to the four
energy domains. Indeed, we set up the following
relationship (Fig.~\ref{f2.6}):

\begin{figure}\centering
\begin{minipage}{50mm}\centering
\epsfig{file=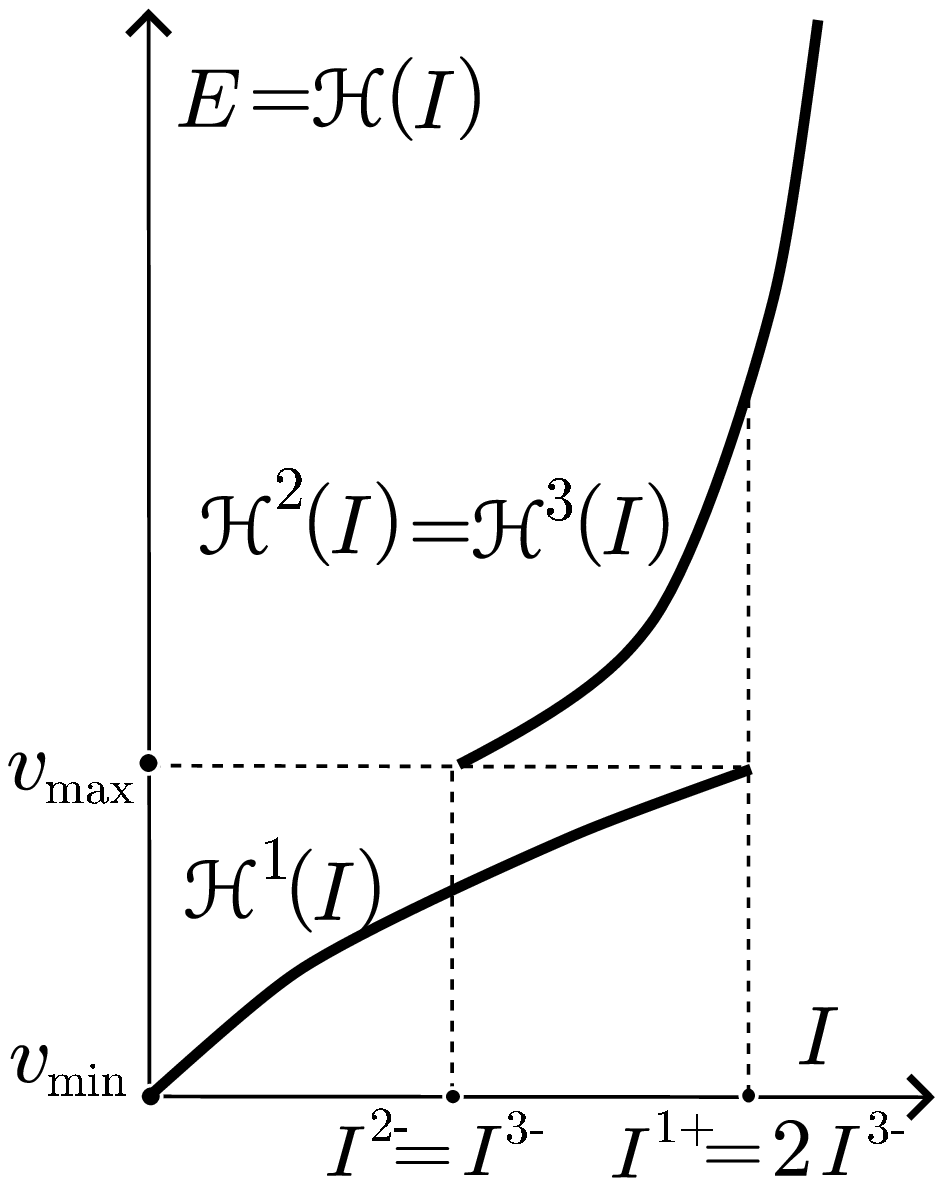,width=40mm}
\caption{Dependence of~the energy on the~action}
\label{f2.5}
\end{minipage}
\begin{minipage}{110mm}\centering
\epsfig{file=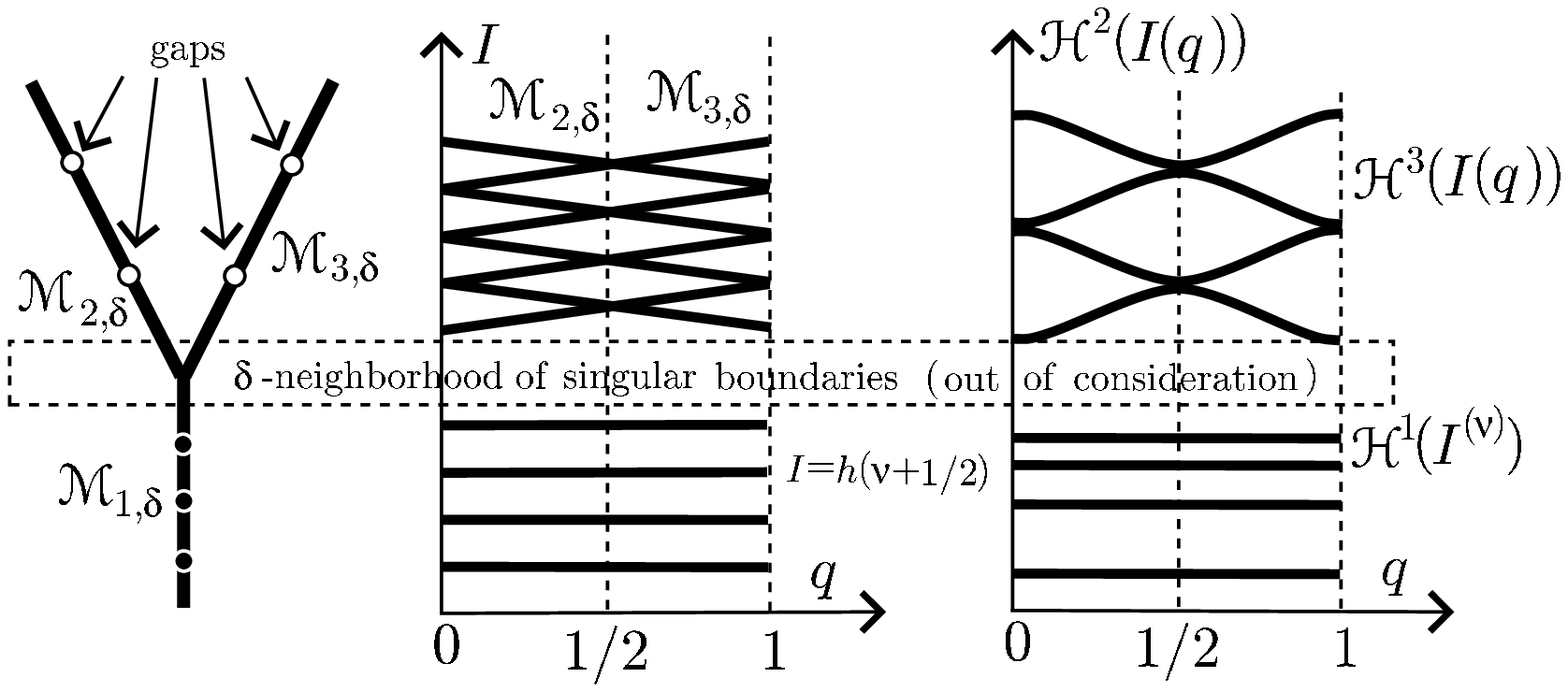,width=105mm}
\caption{Relationship between the Reeb graph
and spectral series\label{f2.6}}
\end{minipage}
\end{figure}

\begin{center}
\begin{tabular}{p{50mm}@{ $\longleftrightarrow$ }p{90mm}}
bottom lower domain &
bottom part of the regime $\MMM_1$; \\
inner lower domain  &
inner part of the regime $\MMM_1$;\\
``transient'' layer &  some
small neighborhood of the boundary between $\MMM_1$,
and $\MMM_{2/3}$;\\
upper domain & the regimes $\MMM_2$ and $\MMM_3$.
\end{tabular}
\end{center}

Keeping in mind the previous explanation concerning the ``transient''
layer  let us introduce new regimes $\MMM_{1,\delta}$,
$\MMM_{r,\delta}$, $r=2,3$  which are the ``old'' regimes
but without
certain $\delta$-neighborhoods of the singular points $I=I^{1+}$,
$I=I^{2-/3-}$;
we will describe the semiclassical quantization in these domains.

Consider first the regime $\MMM_{1,\delta}$ (related to the edge $i_1$).
The~Bohr-Sommerfeld rule \eqref{2.5} in this situation may be rewritten in
the form
\begin{equation}
I=I^{(\nu)}\equiv h \Big(\frac12+\nu\Big)
                                     \label{2.28}
\end{equation}
and gives the ``quantized regime'' or the spectral series corresponding
the the regime $\MMM_{1,\delta}$ (or to the edge $i_1$). The non-negative
integers $\nu$ are chosen in such a way
that $I^{(\nu)}\in [0,I^{1+}-\delta]$. Hence $\nu \sim 1/h$ if $I>\delta>0$.
The~map~\eqref{2.27} together with the closed curves $\Lambda ^1_l(I^{(\nu)})$
gives  the set $E_{1,\nu}=\mathcal{H}^r(I^{(\nu)})$ of ``asymptotic''
eigenvalues \eqref{2.5} and the quasimodes for the operator
$\widehat L$; in fact, one can prove that  for each $L,l\in\NNN$
one can find the numbers (independent of $l$)
\begin{equation}
E^L_{1,\nu}=E_{1,\nu}+O(h^2)
                                  \label{2.29}
\end{equation}
and a~family of
quasimodes $\psi_l^{\nu,L }(x,h)$ of \eqref{2.1} of order~$L$
such that
\begin{equation}
                  \label{supp-lim}
\supp_{h\to 0}\psi_l^{\nu,L}(x)\to \pi_x\Lambda^1_l,
\end{equation}
where $\pi_x$ denotes the projection onto the $x$-plane\footnote{%
Eq.~\eqref{supp-lim} means that $\psi_l^{\nu,L}(x)\to0$
as $h\to0$ for any $x\notin \pi_x\Lambda^1_l$.}.
This general construction
is well known (see e.g. \cite{Fe,MF,KMS}) and may be carried out
using Maslov's canonical
operator $\mathcal K_{\Lambda^1_l(I^{(\nu)})}$ on the curve $\Lambda^1_l(I^{(\nu)})$
(for $I^{(\nu)}>\kappa>0$ one can use the real canonical
operator, in case $I^{(\nu)}\to 0$ it is necessary
to use the complex canonical operator).
For the leading term one has
\begin{equation}
\psi_l^{\nu }(x,h)=\mathcal K_{\Lambda^1_l(I^{(\nu)})}\cdot 1.
                           \label{2.30}
\end{equation}
If $I^{(\nu)}$ is small, then the last formula  for $l=0$ is \eqref{2.14}. If
$I^{(\nu)}>\kappa >0$,
then outside some  small neighborhood
of the intervals $[x_-+2\pi l, x_++2\pi l]=
\pi_x\Lambda ^1_l(I^{(\nu)})$ we have $\psi _l^{\nu }(x,h)=O(h^{\infty})$,
and
for $\psi_0^{\nu }(x,h)$ one has  formula \eqref{2.16} in
the inner points of the interval $(x_-,x_+)$.

One may also express $\psi_l^{\nu }(x,h)$ via
$\psi _0^{\nu }(x,h)$ by the formula
\begin{equation}
\psi _l^{\nu }(x,h)=\psi _0^{\nu }(x-2\pi l,h).
                                \label{2.31}
\end{equation}
Now let us return to quasimodes and the~Bloch conditions~\eqref{2.3}.
We know that~\eqref{2.15} holds in the case at hand,
but its proof uses additional nontrivial asymptotic
considerations, and some of them are not yet available
in multidimensional situations.
But let us give some simple heuristic and purely algebraic
argument how to obtain~\eqref{2.15} with the ansatz
\begin{equation}
\Psi ^{\nu }_0(x,q,h) =\sum_{l=-\infty}^{\infty}
C_l(q,h)\psi ^{\nu }_0(x-2\pi l,h),\quad
                                     \label{2.32}
\end{equation}
where $C_l(q,h)$ are unknown coefficients.
Requiring the Bloch condition we find
\begin{equation*}
\sum _{l=-\infty}^{\infty}C_l(q)\psi ^{\nu }_0(x-2\pi (l-1),h)=
e^{2\pi iq}\sum _{l=-\infty}^{\infty}
C_l(q)\psi ^{\nu }_0(x-2\pi l,h).
\end{equation*}
If the system $\bigl(\psi^{\nu }_0(x-2\pi l,h)\bigr)_{l\in\ZZZ}$
has suitable basis properties, we conclude
\begin{equation}
C_{l+1}(q)=C_l(q)e^{2\pi i q},
                                    \label{2.33}
\end{equation}
hence $C_l=e^{2\pi i q l}C^{\nu}$,
where $C^{\nu}$ is a normalizing constant and $q\in[0,1)$, and we
obtain formulas~\eqref{2.15}, \eqref{2.16}, and as
corollary {\it the structure of the dispersion relation
\eqref{2.21}}. This consideration does not depend on $L$,
the degree of approximation.

\emph{So in this case the used semiclassical
method gives a~$O(h^\infty)$-approximation
of the dispersion relations and
the asymptotics for the Bloch functions.}

Now consider the regimes $\MMM_{r,\delta}$
corresponding to the edges $i_r$, $r=2,3$.
There are  no cycles on
$\Lambda^{2,3}(I)$, and for each $I\in \MMM_{r,\delta}$
and arbitrary large $L$
one can write the following formula for the asymptotic solutions:
\begin{equation}
\psi^{\pm }(x,h,E)=C^{\pm}\left(\frac {\exp\left(\pm \frac {i}{h}
\int_{a}^x\sqrt {E-v(x)}dx\right)}
{\bigl(E-v(x)\bigr)^{1/4}}+O(h)\right),\quad E>v_{\text{max}}+\delta,
                                 \label{2.34}
\end{equation}
where the sign + corresponds to $\MMM_2$,
the sign -- to $\MMM_3$,
and $a$ and $C^\pm$ are some constants.
The function $\psi^\pm$ is associated with the spectral value
\begin{equation}
E^L(I,h)=\mathcal{H}^2(I)+O(h^2)=
\mathcal{H}^3(I)+O(h^2).
                              \label{2.34b}
\end{equation}
Requiring now the Bloch condition for the functions $\psi^\pm$,
we derive the dependence of~$I$ on $q$ as
\begin{equation}
                 \label{I-q}
I^2(q,h)=h(n+q), \quad I^3(q,h)=h(n-q), \quad n\in\ZZZ.
\end{equation}
This dependence also implies the dependence of the energy on the quasimomenta,
\begin{equation}
                 \label{E-q}
E^{2/3,L}(q)=E^L\bigl(I^{2/3}(q,h),h\bigr).
\end{equation}
Recall that the points of the spectrum corresponding
to periodic and anti-periodic
solutions of~\eqref{2.1} lie on the ends of the bands.
Applying this fact to the function~\eqref{2.34}
one immediately obtains the points
\begin{equation}
I^{(\nu)}=h\nu/2, \quad \nu\in\ZZZ
                              \label{2.35}
\end{equation}
from $\MMM_{r,\delta}$, $r=2,3$;
the corresponding energy levels $E^L\bigl(I^{(\nu)}\bigr)$
are therefore $O(h^\infty)$-approximations of the gaps.

Combining now~\eqref{I-q}, \eqref{E-q}, and~\eqref{2.35},
we arrive at~the dispersion relations~\eqref{2.10}--\eqref{2.12}.

Note that points~\eqref{2.35} with even $\nu$
may be obtained by means of the Bohr-Sommerfeld quantization
of the non-contractible closed preimages
of~${\Lambda}^{2,3}(I)$ on the cylinder $Q^2_{p,x}$.
This fact has a rather simple explanation: these
points correspond to periodic solutions of~\eqref{2.1},
and only these Bloch solutions descend to functions
on the cylinder~$Q^2_{p,x}$.
Anti-periodic solutions do not descend to functions on~$Q^2_{p,x}$,
but only to the enlarged cylinder $\widetilde{Q}^2_{p,x}=\RRR_p\times S^1_x$,
$x\in[0,4\pi)$. The Bohr-Sommerfeld quantization rule
on $\widetilde{Q}^2_{p,x}$ then gives exactly the points~\eqref{2.35}.

Figure~\ref{f2.6} shows the relationship between action
variables, quasimomenta and energy.
This picture together with formulas \eqref{2.4}--\eqref{2.6},
\eqref{2.9}--\eqref{2.11}, \eqref{2.15}, \eqref{2.21}
contains the maximal information about the  spectrum of $\widehat L$
which can be derived from additive asymptotics.

\begin{figure}
\centering
\epsfig{file=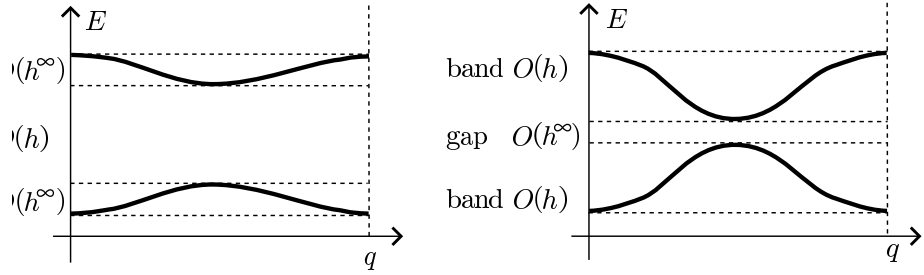,height=40mm}
\caption{Exact dispersion relations\label{f2.7}}
\end{figure}

The precise structure of the dispersion relations is sketched
in~Fig.~\ref{f2.7}; it is not accessible in details by
these methods.

\subsection{The Weil formula}\label{ss2.6}
To conclude this section,
let us suggest a~heuristic method for calculating the number $N(E)$
of bands on the half-line $(-\infty,E)$ (the \emph{Weyl formula}).

If $E\le v_{\text{min}}$, there are no real trajectories of the
Hamiltonian $H(p,x)$, no points on the graph $G$ and $N(E)=0$.

If $E\in(v_{\text{min}},v_{\text{max}}-\delta)$, then the number of bands approximately
coincides with the number of the Bohr-Sommerfeld points
$I_{\nu}$ \eqref{2.28} in the interval $(v_{\text{min}},E)$, i.e. with $I(E)/h$.

If $E\ge v_{\text{max}}+\delta$,  then we have gaps on the edges $i_2$ and~$i_3$,
but by symmetry their projections to the energy axis coincide
and one has to take into account only one edge, say $i_2$,
which gives $N(E)=I^{1+}/h+2\bigl(I(E)-I^{2-}\bigr)/h=2I(E)/h$.

Last two formulas have common geometric
interpretation: $hN(E)$ is approximately equal
to the square of the set $0\le x \le 2\pi$, $H(p,x)\le E$,
i.e. the set covered by the~trajectories of~\eqref{2.24}
with energy not greater than $E$.

\section{Classical averaging}\label{s3}
Now we return to the spectral problem of the magnetic Schr\"odinger
operator~\eqref{1.1}. We will use ideas closed to those
collected in the previous section, but we will start directly
with the classical problem. First we want to show that
the presence of the small parameter $\varepsilon$ renders
the almost integrable system. Basing on this fact,
we give a global geometric classification of the classical motion
in the following sections.

Consider the classical problem in the phase space $\RRR^4_{p,x}$
induced by the operator $\widehat{H}$ and defined by the Hamiltonian~\eqref{1.4}:

\begin{equation}
H=H_0+\varepsilon v(x_1,x_2),\qquad H_0=\frac12(p_1+x_2)^2+\frac12p^2_2.
                            \label{3.1}
\end{equation}
The projections of the trajectories of the~Hamiltonian system with free
Hamiltonian $H_0$ onto the $(x_1,x_2)$-plane are the cyclotron circles,
see e.g. \cite{AF,NY,BP,BZ,Li}, and they induce
new canonical variables in the phase space:
generalized momenta $I_1$, $y_1$ (or $P$, $y_1$)
and generalized positions $\varphi_1$, $y_2$ or ($Q$, $y_2$):
\begin{equation}
\begin{gathered}
x_1=Q+y_1,\quad p_1=-y_2,\quad x_2=P+y_2,\quad p_2=-Q,\\
P=\sqrt{2I_1}\cos \varphi_1,\quad Q=\sqrt{2I_1}\sin \varphi_1,
\end{gathered}
                           \label{3.2}
\end{equation}
such that
\begin{equation*}
dp_1\wedge dx_1+ dp_2\wedge dx_2=dI_1\wedge d\varphi_1 +dy_1\wedge dy_2=
dP\wedge dQ +dy_1\wedge dy_2.
\end{equation*}
The variables $P$, $Q$ (or $I_1$, $\varphi$) describe
fast rotating motion around slow guiding center with coordinates $y_1$,
$y_2$~\cite{Li}.

In these variables, the Hamiltonian~$H$ takes the form
\begin{equation}
H=I_1+\varepsilon v(\sqrt{2I_1}\sin\varphi_1+y_1,\sqrt{2I_1}\cos\varphi_1+y_2),
                          \label{3.3}
\end{equation}
and furnishes probably the simplest example where the averaging methods
(see e.g.\cite{A1,BM,BP,Ne}) can be successfully applied.
The averaging procedure for the Hamiltonian $H$
was first applied by van~Alfven~\cite{AF};
later, it was used in numerous works (usually, not in the variables
$(I_1,\varphi_1,y)$, $y=(y_1,y_2)$, see e.g. \cite{A2,BP,BM,BZ,Li,MS,NY}).
Our goal here is to obtain some elementary formulas
for the averaged Hamiltonian, which are probably new,
and to give a~global interpretation
of the averaged motion basing on the~geometrical
and topological approaches to integrable
Hamiltonian systems developed in~\cite{BF,BF-1,Fom}.
We are also going to show that general result~\cite{Ne}
gives probably the most complete statement
about the averaging for $H$; it seems that the variables
$(I_1,\varphi_1,y)$ are most convenient for
the analysis involved

To simplify further formulas, let us introduce
the~averaged potential $\Bar{v}$.
Expand $v$ into the Fourier series:
\begin{equation}
\begin{aligned}
v(x_1,x_2)& =v(Q+y_1,P+y_2)\\
         & =v(\sqrt{2I_1}\sin\varphi_1+y_1, \sqrt{2I_1}\cos\varphi_1+y_2)\\
         & =\sum_{k=(k_1,k_2)\in\ZZZ^2}
v_k \exp \bigg[ i\Big(
k_1\big(\sqrt{2I_1}\sin\varphi_1+y_1-\frac{2\pi a_{21}}{a_{22}}
(\sqrt{2I_1}\cos\varphi_1+y_2)\big)\\
&\quad\quad+k_2 \frac{2\pi}{a_{22}}\big(\sqrt{2I_1}\cos\varphi_1+y_2\big)
\Big)\bigg].
\end{aligned}
                         \label{3-1}
\end{equation}

Now let us average the potential $v$ with respect
to the angle variable $\varphi_1$:
\begin{equation}
\Bar v(I_1,y_1,y_2)=\frac{1}{2\pi}\int_0^{2\pi} v\,d\varphi_1.
                \label{3-2}
\end{equation}

Taking into account expansion~\eqref{3-1}
and using Bessel's integral representation for the Bessel
functions~\cite[no. 7.3.1]{btm}, one can rewrite~\eqref{3-2} as
\begin{equation}
\begin{aligned}
\Bar v(I_1,y_1,y_2) & = \sum_{k=(k_1,k_2)\in\ZZZ^2}v_k
            J_0(\sqrt{2I_1(k^2_1+(2\pi)^2 (k_2-k_1a_{21})^2/a^2_{22})}\\
&\quad\quad
\times\exp\left[ik_1\Big(y_1-\frac{2\pi a_{21}}{a_{22}}y_2\Big)+
ik_2\frac{2\pi y_2}{a_{22}}\right],
\end{aligned}
                \label{3-3}
\end{equation}
where $J_0$ is the Bessel function of order zero.
Using the~spectral theorem, \eqref{3-3} can be rewritten
in a~more elegant form:
\begin{equation*}
\Bar v(I_1,y_1,y_2)=J_0(\sqrt{-2I_1\Delta})v(y_1,y_2).
\end{equation*}
Here the operator $J_0(\sqrt{-2I_1\Delta})$ is
a pseudo-differential operator~\cite{shubin}.
Note that $\Bar v$ is analytical with respect to $I_1$,
because $J_0$ is an even function.

Let us formulate now our~main result on averaging.
\begin{theorem}                        \label{t3.1}
For any $\kappa>0$ there exist $\varepsilon_0>0$,
positive constants $C$ and $G$, and
a~canonical change of variables
\begin{align*}
P &=\cv{P}+\varepsilon U_1(\cv{P},\cv{Q},\cv{y}_1,\cv{y}_2,\varepsilon),
&\quad
Q &=\cv{Q}+\varepsilon U_2(\cv{P},\cv{Q},\cv{y}_1,\cv{y}_2,\varepsilon),\\
y_1 &=\cv{y}_1+\varepsilon W_1(\cv{P},\cv{Q},\cv{y}_1,\cv{y}_2,\varepsilon),
&\quad
y_2 &=\cv{y}_2+\varepsilon W_2(\cv{P},\cv{Q},\cv{y}_1,\cv{y}_2,\varepsilon),
\end{align*}
defined in the~domain $\cv{I}_1<\kappa$, 
$\varepsilon<\varepsilon_0$ (here and later
$\cv{I}_1=\frac{1}{2}(\cv{P}^2+\cv{Q}^2)$),
such that
\begin{equation*}
H=\cv{H}(\cv{I}_1,\cv{y},\varepsilon)+
e^{-C/\varepsilon}\cv{g}(\cv{P},\cv{Q},\cv{y},\varepsilon).
\end{equation*}
Here $U_{1,2}$, $W_{1,2}$, $\cv{g}$ are real analytic functions
of $\cv{P}$, $\cv{Q}$, $\cv{y}$, and
\begin{equation*}
|U_{1,2}|, |W_{1,2}|, |\cv{g}|+|\nabla_{\cv{y}}\cv{g}|\le G,
\end{equation*}
$\cv{H}$ is a~real analytic function of $\cv{I}_1$ and $\cv{y}$.
The~functions $U_{1,2}$, $W_{1,2}$, $\cv{g}$, $\cv{H}$
are periodic relative~$\cv{y}$ with periods~$(a_1,a_2)$.
In~addition, we have the estimate
\begin{equation*}
\cv{H}(\cv{I}_1,\cv{y},\varepsilon)=
\Bar{H}(\cv{I}_1,\cv{y},\varepsilon)+\varepsilon^2 g(\cv{P},\cv{Q},\cv{y},\varepsilon),
\quad
\Bar{H}(\cv{I}_1,\cv{y},\varepsilon)=
\cv{I}_1+\varepsilon\Bar{v}(\cv{I}_1,\cv{y}),
\end{equation*}
where $|g|+|\nabla_{\cv{y}}g|\le M$ for some
positive constant $M$ independent of~$\varepsilon$.
\end{theorem}

\emph{Proof} of the~theorem follows immediately
from the~general result of Neishtadt~\cite{Ne}
in the domain $\cv{I}_1>\kappa_0>0$.
To~include in~our consideration
the~neighborhood of~$I_1=0$, we need some its modification
based on some special choice of~generating function
of~the~requested transformation.
On the first step, one has to find
a canonical change of variables
$(P,Q,y)\mapsto(P',Q',y')$ that reduces the Hamiltonian to
the form
\begin{equation}
              \label{aver-gen}
H=H'(I_1',y',\varepsilon)+\varepsilon^2 g(P',Q',y',\varepsilon),
\end{equation}
where $I'_1=\bigl((P')^2)+(Q')^2\bigr)/2$, and $g=O(1)$ as $\varepsilon$
tends to~$0$.
Let us try to find this change of variables
using generating function $S(P',Q,y'_1,y_2)=P'Q+y'_1y_2+
\varepsilon s(P',Q,y'_1,y_2)$ from the equations
\begin{equation}
              \label{aver-chan}
P=P'+\varepsilon\frac{\partial s}{\partial Q}, \quad
Q'=Q+\varepsilon\frac{\partial s}{\partial P'}, \quad
y_1=y'_1+\varepsilon\frac{\partial s}{\partial y_2}, \quad
y'_2=y_2+\varepsilon\frac{\partial s}{\partial y'_1}.
\end{equation}
Substituting~\eqref{aver-chan} into~\eqref{aver-gen},
one obtains the following condition on~$s$:
\begin{equation}
               \label{aver-s-cond}
Q\frac{\partial s}{\partial P'}-P'\frac{\partial s}{\partial Q}=
\widetilde{v}(P',Q,y'_1,y_2),
\end{equation}
where
\begin{equation*}
\widetilde{v}(P',Q,y'_1,y_2)=\Bar{v}\Big(\frac{1}{2}\bigl((P')^2+Q^2\bigr),
y'_1,y_2\Big)-v(Q+y'_1,P'+y_2).
\end{equation*}
Introducing polar coordinates $I,\psi$ by the~equalities
$P'=\sqrt{2I}\cos\psi$, $Q=\sqrt{2I}\sin\psi$,
one can rewrite~\eqref{aver-s-cond} in the form $s'_{\psi}=\widetilde{v}$.
General solution of this equation can be written as
$s=\int \widetilde{v}\,d\psi$, but the function $s$ can
be non-analytical relative $P'$ and~$Q$; to avoid this,
one should choose
the integration constant in a~special way, for example,
\begin{multline*}
s(P',Q,y'_1,y_2)=
\frac{1}{2}\bigg(
\int_0^\psi \widetilde{v}(\sqrt{2I}\cos\varphi,\sqrt{2I}\sin\varphi,
y'_1,y_2)\,d\varphi\\
{}+\int_\pi^\psi \widetilde{v}(\sqrt{2I}\cos\varphi,\sqrt{2I}\sin\varphi,
y'_1,y_2)\,d\varphi\bigg)\Bigg|_{\substack{P'=\sqrt{2I}\cos\psi,\\
Q=\sqrt{2I}\sin\psi}}.
\end{multline*}
This procedure is then repeated, and the Neishtadt estimations~\cite{Ne}
are used, see~\cite{BDP} or~\cite{gl} for details.
Note that on the first step described above one has
$H'=I'_1+\varepsilon\Bar{v}(I'_1,y')$.

We illustrate the above consideration in the special
case of example~\eqref{1.6}. Then we find
\begin{equation}
\Bar{v}(I_1,y)=A J_0(\sqrt{2I_1})\cos y_1+
B J_0(\beta \sqrt{2 I_1})\cos (\beta y_2).
                                \label{3.7}
\end{equation}
Properties of the Bessel functions~\cite[no.~7.4]{btm} give the estimations
\begin{align*}
\Bar{v}(I_1,y) &=A(1-\frac{1}{2}I_1) \cos y_1+B(1-\frac{1}{2}\beta^2I_1)
\cos (\beta y_2) +O(I_1^2), & \quad \text{as } I_1\to +0,\\
\Bar{v}(I_1,y) &=A\sqrt{\frac{2}{\pi\sqrt {2I_1}}}\cos (\sqrt {2I_1}-\pi/4)
\cos y_1\\
&\quad{}+\frac{B}{\sqrt{\beta}}\cos (\beta \sqrt {2I_1}-\pi/4)
\cos(\beta y_2)+O(I_1^{-3/4}),& \quad \text{as } I_1\to \infty.
\end{align*}

\section{Classification of~the~averaged motions}\label{s4}

\subsection{A~one-dimensional Hamiltonian system for the~drift}\label{ss4.1}

Since the~function~$\cv{H}$ is a~periodic function of~$\cv{y}$,
it can be viewed as defining a~Hamiltonian system in two different
phase spaces, namely:
\begin{itemize}
\item[\textrm{(1)}] in Euclidean phase space $\Phi =\RRR^4_{\cv{P},\cv{Q},\cv{y}}
=\RRR^4_{p,x}$ and
\item[\textrm{(2)}] in the phase space $\Phi =\RRR^2_{\cv{P},\cv{Q}}\times \TTT^2_{\cv{y}}$.
\end{itemize}
Obviously, these systems are~integrable
and equivalent to the~equations
\begin{gather}
\cv{I}_1= \text{const} \ge 0,
                                 \label{4.1}\\
\Dot{\cv{y}}=\widehat{J}\nabla_{\cv{y}}\cv{H}(\cv{I}_1,\cv{y},\varepsilon), \quad
\widehat{J}=\begin{pmatrix} 0 & -1\\ 1 & 0\end{pmatrix}.
                                 \label{4.2}
\end{gather}

Eq.~\eqref{4.1} defines a~family of ``cyclotron'' circles $S_C(\cv{I}_1)$,
$\cv{I}_1\in[0,\infty)$, in the coordinates $(\cv{P},\cv{Q})$.
The~boundary $\cv{I}_1=0$ of this~family is the~rest point $\cv{P}=\cv{Q}=0$.
For each fixed $\cv{I}_1$, \eqref{4.2} is a~one-dimensional
Hamiltonian system. The~trajectories
of~\eqref{4.2} are the~connected components of the level sets of~$\cv{H}$.
Clearly, the solutions depend also
on $\varepsilon$, the action $\cv{I}_1$, and other parameters, but
we omit this dependence to simplify the notation.
The system~\eqref{4.2} describes the~slow drift of~the~centers
of the~``cyclotron'' circles on the~plane $\RRR_x^2$.

It is now convenient to~describe the trajectories using
the topological and geometric theory of integrable systems 
developed in~\cite{BF}. One may consider $\cv{H}$ as a~Morse-Bott
function on three-dimensional surface $\cv{H}(\cv{P},\cv{Q},\cv{y})=E$
\cite[\S1.8]{BF-1},
or one may consider $\cv{H}$, for each fixed $\cv{I}_1$,
as a~function of~variables $\cv{y}$.
In the~latter case, we suppose that~$\cv{H}$
has only a finite number of non-degenerate
critical points in the elementary cell
(for generic potential~$v$ this property
holds for almost all $\cv{I}_1$), i.e.
$\cv{H}$ is a~Morse function on the torus $\TTT^2$
covered by the plane $\RRR^2_y$,
and \eqref{4.2} is a Hamiltonian system on the torus $\TTT^2$
(or on its covering $\RRR^2_y$). We prefer this second
point of~view and find immediately
a~complete classification of~its trajectories.

\begin{prop}
                      \label{p4.1}
For any trajectory, $\gamma$, of~\eqref{4.2} on $\TTT^2$, one of the~following
assertions holds:

\begin{enumerate}
\item[(a)]
$\gamma$ is a closed contractible curve\textup{;}
\item[(b)]
$\gamma$ is a closed non-contractible curve\textup{;}
\item[(c)]
$\gamma$ is an extremum point of $\cv{H}$\textup{;}
\item[(d)]
$\gamma$ is a saddle point of~$\cv{H}$ or a~separatrix.
\end{enumerate}
\end{prop}

\begin{figure}\centering
\epsfig{file=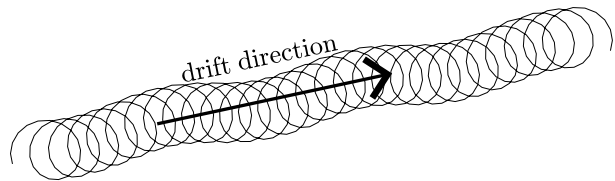,height=30mm}\\
\caption{Drift in the $x$-plane}
                         \label{f4.0}
\end{figure}

In~the~cases (a) and (b), we have periodic trajectories on the~torus.
This means, that for any trajectory $\gamma=\cv{y}(\tau)$ there is a $T>0$
(the period) and $d=(d_1,d_2)\in\ZZZ^2$ such that
$\cv{y}(\tau +T)=\cv{y}(\tau)+d\cdot a$, where
$d\cdot a=d_1a_1+d_2a_2$.
In the case (b) the vector $d$ is non-zero; moreover,
if both components $d_1$ and $d_2$ of the vector $d$
are non-zero, then they are relatively prime.
In the case (a) we have $d=0$.
Obviously, the~vector $d$ is unique for each trajectory
and does not depend on the choice of the trajectory on $\RRR^2_{\cv{y}}$
covering $\gamma$.
Moreover, as different trajectories $\cv{y}(\tau)$ cannot
intersect on the plane $\RRR^2_{\cv{y}}$, for fixed $\cv{I}_1$
exactly two non-zero vectors $d$ with mutually opposite directions
are~possible; we fix one of~them and denote it by $d(\cv{I}_1)$;
if necessary, we write $\pm d(\cv{I}_1)$.
The vector $d$ defines the ``average'' or ``main'' direction of the
motion (\emph{drift}) on the plane $(\cv{y}_1,\cv{y}_2)$ and it is one
of~topological characteristics of~$H$; the meaning of this vector
becomes even more clear if one considers the projection
of the corresponding trajectory in the original space $\RRR^4_{p,x}$
onto the $x$-plane, see Fig.~\ref{f4.0}.
The~ratio $d_1/d_2$ is called \emph{the rotation number},
see e.g. [\S1.6]\cite{BF-1}.

\begin{defin}                 \label{d4.1}
We call $d(\cv{I}_1)$ the drift vector\footnote{A closely connected notion
appears in a more complicated situation in \cite{MN}}
of the~motion.
\end{defin}

\subsection{The~Reeb graph and the~classification of the~drift
motion in~non-degenerate case}

Recall~\cite[Chapter~2]{BF-1} that it is possible to~classify Morse functions
on the~torus by~corresponding foliations, given by level curves,
such that one has a foliation with singularities; the singularities
are caused by critical points of the~Morse function.
There exist infinitely many topologically different types of such foliations
which may be classified by their Reeb graphs.
The complete theory of this classification is elegant but not trivial
(see~\cite{BF}),
and  we~restrict attention here to the~simplest situation
assuming that \emph{$\cv{H}$ is a minimal
Morse function} on the torus $\TTT^2$, i.e.
that $\cv{H}$ has exactly one maximum point $\cv{y}_{\text{max}}$ and
one minimum point $\cv{y}_{\text{min}}$ (and hence two saddle points
$\cv{y}_{\pm}$). We put $g_{\text{max,min}}:=
\cv{H}(\cv{I}_1,\cv{y}_{\text{max,min}})$,
$g_{\pm}:=\cv{H}(\cv{I}_1,\cv{y}_{\pm})$ and suppose
that $g_+\geq g_-$. The~classical motion for a fixed $\cv{I}_1$
is possible when $g_{\text{min}}\le \cv{H}\le g_{\text{max}}$.
Recall that a minimal Morse function is called \emph{simple} if $g_+>g_-$
and \emph{complex} otherwise.

\begin{figure}\centering
\begin{minipage}{50mm}\centering
\epsfig{file=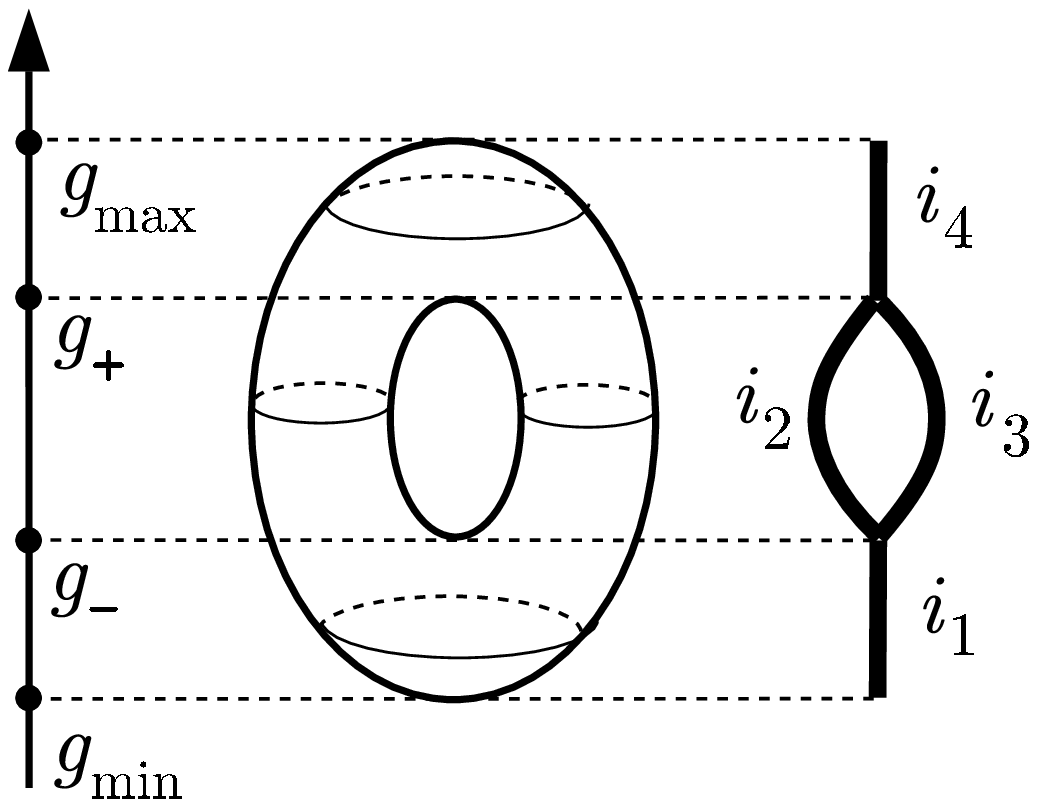,width=40mm,height=35mm}\\
a) Reeb graph
\end{minipage}
\begin{minipage}{50mm}\centering
\epsfig{file=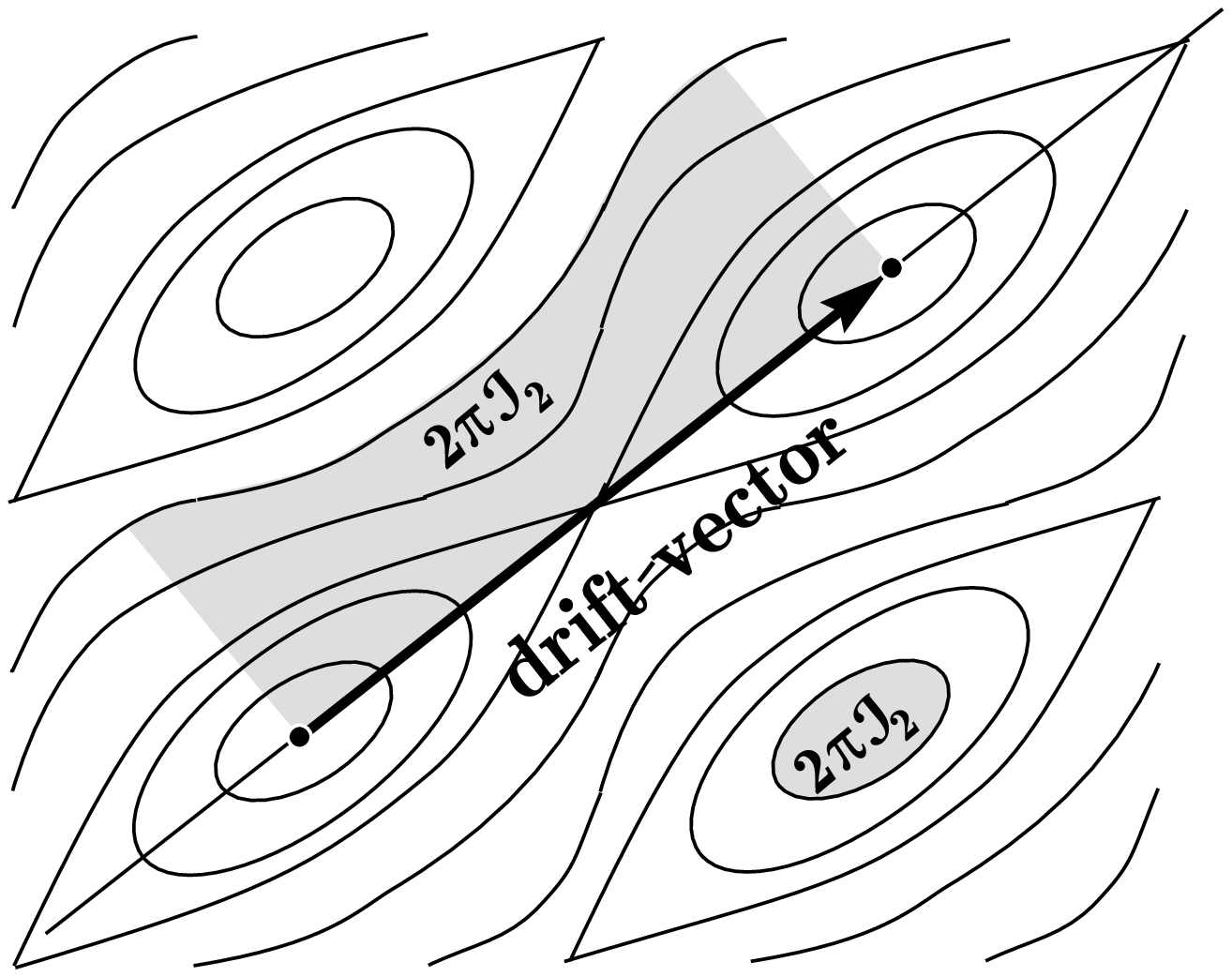,width=42mm,height=35mm}\\
b) Level curves
\end{minipage}
\begin{minipage}{45mm}\centering
\epsfig{file=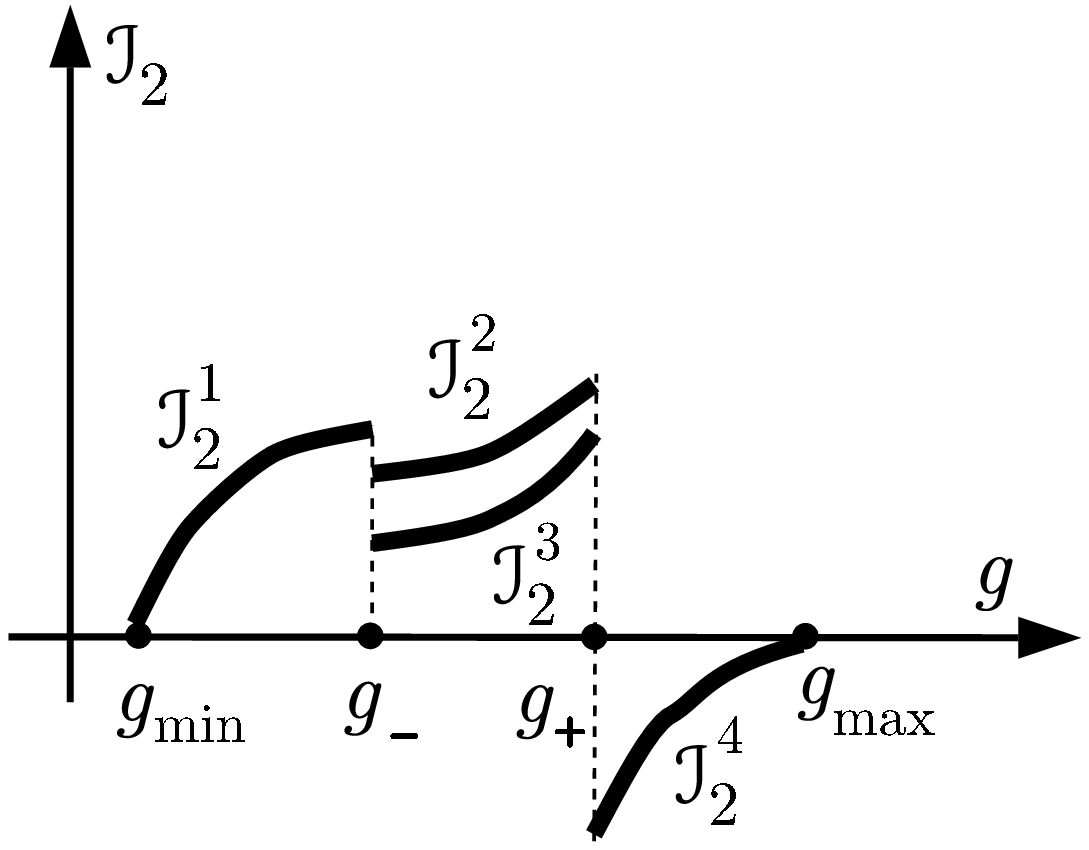,width=40mm,height=35mm}\\
c) Action variables
\end{minipage}
\caption{Characteristics of Morse function\label{f3.1}}
\end{figure}

Thus assume first that for some $\cv{I}_1$ the function $\cv{H}$
is a~simple Morse function.
It is instructive to imagine that $\cv{H}$ is a height function
on torus $\TTT^2$ (as shown in Fig.~\ref{f3.1});
then the~trajectories are the curves of constant height levels
(compare with subsection \ref{ss2.4}).
Consider the three intervals
$(g_{\text{min}},g_-)$, $(g_-,g_+)$, and $(g_+,g_{\text{max}})$.
If~$g\in (g_{\text{min}},g_-)\cup(g_+,g_{\text{max}})$, then the~set $\cv{H}=g$
includes only one connected component, which is a~closed
contractible  curve on $\TTT^2$, diffeomorphic to a~circle.
Hence we are in the case (a) of Proposition~\ref{p4.1},
and this regime of motion is described by~edges $i_1$ and $i_4$
of the~Reeb graph, respectively: each point
(denote it by $g_1$ or $g_4$) on these edges
corresponds to a~contractible trajectory $S(g_{1,4},\cv{I}_1)$
(see Fig.~\ref{f3.1}a). The~corresponding rotation number is equal
to $0/0$ and the~drift vectors are $d(i_{1,4}):=(0,0)$.
Each curve $S(g_{1,4},\cv{I}_1)\subset \TTT^2$
induces a~set of~closed trajectories $S_l(g_{1,4},\cv{I}_1)$,
$l=(l_1,l_2)\in \ZZZ^2$, on the covering (plane) $\RRR^2_{\cv{y}}$.
We will discuss the numbering of
$S_l(g_{1,4},\cv{I}_1)$ by $l$ a little later.

If $g\in (g_-,g_+)$ then the
set $\{\cv{y}\in \TTT^2:\cv{H}=g\}$ consists of~two connected components,
each of~them being again a~trajectory on the torus, diffeomorphic to a circle,
but now they are both non-contractible such that we obtain
a non-trivial rotation number $d_1/d_2$.
We in the case (b) of~Proposition~\ref{p4.1},
and each trajectory is characterized by the points~$g_2\in i_2$
and $g_3\in i_3$ on the two edges $i_2$ and $i_3$ of the~Reeb graph;
denote these trajectories on the~torus $\TTT^2$
by $\widetilde S(g_{2,3},\cv{I}_1)$. They induce a set of open
trajectories $\widetilde S_k(g_{2,3},\cv{I}_1)$, $k\in \ZZZ$,  on the
covering space $\RRR^2_y$. The~numbering of~these trajectories
we also discuss later. The~drift vectors $d=d(i_2)$
and~$d=d(i_3)$ corresponding to~the~edges~$i_2$ and $i_3$ have
opposite directions;
we use the notation $d=d(\cv{I}_1)$ for $i_2$ and $-d(\cv{I}_1)$ for $i_3$.

Points on the~Reeb graph corresponding to the extreme levels
$g_{\text{min}}$ and $g_{\text{max}}$ are called \emph{end points};
they correspond to the~stable rest points
of the Hamiltonian system~\eqref{4.2}. The points corresponding
to the critical levels $g_\pm $ correspond to separatrices,
including the~saddle points $\cv{Y}_\pm$. Thus each
interior point of~any edge defines a closed trajectory
or a closed oriented curve on the torus $\TTT^2$
where the orientation is given by the natural parameter $t$ (the time).
We may parameterize the edges of the~Reeb graph
by the~variable~$g$ given by the value of~$\cv{H}$; this defines
one-to-one maps $g:[g_{\text{min}},g_-)\to i_1$, $g:[g_-,g_+]\to i_2$,
$g:[g_-,g_+]\to i_3$, and $g:(g_+,g_{\text{max}}]\to i_4$.

\subsection{Action variables and parameterization
of the~drift trajectories}\label{ss4.3}

One may also parameterize points on the edges of the~Reeb graph
by action variables
\begin{equation}
\cv{I}_2=
\frac{1}{2\pi} \int_\sigma^{\sigma+T}\cv{Y}_1(\tau)d\cv{Y}_2(\tau)-
\frac{\cv{Y}_2(\sigma)(d(i_j)\cdot a)_1}{2\pi}-
\frac{(d(i_j)\cdot a)_1(d(i_j)\cdot a)_2}{4\pi}
                                 \label{4.3}
\end{equation}
with sign prescribed by the natural orientation, where $\sigma$
is an~arbitrary real number.

As this definition of the~action variable
in the phase space $\TTT^2$ is somewhat different
from the one in $\RRR^2$, let us explain formula~\eqref{4.3}.

The~second and the~third term are present only
for the edges $i_2$ and~$i_3$.
The~geometric interpretation of $|\cv{I}_2|$ for $i_{1,4}$ is standard:
$2\pi|\cv{I}_2|$ is
the square of the~domain in~$\RRR^2_{\cv{y}}$ bounded
by the~corresponding closed trajectory (see Fig.~\ref{f3.1}b).
Of~course, here the~action variables do not depend on
parallel transport of the coordinate system on $\RRR^2_{\cv{y}}$
and on the~choice of the closed curve on~$\RRR^2_{\cv{y}}$.
It is not difficult to show that $\cv{I}_2$ is positive
for $i_1$ and negative for $i_4$,
that $|\cv{I}_2|\le a_{22}=a_{11}a_{22}/(2\pi)$, and that $\cv{I}_2=0$
in the~end points $g_{\text{min,max}}$ of~$i_{1,4}$.

The geometric interpretation of $\cv{I}_2$ for the edges
$i_{2,3}$ is as follows. Denote by $L_d$ the~straight line passing
through the~origin on the~plane $\RRR^2_{\cv{y}}$
in the direction of $d\cdot a$.
Consider one of the lifts
$\widetilde S_0 =\bigl(\cv{y}=\cv{y}(\tau)\bigr)$ on $\RRR^2_{\cv{y}}$
of the~trajectory $\widetilde S$ on $\TTT^2_{\cv{y}}$.
Let us fix two points $\cv{y}(\sigma)$
and $\cv{y}(\sigma)+d(i_{2,3})\cdot a$ on this curve and project them
onto $L_d$, such that
we obtain some~curved trapezium.
The square of this trapezium is equal to $|2\pi \cv{I}_2|$,
where $\cv{I}_2$ is defined by \eqref{4.3} (see Fig.~\ref{f3.1}b).
This interpretation allows us to derive some simple properties
of~the action variable.

In contrast to the previous case we see that now $\cv{I}_2$
depends on the choice of a lift on the plane $\RRR_{\cv{y}}^2$
of the trajectory (but it does not depend on this choice
modulo $a_{11}a_{22}/2\pi)$, and it also depends on translations
of the~coordinates.

Let us fix next a certain continuous family of trajectories
on the plane $\RRR^2_{\cv{y}}$ corresponding to the edges $i_j$.
Using the geometric interpretation of~$\cv{I}_2$
it is easy to show that $\cv{I}_2$ increases along
each edge $i_j$, such that we obtain
a parameterization of all trajectories on the torus by means of
the action variable $\cv{I}_2$, associated with the Reeb graph.
If necessary, we write $\cv{I}_2^j$
to indicate that this action variable is associated with the edge $i_j$.

Obviously, $\cv{I}_2$ admits upper and lower limits along each edge.
These limits depend also on $\cv{I}_1$ and given by
\begin{equation*}
\cv{I}_2^{1+}(\cv{I}_1)=\lim_{g\to g_--0}\cv{I}_2^1,\quad
\cv{I}_2^{4-}(\cv{I}_1)=\lim_{g\to g_++0}\cv{I}_2^4,\quad
\cv{I}_2^{2\pm/3\pm}(\cv{I}_1)=\lim_{g\to g_{\pm}\mp 0}\cv{I}_2^{2/3}.
\end{equation*}

One may calculate all the quantities $\cv{I}_2^{j\pm}$ as
\begin{equation*}
\cv{I}_2^{j\pm}(\cv{I}_1)=\int_{\gamma^\pm} \cv{y}_1\,d\cv{y}_2,
\end{equation*}
where $\gamma^\pm$ is a~separatrix connecting
the corresponding saddle point $\cv{y}_{\pm}(\cv{I}_1)$ with the saddle
point $\cv{y}_{\pm}(\cv{I}_1)+d$ on the plane $\RRR^2_{\cv{y}}$.
The numbers $\cv{I}_2^{1+}(\cv{I}_1)$ and
$\cv{I}_2^{4-}(\cv{I}_1)$ are well defined,
whereas the~numbers $\cv{I}_2^{2\pm/3\pm}(\cv{I}_1)$
are defined only up to~$a_{11}a_{22}/2\pi$.
If one fixes one of~them, say, $\cv{I}_2^{2-}(\cv{I}_1)$,
then the~others can be uniquely fixed by the ``Kirchhof law''
\begin{equation*}
\cv{I}_2^{1+}(\cv{I}_1)=\cv{I}_2^{2-}(\cv{I}_1)+\cv{I}_2^{3-}(\cv{I}_1),\qquad
\cv{I}_2^{2+}(\cv{I}_1)+\cv{I}_2^{3+}(\cv{I}_1)=\cv{I}_2^{4-}(\cv{I}_1)
+\frac{1}{2\pi}a_{11}a_{22}.
\end{equation*}
We fix $\cv{I}_2(i_{3,4})$ in this way; the~choice
of~$\cv{I}_2^{2-}(\cv{I}_1)$ will be explained later.
But for any choice of the action $\cv{I}_2$
the~following inequalities and equalities are true:
\begin{gather}
0<\cv{I}_2-\cv{I}_2^{2-/3-}(\cv{I}_1)<\frac{1}{2\pi}a_{11}a_{22},
                                    \label{4.4}\\
\cv{I}_2^{1+}(\cv{I}_1)+\Big(\cv{I}_2^{2+}(\cv{I}_1)-\cv{I}_2^{2-}(\cv{I}_1)\Big)+
\Big(\cv{I}_2^{3+}(\cv{I}_1)-\cv{I}_2^{3-}(\cv{I}_1)\Big)+
\cv{I}_2^{4-}(\cv{I}_1)=\frac{1}{2\pi}a_{11}a_{22}.
                                    \label{4.5}
\end{gather}

Now we describe the numbering of the closed trajectories
$S_l(g_{1,4},\cv{I}_1)$. We define the multiindex $l$ as follows:
Let us fix some extreme point $\cv{y}_{\text{min,max}}$
of $\cv{H}$ in $\RRR^2_{\cv{y}}$; we give the~number
$l=(0,0)$ to this point.
It is clear that this choice determines the numbering of other
extreme points by $\cv{y}^l_{\text{max,min}}:=\cv{y}_{\text{min,max}}+l \cdot a$,
and the numbering of the corresponding trajectories  $S_l(g_{1,4})$,
depending continuously on $\cv{I}_2$ (they also depend on
$\cv{I}_1$, see subsection~\ref{ss4.5}).
Indeed, if $S_0(g_{1,4})$ is defined by
the equation $\cv{y}=\cv{y}(\tau,\cv{I})$, then the other
trajectories are
$S_l(g_{1,4},\cv{I}_1):= \cv{y}=\cv{y}(\tau,\cv{I})+l\cdot a$.

In contrast to the case (a), we enumerate the~curves
$\widetilde S_k$ by a single index $k\in \ZZZ$.
Fix a~vector $f=(f_1,f_2)\in \ZZZ^2$, conjugate
to $d$, i.~e. $d_1f_1+d_2f_2=1$, which always exists.
Then we fix some open trajectory $\widetilde S_0$ corresponding to a~certain
point from the edge $i_2$ and give it the index $k=0$.
According to the ``Kirchhof law'' for the action variables,
we have to give this index also to the full family of open
trajectories associated with both edges $i_2$, $i_3$ depending
continuously on the corresponding action variable $\cv{I}_2$.
If these trajectories are given by the equation
$\widetilde S_0(g_{2,3},\cv{I}_1):\cv{y}=\cv{y}(\tau,\cv{I})$,
then the open trajectories with index $k$
are $\widetilde S_k(g_{2,3},\cv{I}_1): \cv{y}=\cv{y}(\tau,\cv{I})-
k\widehat{J} f\cdot a$.

Thus we see that the trajectories of the system on the torus and on
the plane are parameterized by action variables $\cv{I}_1$
and $\cv{I}_2$, indices $l\in\ZZZ^2$ or $k\in\ZZZ$,
and the edges of the Reeb graph; we include
all these parameters to the notation in the next subsection.
Finally, we have Figs.~\ref{f3.1}b and \ref{f3.1}c for the trajectories
on $\RRR^2_{\cv{y}}$ (generally speaking, the curves $\cv{I}_2^2$ and
$\cv{I}_2^3$ in Fig.~\ref{f3.1}c may coincide).

\subsection{The Reeb graph in degenerate cases}\label{ss4.4}
Now consider now the case when $\cv{H}(\cv{I}_1,\cdot)$ is not
a simple Morse function.

There are two possible cases.
Firstly, $\cv{H}(\cv{I}_1,\cdot)$ can be a complex Morse function.
Denote the corresponding value of $\cv{I}_1$
by $\cv{I}^{1}_1$, and if necessary add the
subindex $\alpha$ for numbering of these critical values.

\begin{figure}[t]\centering
\begin{minipage}{50mm}\centering
\epsfig{file=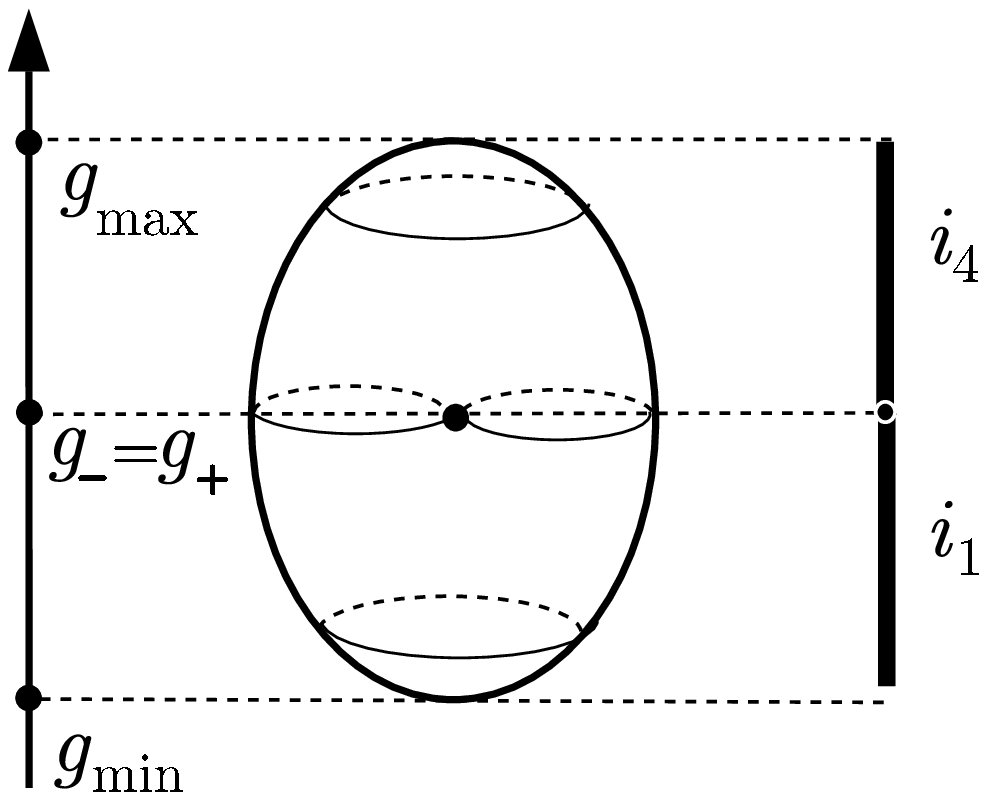,width=40mm,height=35mm}\\
a) Reeb graph
\end{minipage}
\begin{minipage}{50mm}\centering
\epsfig{file=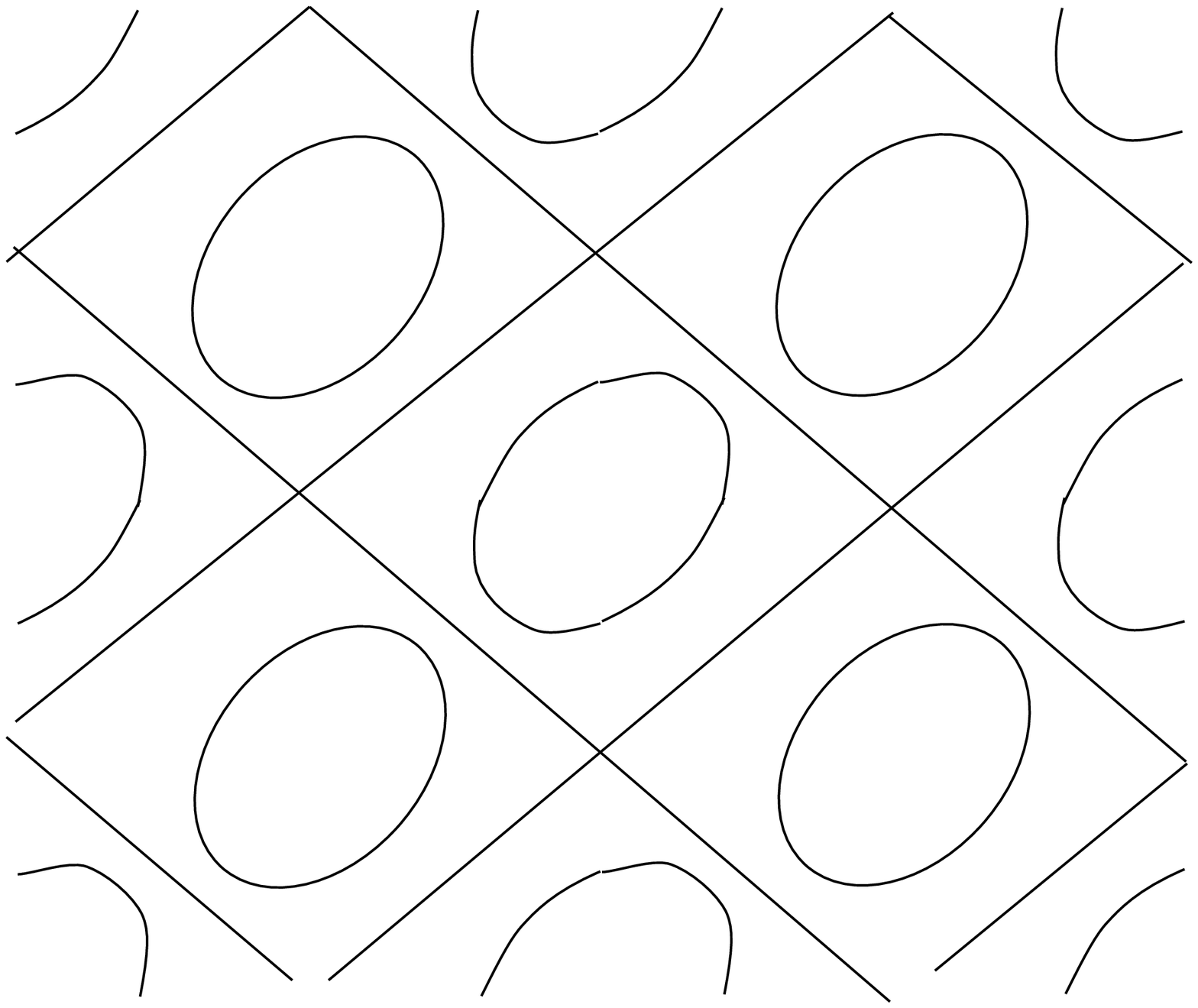,width=42mm,height=35mm}\\
b) Level curves
\end{minipage}
\begin{minipage}{45mm}\centering
\epsfig{file=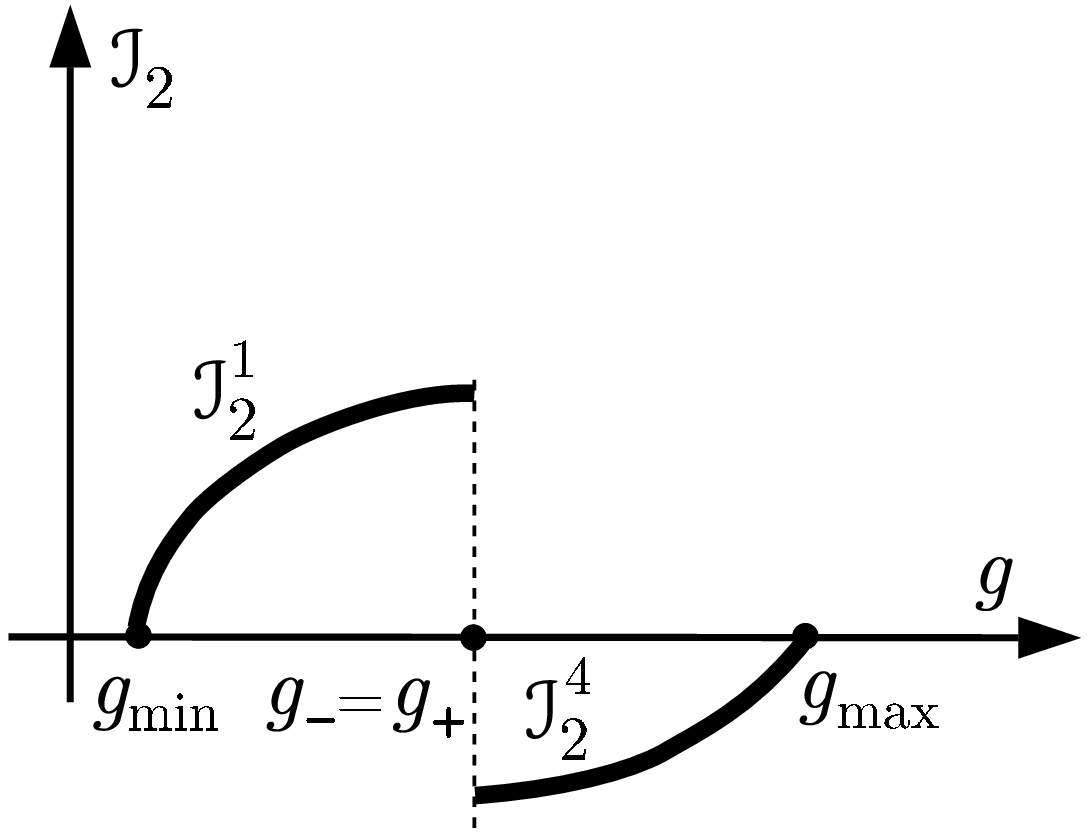,width=40mm,height=35mm}\\
c) Action variables
\end{minipage}
\caption{Characteristics of Morse function: The degenerated case~I.\label{f3.2}}
\end{figure}

The regime of regular motion consists of contractible curves
only, and the Reeb graph has the form described in Fig.~\ref{f3.2}a.
This graph may be considered as a~limit of the~previous
case as $g_- \to g_+$ such that
and the edges $i_2$ and $i_3$ contract to a common point.
The action variables are sketched in~Fig.~\ref{f3.2}c,
and the phase picture for the trajectories on the plane
$\RRR^2$ is shown in Fig.~\ref{f3.2}b.

\begin{figure}[b]\centering
\begin{minipage}{50mm}\centering
\epsfig{file=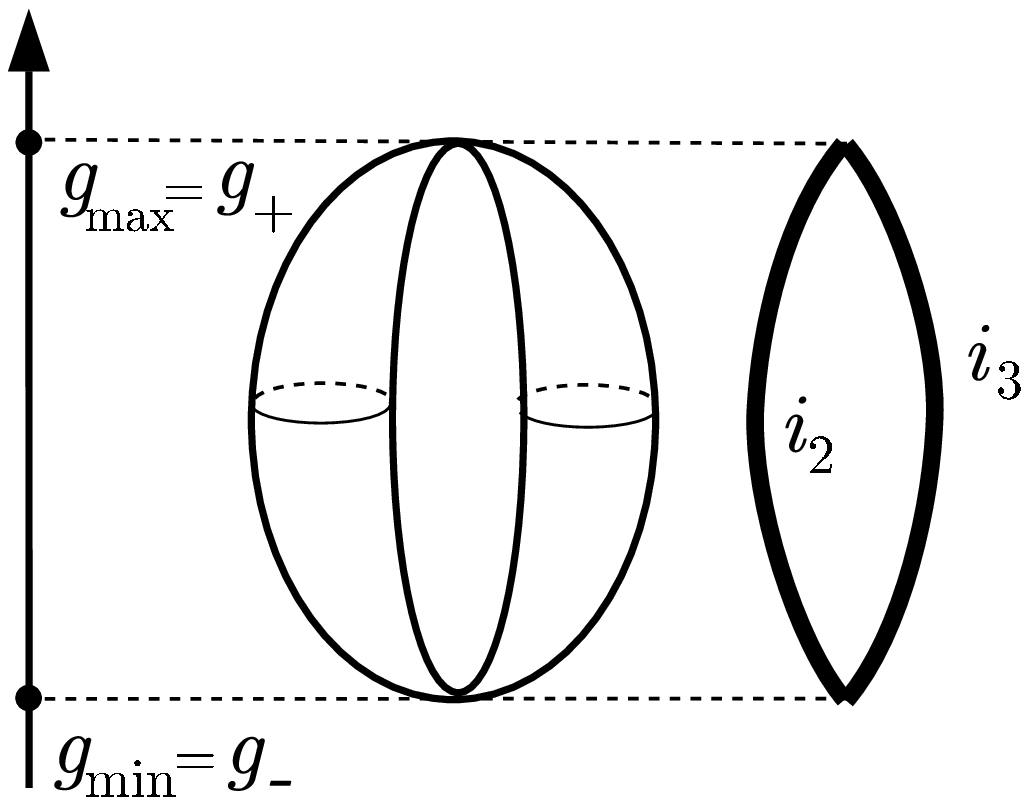,width=40mm,height=35mm}\\
a) Reeb graph
\end{minipage}
\begin{minipage}{50mm}\centering
\epsfig{file=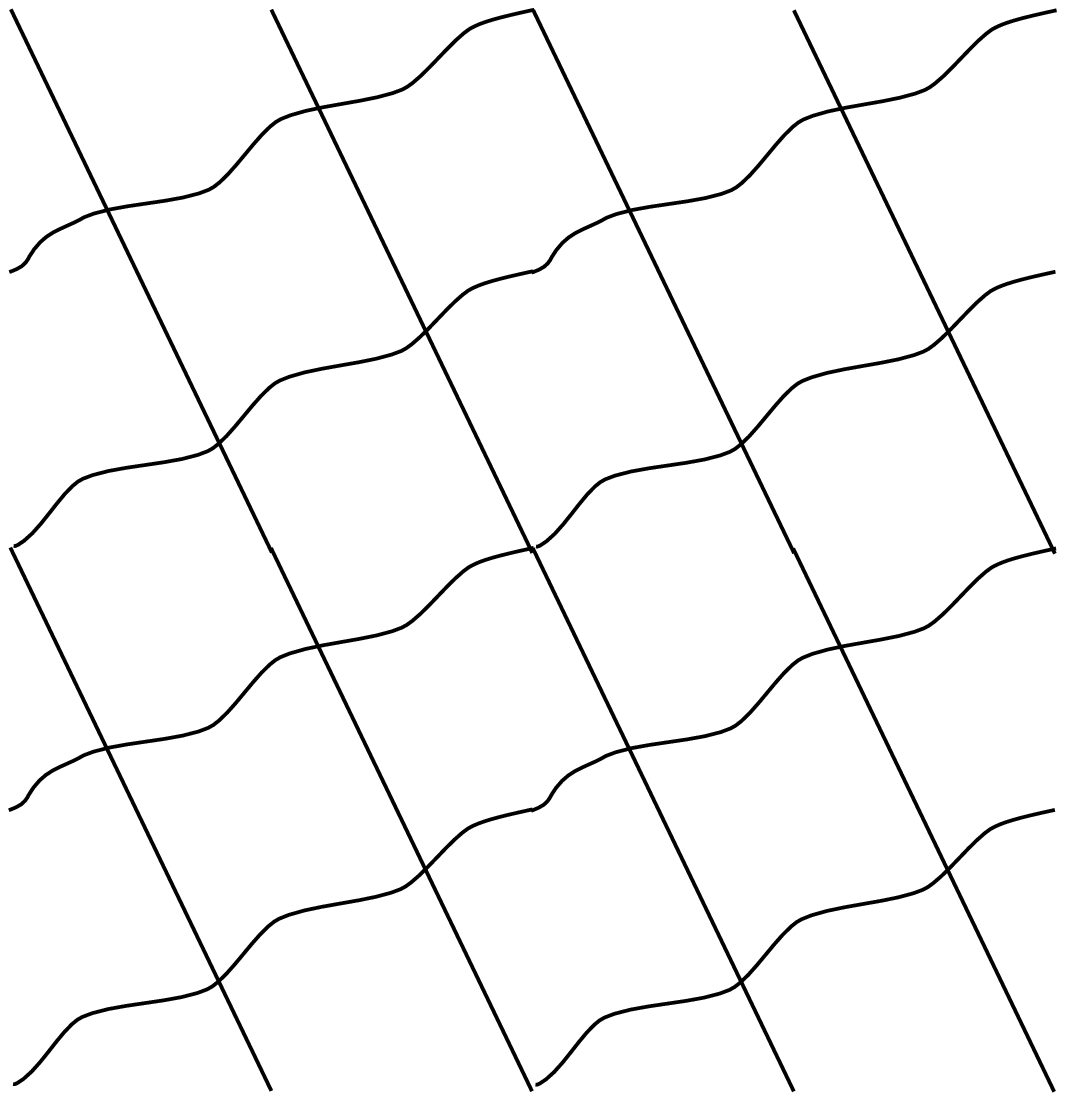,width=42mm,height=35mm}\\
b) Level curves
\end{minipage}
\begin{minipage}{45mm}\centering
\epsfig{file=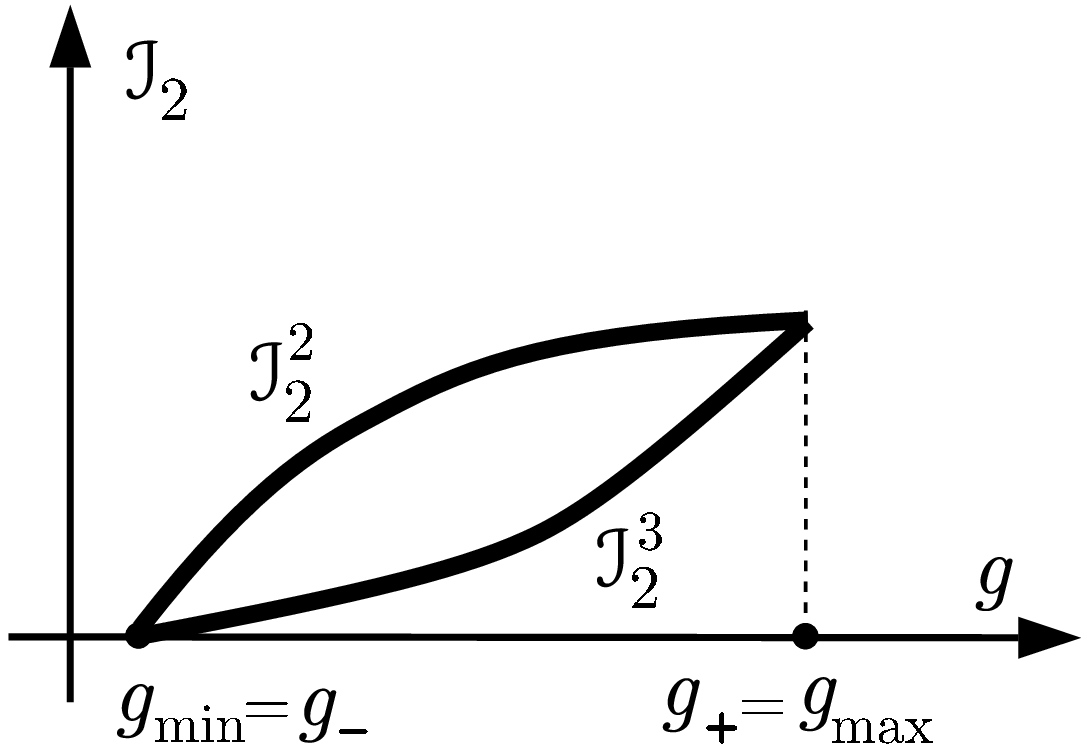,width=40mm,height=35mm}\\
c) Action variables
\end{minipage}
\caption{Characteristics of Morse function: Degenerated case~II\label{f3.3}}
\end{figure}

Another limit case (Fig.~\ref{f3.3}) occurs
if $g_- \to g_{\text{min}}$ and $g_+\to g_{\text{max}}$;
all contractible trajectories disappear,
and we are left with non-contractible trajectories only.
In this case, $\cv{H}$ is not a~Morse function, but we can still
assign a~Reeb graph to~this situation (see Fig.~\ref{f3.3}a): we keep
only the edges $i_2$ and $i_3$, and the action variables
$\cv{I}^1_2$ and $\cv{I}^4_2$ show a similar behavior (Fig.~\ref{f3.3}c).
We denote the corresponding values
of the~actions $\cv{I}_1$ by $\cv{I}^{2}_1$ and add,
if necessary, the subindex $\alpha$ for numbering of~these points.

\subsection{The description of the~averaged motion in 4-D phase space}\label{ss4.5}

Using the~above considerations we now represent the global structure
of the classical motion defined by $\cv{H}$ under assumption that
the domain of the motion on the half-plane $(\cv{I}_1\ge 0, E)$
is separated into such (connected) subdomains,
that the behavior of~trajectories
of~the corresponding Hamiltonian systems for each of~these subdomains
is topologically equivalent and have the~same rotation
number~(see Fig.~\ref{f1.1}).
As before, we call these domains \emph{regimes}.
The~interior points of~each regime correspond to closed
trajectories on tori $\TTT^2$; the boundary of the regimes
is formed by the critical manifolds
of the function $\cv{H}$ and by the left boundary $\cv{I}_1=0$.

A regime is called of \emph{boundary} type and is denoted
by $\MMM_r$ if it a~certain part (of non-zero lenfth)
of its boundary consists of extreme points of~$\cv{H}$).
On the plane $\RRR^2_{\cv{y}}$ the~corresponding
level curves of the function $\cv{H}$ are families
of closed trajectories, their preimages on the torus
$\TTT^2$ are contractible trajectories with rotation number $0/0$, and they
have no special direction.
If we return back to the original four-dimensional phase space
$\RRR^4_{p,x}$, for each interior point in these regimes
we get a~family of invariant Lagrangian manifolds
of~Hamiltonian $\cv{H}$. They are topological products
of the~cyclotron circles  $S_C(\cv{I}_1)$  and the~closed curves $S_l(g_{1,4},\cv{I}_1)$,
and they are diffeomorphic to two-dimensional tori
(we call them \emph{Liouville tori}).
We parameterize each point in $\MMM_r$
by the action variables $\cv{I}_1$ and $\cv{I}_2$, which belong
to a certain domain on the plane $\RRR^2_{\cv{I}}$;
to simplify the notation we denote these domains also by
$\MMM_r$.

Thus each interior point $\cv{I}=(\cv{I}_1,\cv{I}_2)\in \MMM_r$
indicates a~discrete family of invariant Lagrangian tori
$\Lambda_l^r(\cv{I}_1,\cv{I}_2)$\footnote{Of course, these manifolds
depend also on $\varepsilon$; we will include this fact into notation later.}
in the original phase space $\RRR^4_{p,x}$
numbered by a multiindex $l=(l_1,l_2) \in \ZZZ^2$ given by
the numbering of the curves $S_l(g_{1,4},\cv{I}_1)$.

If $\Lambda_0^r(\cv{I})$ is defined by the equations
\begin{equation}
x_{1,2}=X_{1,2}(\cv{I},\varphi),\quad
p_{1,2}=P_{1,2}(\cv{I},\varphi),
                             \label{4.6}
\end{equation}
where $\varphi=(\varphi_1,\varphi_2)$
are angle variables conjugate to
$\cv{I}=(\cv{I}_1,\cv{I}_2)\in \MMM_r$, then
\begin{equation*}
\Lambda_l^r(\cv{I})= \big\{x=X(\cv{I},\varphi )+l \cdot a,\quad
p_1=P_1(\cv{I},\varphi)+(l \cdot a)_1,\quad p_2=P_2(\cv{I},\varphi)\big\}.
\end{equation*}
The action variables do not depend on $l$, and they are
defined by \eqref{4.6} with $d=0$.

The remaining regimes are called \emph{interior} regimes.
We denote them by $\widetilde{\MMM}_r$ and
use the symbol $\widetilde{\MMM}_r$ also for the~domain
of the~corresponding action variables on the
$(\cv{I}_1,\cv{I}_2)$-plane and $(\cv{I}_1,E)$-plane.
On the plane $\RRR^2_{\cv{y}}$,
the~level sets of $\cv{H}$ are families of open curves
with the main vector $d^r=(d_1,d_2)$,
and their preimages on the torus $\TTT_2$ are non-contractible
closed trajectories with rotation number $d_1/d_2$.
In the original phase space $\RRR^4_{p,x}$, these trajectories are
covered by a~discrete set of the families of invariant two-dimensional
Lagrangian manifolds $\widetilde\Lambda^r_k(\cv{I})$,
$\cv{I}=(\cv{I}_1,\cv{I}_2)\in \widetilde{\MMM}_r$,
of~$\cv{H}$. They are products
of the ``cyclotron'' circles $S_C(\cv{I}_1)$ and the open curves
$\widetilde S_k(g_{1,4},\cv{I}_1)$ and are diffeomorphic to
two-dimensional cylinders; for brevity we call $\widetilde\Lambda_k^r(\cv{I})$
\emph{Liouville cylinders}.

The cylinders $\widetilde\Lambda_k^r(\cv{I})$ depend smoothly on
$\cv{I}=(\cv{I}_1,\cv{I}_2)\in \widetilde{\MMM}_r$,
their numbering by the index $k$ is induced by the numbering
of the curves $\widetilde S_k(g_{1,4},\cv{I}_1)$. Hence we have
\begin{equation*}
\widetilde \Lambda^r_k(\cv{I})=
\big\{x=X(I,\varphi)-k\widehat{J}f\cdot a,\quad
p_1=P_1(\cv{I},\varphi)+ k \widehat{J} f \cdot a,\quad
p_2=P_2(\cv{I},\varphi,\varepsilon)\big\},
\end{equation*}
where the~vector functions $P$ and $X$  define the Liouville
cylinder $\widetilde{\Lambda}^{r}_0(\cv{I})$ by \eqref{4.6},
$\varphi=(\varphi_1,\varphi_2)$
are the angle variables, and $\cv{I} \in \widetilde{\MMM}_r$.

We can give  formulas for the action variables $\cv{I}_1$ and
$\cv{I}_2$ directly on the tori and cylinders by:
\begin{gather}
\cv{I}_1=\frac{1}{2\pi}
\int_{\varphi_1}^{\varphi_1 +2\pi }p\,dx|_{\varphi_2=\text{const}},\notag\\
\cv{I}_2=\frac{1}{2\pi }\Big(\int_{\varphi_2} ^{\varphi_2 +2\pi }p\,dx
|_{\varphi_1=\text{const}}+
x_1(d^r\cdot a)_2+
\frac{(d^r\cdot a)_1(d^r\cdot a)_2}{2} \Big),
                                \label{4.7}
\end{gather}
where $(p,x)$ belongs to $\Lambda^l_k(\cv{I})$ or to~$\widetilde\Lambda^r_k(I)$,
respectively. Note that in the latter case $\cv{I}_2$ depends on index $k$,
but in what follows we fix $\cv{I}_2$ by setting $k=0$ in the definition
and using this action for parameterization
of all $\widetilde{\Lambda}^r_k(\cv{I})$. We also assume that the families
$\Lambda^r_l(\cv{I})$ and  $\widetilde{\Lambda}^r_k(\cv{I})$
depend smoothly on $\cv{I}=(\cv{I}_1,\cv{I}_2)$ in the whole regime.
Then, by fixing a certain cylinder from the interior
regime associated with the edge $i_2$ of the Reeb graph
and giving it the index $k=0$, we determine by the ``Kirchhof law''
the action the cylinders corresponding to the edge $i_3$.

It is well known that Lagrangian manifolds have the integer-valued homotopic
invariants which are called Maslov indices and connected with the cycles
on these manifolds. Obviously the Betty number (the
rank of the cohomology group, or the number of basis cycles)
of any Liouville torus $\Lambda^r_l$ is equal to two, hence $\Lambda^r_l$
have two Maslov indices and the Betty number of any Liouville cylinder
$\widetilde{\Lambda}^r_k$ is equal to one, hence  $\widetilde{\Lambda}^r_k$ has one Maslov index.
Standard calculations lead to the~following simple fact.

\begin{prop}                           \label{p4.2}
The Maslov index of the cycles $\gamma_{1,2}=
(\varphi_{2,1}=\text{const})$
on any torus $\Lambda^r_l(\cv{I})$ is equal to~$2\mod 4$.
The Maslov index of the cycle $\gamma =(\varphi_2 =\text{const})$
on any cylinder
$\widetilde \Lambda ^r_k(\cv{I})$ is also equal to $2\mod 4$.
\end{prop}

Each non-degenerate point $\cv{I}=(\cv{I}_1,0)$
of the extreme boundaries of the regimes
$\MMM_r$ defines in $\RRR^4_{p,x}$ a degenerate torus,
namely a closed trajectory, which is an isotropic manifold.
In this case we have only
cyclotron motion the drift is absent.
The other non-degenerate boundaries between $\MMM_r$ and
$\widetilde{\MMM}_r$ define separatrices
of a one-dimensional Hamiltonian system with the Hamiltonian $\cv{H}$,
which generates in $\RRR^4_{p,x}$ a two-dimensional
invariant singular manifold of the Hamiltonian $\cv{H}$.

The critical points on the boundaries of the regimes induce
degenerate singular invariant manifolds;
these manifolds together with their neighborhoods
are called \emph{atoms}. Generally
speaking, there exist infinitely many topological types of atoms, one
can find some classifications some of them in \cite{BF,BF-1}. We restrict our
consideration to the simplest situation described in the previous sections,
then all critical values on the axis  $\cv{I}_1$
are of the form $\cv{I}^{1/2}_{1,\alpha}$. The Morse function $\cv{H}$
changes its type, when $\cv{I}_1$ crosses these values
(in our simplest case only rotation number changes;
in more complicated problems a new Reeb graph can arise).

The left boundary $\cv{I}_1=0$ plays a special role in quantum applications,
as it corresponds to the so-called \emph{low Landau bands}.
In the original phase space the corresponding trajectories belong
the two-dimensional invariant subspace $\cv{I}_1=0$.
This subspace presents only slow drift and
the ``cyclotron'' motion is absent. All previous considerations concerning
$\cv{H}$ remain valid, but now the ``limit'' Liouville tori
$\Lambda_l^r(0,\cv{I}_2)$ in $\RRR^4_{p,x}$ are just closed curves,
and the ``limit'' Liouville cylinders $\widetilde\Lambda_k^r(0,\cv{I}_2)$
are open curves with drift vector $d$;
these curves are isotropic manifolds also.

The critical points $(\cv{I}_1=0$, $g=g_{\pm})$
can be considered as zero-dimensional singular manifolds.
In the degenerate case $g_+(0)=g_-(0)$, there
exist only two boundary regimes; this case is not generic,
but it appears in connection with the Harper equation (see below).
The other degenerate case is $g_{\text{min}}(0)=g_-(0)$
and $g_{\text{max}}(0)=g_+(0)$. It appears, for instance,
when $v$ depends only on one variable. It seems
that in this case one can separate the variables in the
original spectral problem.

The angle points $\cv{I}=(0,0)$ in the left boundary correspond
to the stable rest points of both the averaged and the original
Hamiltonian $\cv{H}$ and $H$.

At last we remark that there is no reasonable definition of the Maslov index
for an individual isotropic manifold $\Lambda\in \RRR^{2n}_{p,x}$,
if $\dim \Lambda <n$, see e.~g.~\cite{M2,BD,DO,BB}.
But if this manifold arises as the limit of a family
of Lagrangian manifolds, one can associate with this manifold
a Maslov index and make its use in the semiclassical approximation.
Obviously, this applies to the problem under consideration.

\subsection{Example}\label{ss4.6}
We illustrate the considerations of this
section by the example \eqref{1.6}~--\eqref{3.7}.
Let us consider first $\Bar{H}$ instead of $\cv{H}$, then
\begin{gather*}
g_{\text{min}}=-\Big(A\big|J_0(\sqrt{2\cv{I}_1})\big|+
B\big|J_0(\beta \sqrt{2\cv{I}_1})\big|\Big),\quad
g_{\text{max}}=A\big|J_0(\sqrt{2\cv{I}_1})\big|+
B\big|J_0(\beta \sqrt{2\cv{I}_1})\big|,\\
g_{\pm}=\pm \Big|A\big|J_0(\sqrt {2\cv{I}_1})\big|-B\big|J_0(\beta \sqrt{2\cv{I}_1})\big|\Big|.
\end{gather*}

The first series of critical points, $\cv{I}_{1,\alpha}^{1}$,
$\alpha=1,2,\dots$, is obtained from the equations
\begin{equation*}
J_0(\cv{I}_{1,\alpha}^{1})=0 \quad\text{and}\quad
J_0(\beta \cv{I}_{1,\alpha}^{1})=0.
\end{equation*}
The second series, $\cv{I}_{1,\alpha}^{2}$, $\alpha=1,2,\dots$,
consists of the solutions of the equations
\begin{equation*}
A\big|J_0(\cv{I}_{1,\alpha}^{2})\big|=
        B\big|J_0(\beta \cv{I}_{1,\alpha}^{2})\big|.
\end{equation*}
On the plane $(\cv{I}_1,E)$, the boundary regimes are
the sets (see Fig.~\ref{f1.1})
\begin{gather*}
\cv{I}_1+\varepsilon g_{\text{min}}(\cv{I}_1)<E<\cv{I}_1+\varepsilon g_-(\cv{I}_1),\qquad
\cv{I}_{1,\alpha}^{1}<\cv{I}_1<\cv{I}_{1,\alpha+1}^{1},\\
\intertext{and}
\cv{I}_1+\varepsilon g_+(\cv{I}_1)<E<\cv{I}_1+\varepsilon g_{\text{max}}(\cv{I}_1),\qquad
\cv{I}_{1,\alpha}^{1}<\cv{I}_1<\cv{I}_{1,\alpha+1}^{1};\\
\intertext{the interior regimes are the sets}
\cv{I}_1+\varepsilon g_-(\cv{I}_1)<E<\cv{I}_1+\varepsilon g_+(\cv{I}_1),\qquad
\cv{I}_{1,\alpha}^{2}<\cv{I}_1<\cv{I}_{1,\alpha+1}^{2}.
\end{gather*}
Note that the rotation number changes when $\cv{I}_1$ crosses
these critical points.

The drift vectors of the interior regimes are
\begin{equation*}
d=\begin{cases}(1,0) & \text{if
$A\big|J_0(\sqrt{2\cv{I}_{1,\alpha}^{1}})\big|>
B\big|J_0(\beta \sqrt{2\cv{I}_{1,\alpha}^{1}})\big|$},\\
(0,1) & \text{if
$A\big|J_0(\sqrt{2\cv{I}_{1,\alpha}^{1}})\big|<
B\big|J_0(\beta \sqrt{2\cv{I}_{1,\alpha}^{1}})\big|$.}
\end{cases}
\end{equation*}
A simple calculation gives
\begin{gather}
\cv{I}^{1+}_2(\cv{I}_1)=-\cv{I}^{4-}_2(\cv{I}_1)=
2\cv{I}^{2+}_2(\cv{I}_1)=2\cv{I}^{3+}_2(\cv{I}_1)=
\frac{2}{\pi\beta }\int_{0}^{\Gamma(\cv{I}_1)}\frac{1}{\xi}
\log \left(\frac{1+\xi}{1-\xi}\right)d\xi,\notag\\
\cv{I}^{2+}_2(\cv{I}_1)=\cv{I}^{3+}_2(\cv{I}_1)=
\frac{\pi}{\beta}-\frac{1}{2}\cv{I}^{1+}_2(\cv{I}_1),
                               \label{4.8}
\end{gather}
where
\begin{equation*}
\Gamma(\cv{I}_1)=
\left(\frac{A|J_0(\sqrt{2\cv{I}_1})|}{B|
J_0(\beta \sqrt{2\cv{I}_1})|}\right)^{\pm 1},
\end{equation*}
and the~sign in the~exponential is such that $\Gamma \le 1$.

Using $\cv{H}$ gives a~discrepancy $O(\varepsilon^2)$
in all above estimations.

\section{Almost invariant manifolds of the original Hamiltonian}\label{s5}

\begin{defin}                 \label{d5.1}
Let $\Lambda=\{p=P(\varphi,\varepsilon),\quad x=X(\varphi,\varepsilon)\}
\subset \RRR^4_{p,x}$ be either a two-dimensional Lagrangian manifold,
diffeomorphic to a two-dimensional torus (or cylinder), or
a smooth closed (or open) curve.
Let $C>0$. We say that $\Lambda$ is an almost invariant
manifolds of the Hamiltonian $H(p,x,\varepsilon)$ up to
$O(e^{-C/\varepsilon})$ if
\begin{equation}            \label{5.1}
H|_{\Lambda}=\text{const}+O(e^{-C/\varepsilon}),
\end{equation}
and if a vector $\omega$ exists, such that
\begin{equation}           \label{5.2}
\parbox{140mm}{
$\big(P(\varphi+\omega t,\varepsilon),
X(\varphi+\omega t,\varepsilon)\big)$ satisfies
the Hamiltonian system up to $O(e^{-C/\varepsilon})$,
uniformly in $t\in\RRR$.
}
\end{equation}
We call a family of two-dimensional almost invariant tori
(respectively, cylinders) depending smoothly on action
variables $\cv{I}\in\MMM$ \emph{almost Liouville}
tori (respectively, cylinders) of the Hamiltonian $H$.
\end{defin}

According to this definition, the manifolds
$\Lambda_l^r(\cv{I})$ and $\widetilde{\Lambda}_k^r(\cv{I})$ constructed
in the previous section are almost Liouville tori or cylinders
of the Hamiltonian $H$.

Of course, not all the applications need
a construction of this accuracy.
In some cases, it is reasonable to construct an averaged
Hamiltonian $H^K$ up to $O(\varepsilon^{K+1})$
and to obtain ``almost invariant'' manifolds
of $H$  with a discrepancy $O(\varepsilon^{K+1})$
(this means that in~\eqref{5.1} and~\eqref{5.2}
one has $\varepsilon^{K+1}$ instead of $e^{-C/\varepsilon}$).
In particular, $\Bar{H}$ is the~averaged Hamiltonian
up to~$O(\varepsilon^2)$, and its invariant manifolds
are almost invariant manifolds of~$H$
but only up to $O(\varepsilon^2)$.

Let us fix a sufficiently small $\delta$ and consider a boundary
or an inner regime without $\delta$-neighborhood
of the boundary formed by separatrices.
If the regime includes the left boundary $\cv{I}_1=0$, then
a neighborhood of this boundary (with a neighborhood of the singular
points removed) also belongs to the non-singular part of this regime.
Let us denote these non-singular parts of $\MMM_r$
or $\widetilde {\MMM}_r$ by $\MMM_{r,\delta}$
or $\widetilde{\MMM}_{r,\delta}$, respectively.
The corresponding regimes for to $H^K$ we denote by
$\MMM^K_{r,\delta}$ and $\widetilde{\MMM}^K_r$;
they coincide with $\MMM_{r,\delta}$
and $\widetilde{\MMM}_{r,\delta}$ up to $O(\varepsilon^K)$,
and they have the same drift vectors.
Moreover,  the corresponding almost invariant tori and cylinders,
$\Lambda^{r,K}_l$ and $\widetilde \Lambda^{r,K}_l$ have the same structure
as $\Lambda^{r,K}_l$ and $\widetilde \Lambda^{r,K}_l$, and, in particular,
their Maslov indices coincide.

\section{Semiclassical spectral series}

Now we are going to use the almost invariant manifolds described
in the previous sections for constructing the spectral asymptotics
for $\Hat H$, like it was done in subsection~2.5.
Let us emphasize again that in even in the simplest multidimensional
problems there is usually no global asymptotic formula for the spectrum;
it is useful to divide the spectrum into
several parts, such that the asymptotic behavior of the spectrum
is preserved in each of these parts.
These parts together with the corresponding formulas 
for the spectrum are usually referred to as \emph{spectral series}.
According the fundamental \emph{correspondence principle}
of the quantum mechanics (connected with names of Bohr, Sommerfeld,
Einstein, Ehrenfest, Brillouin, and others),
the classification of spectral series is connected
with the motions of the classical Hamiltonian system [27, 46, 60, 66, 67].
As a mathematical expression of this principle we use
the method of the canonical operator developed by Maslov.

Now we are going to show how the correspondence
pPrinciple appears in the problem under study.

\begin{defin}  \label{d_6_1}
A pair $(\psi,E)$ with $\psi(x,h,\varepsilon)\in C^\infty(\R^2)$
and $E(h,\varepsilon)\in\R$ is called a \emph{quasimode
of the~operator $\Hat H$ with error $O(h^L+\varepsilon^K)$}
if for any compact set $\Omega\subset\R^2$ there are
positive numbers $A$ and $B$ satisfying
$\|(\Hat H-E)\Psi\|_{L_2(\Omega)}\le A h^L+
B\varepsilon^K$.
\end{defin}

From now on we fix some positive integer numbers $K$ and $L$;
in the rest of the section we describe the construction
of quasimodes for $\Hat H$ with error $O(h^L+\varepsilon^K)$.

\subsection{Quasimodes associated with the almost Liouville tori}
Consider a certain boundary regime $\mathcal M_{r,\delta}$
and the corresponding family of the almost Liouville
tori $\Lambda_l^{r}(\cv{I},\varepsilon)$.
As it was mentioned above, the asymptotics of the spectrum
of $\Hat{H}$ is defined by means of the canonical operator.

For each fixed $h>0$ let us choose a discrete
subset of the values of the action variables $\cv{I}_1$ and $\cv{I}_2$,
by the rules .
\begin{gather}
                    \label{e_6_1}
\cv{I}_1=\cv{I}_1^\mu(h):=(\frac12+\mu)h,\\
                    \label{e_6_2}
\cv{I}_2=\cv{I}_2^\nu(h):=(\frac12+\nu)h,
\end{gather}
here $\mu$ and $\nu$ are integer numbers such that
\begin{equation}
                    \label{e_6_3}
\big(\cv{I}_1^{\mu}(h), \cv{I}_2^{\nu}(h)\big)\in{\mathcal M}_{r,\delta}.
\end{equation}
The conditions~\eqref{e_6_1} and~\eqref{e_6_2} are nothing
but the necessary and sufficient condition for constructing
the canonical operator on the tori
$\Lambda^{r}_l\big(\cv{I}^\mu_1(h),\cv{I}_2^\nu(h),\varepsilon\big)$.

\begin{prop}
      \label{p_6_1}
For any $(\mu,\nu)\in \Z_+\times\Z$ satisfying~\eqref{e_6_3}
there exist quasimodes
$\big(\psi_{r,l}^{\mu,\nu}(x,h,\varepsilon),E^{\mu,\nu}_r(h,\varepsilon)\big)$
of the~operator $\Hat H$ with error $O(h^L+\varepsilon^K)$;
these quasimodes can be given by the equalities
\begin{gather*}
\psi_{r,l}^{\mu,\nu}=\mathcal{K}_{\Lambda^{r}_l\big(\cv{I}^\mu_1(h),\cv{I}_2^\nu(h),\varepsilon\big)}
\chi_{r,l}^{\mu,\nu},\\
\chi_{r,l}^{\mu,\nu}=1+O(h)\in
C^{\infty}\Big(\Lambda^r_l\big(\cv{I}^\mu_1(h),\cv{I}^\nu_2(h),\varepsilon\big)\Big),\\
E^{\mu,\nu}_r=\cv{H}^r\big(\cv{I}^\mu_1(h),\cv{I}^\nu_2(h),\varepsilon\big)+O(h^2).
\end{gather*}
All the functions $\psi_{r,l}^{\mu,\nu}$ belong to~$L^2(\R^2_x)$ and
they can be expressed through each other as
\textup{(}cf.~\textup{(2.36)):}
\begin{equation}
     \label{e_6_4}
\psi_{r,l}^{\mu,\nu}(x,h,\varepsilon)=\psi_{r,0}^{\mu,\nu}
(x-l\cdot a,h,\varepsilon) e^{-\frac{i}{h} l_2a_{22}x_1};
\end{equation}
these functions are asymptotically localized near the projections
$\pi_x\Lambda^{r}_l\big(\cv{I}^\mu_1(h),\cv{I}^\nu_2(h),\varepsilon\big)$
of the corresponding tori onto the~$x$-plane as $h$ tends to $0$, i.~e.
\begin{equation*}
\lim_{h\to 0}\psi_{r,l}^{\mu,\nu} =0
\text{ as } x\notin
\pi_x\Lambda^{r}_l\big(\cv{I}^\mu_1(h),\cv{I}^\nu_2(h),\varepsilon\big)
\end{equation*}
One also has the estimate
$\dist\big(E^{\mu,\nu}_r(h,\varepsilon),\spec\Hat{H}\big)=O(h^L+\varepsilon^K)$.
\end{prop}

\begin{remark}
The numbers $E^{\mu,\nu}_r$ as well as the functions
$\psi_{r,l}^{\mu,\nu}$ depend also on $K$ and $L$;
now we omit this dependence to simplify the notation,
but sometimes we will write $E^{\mu,\nu}_{r,K,L}$
instead of $E^{\mu,\nu}_r$ to emphasize this dependence.
\end{remark}

The proof directly follows from the general properties
of the canonical operator; we describe the scheme of the proof
in the Appendix.

Denote by
$\Sigma^{r,\delta}_{K,L}(h,\varepsilon)$
the~union of~all possible points
$E^{\mu,\nu}_r(h,\varepsilon)$. We call this set
\emph{the~semiclassical spectral series
up to~$O(h^L+\varepsilon^K)$ corresponding
to the~boundary regime $\mathcal{M}_{r,\delta}$.}

\subsection{Quasimodes associated with
the almost Liouville cylinders}

Let us consider now a certain interior regime $\Tilde{\mathcal M}_{r,\delta}$
and the corresponding  family of the almost Liouville cylinders
$\Tilde{\Lambda}_l^{r}(\cv{I}_1,\cv{I}_2,\varepsilon)$.

To construct the canonical operator on these cylinders,
we quantize only the action variable $\cv{I}_1$ by the rule~\eqref{e_6_1},
and $\cv{I}_2$ remains free.
The integer numbers $\mu$ in~\eqref{e_6_1} are such that
\begin{equation}
                   \label{e_6_5}
\big(\cv{I}_1^{\mu}(h),\cv{I}_2\big) \in  \cv{M}_{r,\delta}.
\end{equation}

\begin{prop}
                   \label{p_6_2}
For any $(\mu,\cv{I}_2)\in\Z_+\times\R$ satisfying~\eqref{e_6_5}
there exist quasimodes
$\big(\Tilde \psi_{r,k}^{\mu}(x,\cv{I}_2,h,\varepsilon),
\Tilde E^{\mu}_r(\cv{I}_2,h,\varepsilon)\big)$ of~$\Hat{H}$
with error $O(h^L+\varepsilon^K)$;
this quasimode is defined by
\begin{gather*}
\Tilde \psi_{r,k}^{\mu}=
\mathcal K_{\Tilde \Lambda^{r}_k\big(\cv{I}_1^\mu(h),\cv{I}_2,\varepsilon\big)}
\Tilde \chi_{r,k}^{\mu},\\
\Tilde \chi_{r,k}^{\mu}=1+O(h)\in
C^{\infty}\Big(\Tilde \Lambda^{r}_k\big(\cv{I}_1^\mu(h),\cv{I}_2,\varepsilon)\big)\Big),\\
\Tilde E^{\mu}_{r}(\cv{I}_2,h,\varepsilon)=
\cv{H}\big(\cv{I}_1^\mu(h),\cv{I}_2,\varepsilon\big)+O(h^2).
\end{gather*}
The~functions $\Tilde\psi_{r,k}^{\mu}$ belong to $L^2_{\mathrm{loc}}(R^2_x)$;
they can be expressed through each other by the equality
\begin{equation}
                 \label{e-temp-3}
\Tilde\psi_{r,k}^{\mu}(x,\cv{I}_2,h,\varepsilon)=\Tilde\psi_{r,0}^{\mu}(x+k(Jf)\cdot a,\cv{I}_2,h,\varepsilon)
e^{\frac{i}{h} f_1a_{22}x_1},
\end{equation}
and enjoy the property
\begin{equation}
                \label{e-temp-2}
\Tilde\psi_{r,k}^{\mu}(x+{d}\cdot a,\cv{I}_2,h,\varepsilon)=
\Tilde\psi_{r,k}^{\mu}(x,\cv{I}_2,h,\varepsilon)
e^{i(2\pi \cv{I}_2-({d}\cdot a)_2 x_1-({d}\cdot a)_1({d}\cdot a)_2/2)},
\end{equation}
where $d$ is the drift vector associated with the~section
$\Tilde {\mathcal{M}}_{r,\delta}$, and
the vector $f=(f_1,f_2)$ is dual to $d$, i.e. $d_1 f_1+d_2 f_2=1$. 
The functions
$\Tilde\psi_{r,k}^\mu$ are asymptotically localized near the projections
$\pi_x\Tilde\Lambda^{r}_k\big(\cv{I}^\mu_1(h),\cv{I}_2,\varepsilon\big)$
of the~corresponding cylinders onto the plane $x$ as $h$ tends to $0$.

For the~numbers $\Tilde{E}^{\mu}_r$ we have the estimate
$\dist(\Tilde E^{\mu}_{r}(\cv{I}_2,h,\varepsilon),\spec\Hat{H})=
O(h^L+\varepsilon^K)$.
\end{prop}

In~contrast to the~boundary regimes,
the~points $\Tilde{E}^\mu_r(\cv{I}_2,h,\varepsilon)$
form vertical intervals in the~domains
$\Tilde{\mathcal{M}}_{r,\delta}$ on the plane $(E,\cv{I}_1)$.
Denote by~$\Tilde{\Sigma}^{r,\delta}_{K,L}(h,\varepsilon)$
the union of~all these intervals.
We call this set \emph{the semiclassical
up to~$O(\varepsilon^K+h^L)$}
series corresponding to the interior regime
$\Tilde{\mathcal{M}_{r,\delta}}$.

\subsection{Semiclassical spectrum}
The union of the semiclassical series $\Sigma^{r,\delta}_{K,L}(h,\varepsilon)$
and $\Tilde\Sigma^{r,\delta}_{K,L}(h,\varepsilon)$
corresponding to all regimes will be called
\emph{the semiclassical up to $O(h^L+\varepsilon^K)$
spectrum of the~operator $\Hat{H}$}.
We denote this set by~$\Sigma_{K,L}(h,\varepsilon)$.
An example of the structure of the semiclassical spectrum
is shown in Fig.~1.3.

The statements of this~section assert only that
for arbitrary $K$ and $L$ one can find points from
the spectrum of~$\Hat{H}$ in $O(h^{L}+\varepsilon^{K})$-neighborhood
of the set $\Sigma_{K,L}(h,\varepsilon)$. 
Now a natural question arises: what kind of relationship between the~exact
and semiclassical spectrum of the~operator $\Hat H$ exist?
In particular: does the~whole spectrum
of~$\Hat{H}$ belong to a certain
$O(h^L+\varepsilon^K)$-neighborhood of~$\Sigma_{K,L}$?
At the moment the~answer is unknown, but we hope that it is positive.
Our expectations are based, in particular, on the existence
of such relationship between the semiclassical and
the exact spectrum for the Sturm-Liouville problem (Section~2).

From the other side, each point of the semiclassical spectrum
``asymptotically'' has infinite degree of degeneracy
in the sense that one can construct
infinitely many linear independent quasimodes with the same
semiclassical energy; if $\varepsilon$ is small enough, then
there are no isolated points in $\Sigma_{K,L}$.
As the exact spectrum does not have such properties
(at least for rational values of the flux $\eta$),
it is clear that the~semiclassical and the~exact spectrum
do not coincide.
Nevertheless, the semiclassical spectrum gives some
information about the exact one.
Before discussing this question,
in Section $7$ we consider  the situation when $\eta$
is a rational number and try to understand the meaning
of the magneto-Bloch conditions (1.3) and (1.4)
for the semiclassical analysis.

At last let us note that to~construct semiclassical spectrum
up to~$O(h^L+\varepsilon^K)$ it is enough to~know
only the~averaged Hamiltonian $H^K$, see Section~5.

\subsection{Relationship between $h$ and $\varepsilon$,
and the widths of the~Landau bands}
Up to now, we did not assume yet any relationship between
the~parameters $h$ and $\varepsilon$,
but in concrete physical problems
such relationship may appear, say, $h=\varepsilon^\kappa$, $\kappa>0$.
The energy levels $E^{\mu,\nu}_r(h,\varepsilon)$
and $\Tilde{E}^\mu(\cv{I}_2,h,\varepsilon)$ depend on
the parameters  $h$ and $\varepsilon$ in a~regular way, and
this means that increasing of~$K$ and $L$ gives the~correction
to the~energy levels corresponding to smaller numbers $K$ and $L$.
We cannot say the same about the
functions $\psi_{r,l}^{\mu,\nu}$ and $\Tilde{\psi}_{r,k}^{\mu}$,
because the ratio like $\varepsilon/h$ appears
in the formulas for these functions.
This fact does not allow to~use
the~Raleigh-Schr\"odinger perturbation theory based
on the~small parameter $\varepsilon$ (as it was done e.~g. in~[63]
for the case of fixed $h$).

The formulas for the semiclassical spectrum describe the well-known
broadening of the~Landau levels $E_\mu=\cv{I}_1^\mu(h)$
(these numbers are infinitely degenerated eigenvalues
of $\Hat H$ for $\varepsilon=0$)
implied by the~appearance of the~electric field.
For each $\mu$, the union of all the numbers
$E^{\mu}_r(h,\varepsilon)$ and $E^{\mu,\nu}_r(\cv{I}_2,\varepsilon,h)$
for all possible values of $\mu$, $\nu$, and $\cv{I}_2$
will be called
the \emph{$\mu$th semiclassical Landau band}
and denoted by $L_{\mu}(h,\varepsilon)$.

If $\varepsilon(g_{\text{max}}(\cv{I}_1^\mu)-g_{\text{min}}(\cv{I}_1^\mu))<h$,
then the semiclassical Landau bands do not intersect,
and one can calculate their widths:
\begin{equation}
                 \label{diam-est}
\diam L_\mu(h,\varepsilon)=\varepsilon (g_{\text{max}}(\cv{I}_1^\mu)-
g_{\text{min}}(\cv{I}_1^\mu))+O(\varepsilon^2+h^2).
\end{equation}
For the~example (1.7) we have
\begin{equation*}
\diam L_\mu(h,\varepsilon)=2\varepsilon \big(A|J_0(\sqrt{2\cv{I}_1^\mu})|+
B|J_0(\beta \sqrt{2\cv{I}_1^\mu})|+O(h^2+\varepsilon^2)\big).
\end{equation*}

For small and large values $\cv{I}^\mu_1$
one can use the estimates for the Bessel functions
(see Section 3); in particular, for large $\cv{I}_1^\mu$ we have
$\diam L_\mu(h,\varepsilon)\approx(\cv{I}_1^\mu)^{-1/4}$.
For the~example (1.7) we have
\begin{align*}
\diam L_\mu(h,\varepsilon)&\approx \varepsilon \big(A(2-\cv{I}_1^\mu)+
2B\beta (2-\cv{I}_1^\mu)\big) & \text{for small }\cv{I}_1^\mu,\\
\diam L_\mu(h,\varepsilon)&\approx
\varepsilon\big(A\sqrt {\frac{2}{\pi\sqrt {2\cv{I}_1^\mu}}}
(|\cos (\sqrt {2\cv{I}_1^\mu}-\pi/4)|+{}\\
&~~{}+{\frac{B}{\sqrt {\beta}}}|\cos (\beta \sqrt {2\cv{I}_1^\mu}-\pi/4)|)\big)
& \text{for large }\cv{I}^\mu_1.
\end{align*}

Note also that numerical considerations~[51] show that the flux-energy
diagram for the~operator $\Hat H$ for each Landau band
in the plane $(E,\eta)$ looks like a~butterfly (``Hofstadter butterfly'').
Our assumptions about the smallness of $h$ and $\varepsilon$
mean that we consider the~asymptotics of the spectrum
corresponding to the~upper part of~the~Hofstadter butterfly.

In this paper, we do not consider the~structure of the~spectrum in
the~neighborhood of the singular boundaries;
the standard semiclassical approximation does not work there.
It is clear that this asymptotics depends of the type of singularity,
and in some cases is based of the parabolic cylinder functions.
We will study this question in forthcoming papers.

\subsection{Quantum averaging and the 
Harper-type equations}\label{s-harper}
Basing on the Correspondence Principle, one can expect that
there exist some quantum analogies of the averaging procedure
(or, more generally, of the canonical transformation)
in the classical mechanics.
The study of such correspondence is developed now in
several directions.

One can consider the Schr\"odinger equation as an infinite-dimensional
Hamiltonian system and use for its solution different nontrivial
generalizations of the~classical averaging methods
(see [11, 58, 85, 86]).
This view on the Schr\"odinger equation with operator $\Hat{H}$
was used in [11], and this study is connected with non-commutative
analysis.

Another interpretation of the quantum averaging is based on the construction
of an operator corresponding to the canonical change of variables,
and this approach exploit the idea that a canonical transformation
in the classical mechanics implies (with some accuracy)
a unitary transformation in the quantum mechanics.
This procedure approximately reduces the~original spectral problem
to the set of low-dimensional ones [54] and
requires approximate solutions of some~quantization problems.
Such an approach for the problem under consideration
gives the results formulated above.
We have obtained these results by direct applying of the
averaging and semiclassical methods;
now let us try now to find a correspondence between the spectral
problem for $\Hat{H}$ and one-dimensional spectral problems
basing on this interpretation of quantum averaging.


Roughly speaking, according to~[54], to construct
the first nontrivial approximation (with
respect to parameters $h$ and $\varepsilon$)
to the solutions of the equation $\Hat H \Psi =E\Psi $
one has to solve approximately the equation
\begin{equation}
                \label{harp-1}
\cv{H}\Big(
\frac{1}{2}(-h^2\frac{\partial^2}{\partial \cv{Q}^2}+\cv{Q}^2),
-ih\frac{\partial}{\partial\cv{y}_2},\cv{y}_2,\varepsilon\Big)
\Phi(\cv{Q},\cv{y}_2)=E\Phi(\cv{Q},\cv{y}_2),
\end{equation}
where all the operations are ordered according to the Weyl rule.
The operator in the left-hand side commutes with
the harmonic oscillator
$\frac{1}{2}(-h^2{\partial^2}/{\partial \cv{Q}^2}+\cv{Q}^2)$;
this observation lets us write the solutions of~\eqref{harp-1}
in the form $\Phi(\cv{Q},\cv{y}_2)=\psi_\mu(\cv{Q})w_\mu(\cv{y}_2)$,
where $\psi_\mu$ is the $\mu$th eigenfunctions of the harmonic oscillator,
and equation~\eqref{harp-1} is reduced to the family of equations
\begin{multline}
                 \label{harp-2}
\cv{H}\big(\cv{I}_1^\mu(h),
-ih\frac{\partial}{\partial\cv{y}_2},\cv{y}_2,\varepsilon\big)
w_\mu(\cv{Q},\cv{y}_2)=E w_\mu(\cv{y}_2),\quad \mu\in\Z_+.
\end{multline}
In particular, if we take $\Bar{H}$ instead of $\cv{H}$
for the example~(1.7), then the equations~\eqref{harp-2}
will have the form
\begin{multline*}
A J_0(\sqrt{2\cv{I}_1^\mu})\dfrac{w_\mu(\cv{y}_2+h)+w_\mu(\cv{y}_2-h)}{2}\\
{}+B J_0(\beta\sqrt{2\cv{I}_1^\mu})\cos \beta \cv{y}_2\, w_\mu(\cv{y}_2)=
\frac{E-\cv{I}_1^\mu}{\varepsilon}w_\mu(\cv{y}_2),
\end{multline*}
i.e. they form a family of the Harper equations.
Using this analogy, the equations~\eqref{harp-2}
are usually called the \emph{Harper-type} equations.
Approximate solutions of each of them can be found
using the usual one-dimensional WKB method.
The description of the spectral asymptotics
for equations of such kind using the Reeb graphs
was obtained in~[39].
Thefore, from this point of view, each semiclassical
Landau band is described by a certain Harper-type equation;
this equation depends on the band, and, therefore,
different Landau bands can have different asymptotic
structure. As follows from the preceding,
the asymptotics of $\mu$th Landau band can be obtained
by the quantization of the Reeb graph for the function
$\cv{H}\big(\cv{I}_1^\mu(h),\cv{y}_1,\cv{y}_2,\varepsilon\big)$
considered as a function on the torus $\mathbb{R}^2/(a_1,a_2)$,
what can be easily seen from Fig.~1.3.

\section{The spectral asymptotics in the case of rational flux}

Consider now the case when the flux $\eta:=a_{22}/h$ is a rational number,
$\eta=N/M$, where $N$ and $M$ are mutually prime integer numbers
and $M>0$. As mentioned in Section~1, in this case to each point
from the spectrum of $\Hat{H}$ one can assign a family
of eigenfunctions satisfying the magneto-Bloch conditions~(1.3) and~(1.4).
Crearly, the functions $\psi_{r,l}^{\mu,\nu}$
and $\Tilde\psi_{r,k}^{\mu,r}$ do not satisfy these conditions,
vbut, like in subsection 2.5,
we can use them as a~base for construction
of~quasimodes $\Psi^{r,j}_{\mu,\nu}$, $j=0,\ldots,M-1$,
satisfying~(1.3) and~(1.4) (cf.~[73, 87]).
For convenience, we call the set of $M$
quasimodes satisfying the magneto-Bloch conditions
as a \emph{family of magneto-Bloch quasimodes}.
We expect that this procedure will improve a detalization
of the asymptotics to the spectrum.

\subsection{Magneto-Bloch quasimodes corresponding
to the boundary regimes}

Let us consider a certain boundary regime $\cv{M}_{r,\delta}$
and the corresponding semiclassical spectral series
$\Sigma^{r,\delta}_{K,L}(h,\varepsilon)$.
\begin{prop}
                      \label{p_7_1}
For any $(q_1,q_2)\in [0,1/M)\times[0,1)$ and any
$(\cv{I}_1^\mu,\cv{I}_2^\nu)\in \cv{M}_{r,\delta}$
there exist exactly $M$ families of magneto-Bloch quasimodes
of the form \textup{(}cf.~\textup{(2.18))}
\begin{equation}
      \label{e_7_1}
\begin{gathered}
\Psi^{r,s,j}_{\mu,\nu}(x,h,\varepsilon,q)=
\sum_{l\in \Z^2}C^{s,j}_l(q_1,q_2,h)\psi_{r,0}^{\mu,\nu}(x-l\cdot a,h,\varepsilon)
e^{-\frac i h x_1l_2a_{22}},\\
s,j=0,\ldots,M-1,
\end{gathered}
\end{equation}
such that all the functions $\Psi^{r,s,j}_{\mu,nu}$
are linearly independent.
The coefficients $C^{s,j}_l(q)$ in~\eqref{e_7_1}
may be chosen in the~following form \textup{(}cf.~\textup{(2.38)):}
\begin{equation}
C^{s,j}_{l_1,l_2}(q_1,q_2,h)=
\left\{
\begin{array}{l}
\exp\Big[-2\pi i (q_1l_1+q_2 n) +2\pi i \eta l_1 j-i\eta l_2 a_{21}/2\Big],\\
\qquad\qquad \text{ if } l_2+j-s+nM=0, \quad n\in\Z,\\
0,  \text{ otherwise.}
\end{array}\right.
             \label{Cqq}
\end{equation}
Here $l=(l_1,l_2)\in\Z^2$, the index
$s=0,\ldots,M-1$ indicates the family of magneto-Bloch
quasimodes, the index $j=0,\ldots,M-1$ indicates
the number of a member in each of these families.
\end{prop}

To prove this Proposition one has to substitute the sum~\eqref{e_7_1}
into~(1.3) and~(1.4), to~equate the coefficients of
$\psi_0^{\mu,\nu,r}(x-l\cdot a,h,\varepsilon)$, and to study
the infinite linear system obtained.

We see that for each fixed value
of the quasimomentum $q$ there are $M^2$ magneto-Bloch
quasimodes corresponding to the same spectral value;
in the other words, we have a degeneracy of degree $M^2$.
We try to give an interpretation of this fact in Section~8.

\begin{figure}\centering
\includegraphics[height=40mm]{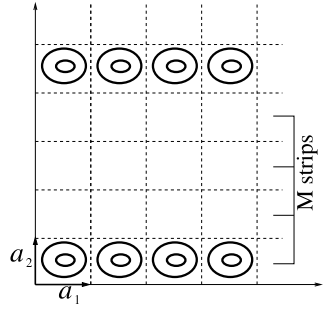}
\caption{Asymptotic supports of magneto-Bloch
quasimodes $\Psi^{r,s,j}_{\mu,\nu}$ on the $x$-plane}\label{f71}
\end{figure}

Let us describe the structure of the functions
$\Psi^{r,s,j}_{\mu,\nu}$.
If the number $\cv{I}_1^\mu$ is small enough, then
the~asymptotic support of~each of~them consists of family of annulas.
These annulas form strips separated
by array of $M-1$ ``empty'' strips (see Fig.~\ref{f71}), where the corresponding quasimodes
have order $O(h^\infty)$.
If $\eta$ tends to an irrational number, then $M\to \infty $, and
only one strip is kept.
This means, probably, that each of generalized eigenfunctions
in the irrational flux case is asymptotically localized
in such isolated strips.
The diameters of the annulas depend on $\cv{I}_1^\nu$
(i.e., on the index of the Landau band); if the number $\cv{I}_1^\mu$
is large, then these annulas may intersect, and, probably,
can cover the whole plane $\R^2_x$.

\subsection{Magneto-Bloch quasimodes corresponding
to the interior regimes}

Now consider the quasimodes associated with the almost Liouville
cylinders. Like in the previous subsection,
the functions $\Tilde\psi_{r,k}^{\mu}$ do not satisfy
the magneto-Bloch conditions, and we
again use them as a base for constructing magneto-Bloch quasimodes,
i.e. we put 
\begin{equation}
      \label{e_qimr}
\Tilde \Psi^{j}_{\mu}(x,h,\varepsilon,q)=\sum_{k\in \Z}C^{j}_k(q_1,q_2,h)\ 
\Tilde\psi_{k}^{\mu}(x,\cv{I}_2,\varepsilon,h), \quad q=(q_1,q_2).
\end{equation}
To obtain the expression for the coefficients $C^j_k$,
let us substitute the expressions~\eqref{e_qimr} into (1.3) and~(1.4),
take into account the property~\eqref{e-temp-2},
and then equate the coefficients of
$\Tilde \psi_{r,k}^{\mu}(x+k(Jf)\cdot a,\cv{I}_2,\varepsilon,h)$
for all $k$. This system has the following form:
\begin{equation}
                            \label{cjk-d}
\left\{\begin{array}{l}
\displaystyle C^{j}_k\exp\Big[2\pi\frac{i}{h}\cv{I}_2
+2\pi i k\eta -\frac{1}{2}i(2\pi d_1+a_{21} d_2)d_2\Big]\\
\displaystyle \quad{}=C^{(j+d_2)\tmod M}_k \exp\Big[-2\pi i d_1(q_1-j\eta)
-i\eta d_2^2 a_{21}/2+2\pi i L'q_2 \Big],\\
\displaystyle \qquad{}-L'\le\frac{d_2+j}{M}\le(-L'+1)-\frac{1}{M},
\quad L'\in\mathbb{Z},\\[\smallskipamount]
\displaystyle C^j_{k-1}\exp\Big[i(k-1) f_1\eta(-2\pi f_2+a_{21} f_1)\Big]\\
\displaystyle \quad{}=C^{(j+f_1)\tmod M}_k\exp\Big[2\pi i f_2(q_1-j\eta)
	-i\eta f_1^2 a_{21}/2+2\pi i L''q_2\Big],\\
\displaystyle \qquad{}-L''\le\frac{f_1+j}{M}\le(-L''+1)-\frac{1}{M},
\quad L''\in\mathbb{Z}
\end{array}\right.
\end{equation}
(here and later by $A\tmod M$ we mean the remainder
after the integer division of $A$ by $M$).                                                          
Obviously, the system depends crucially on the drift vector $d$
of the regime.
Let us consider first the case $d=(\pm1,0)$,
then the system for the coefficients takes the form
\begin{equation}
                 \label{cjk-1}
\begin{gathered}
C^{j,\pm}_k e^{\pm 2\pi\frac{i}{h}(\cv{I}^\pm_2+k a_{22})}
=C^{j,\pm}_k e^{-2\pi i(q_1-j\eta)}, \quad k\in\mathbb{Z},
\quad j=0,\ldots,M-1,\\
C^{(j+1)\tmod M,\pm}_{k\pm 1} =
C^{j,\pm}_ke^{i\eta a_{21}/2\pm ik\eta a_{21}} \sigma_j,\\
k\in\mathbb{Z},\quad j=0,\ldots,M-1,
\quad\sigma_{M-1} =e^{2\pi i q_2},
 \quad \sigma_j=1,\quad j\ne {M-1},
\end{gathered}
\end{equation}
where the index $\pm$ corresponds to $d=(\pm1,0)$.
We see that all the coefficients are uniquely determined
by arbitrary chosen numbers $C^{j,\pm}_0$, $j=0,\ldots,M-1$,
and we have, therefore,
at most $M$ families of magneto-Bloch functions.
For $s$th family, $s=0,\ldots,M-1$, set $C^{s,j,\pm}_0=\delta_{sj}$.
It is easy to see that the equality
\begin{equation}
\cv{I}_2^\pm=\cv{I}^\pm_2(n^\pm,q_1,h)=
h\Big(\frac{n^\pm}{M}\mp q_1\Big),\quad n^\pm\in\mathbb{Z},
\end{equation}
where $\cv{I}^\pm_2$ is the action variable corresponding to
the drift vector $d=(\pm1,0)$,
is a necessary condition for the existence of solutions
for~\eqref{cjk-1}.
Obviously, the coefficients $C^{s,j,\pm}_k$ can be obtained
from one set, say, $C^{0,j,\pm}_k$, by the shift of the index $j$
(and this means that really we have only
one family of magneto-Bloch quasimodes).
Therefore, the resulting coefficients can be chosen in the form
\begin{equation}
C^{j,\pm}_k=\begin{cases}
e^{i\eta k^2 a_{21}/2+2\pi i n q_2}, & \text{ if } j\mp k+
nM=0,\\
&\phantom{\text{ if }}n=n^\pm\Tilde{N}+\Tilde{n}M,\quad \Tilde{n}\in\mathbb{Z},\\
0, & \text{ otherwise},
\end{cases}
\end{equation}
where $\Tilde{N}$ is an integer number such that for some
$\Tilde{M}\in\mathbb{Z}$ one has $\Tilde{N}N+\Tilde{M}M=1$.                            

\begin{figure}
\centering
\begin{minipage}{60mm}\centering
\includegraphics[width=45mm]{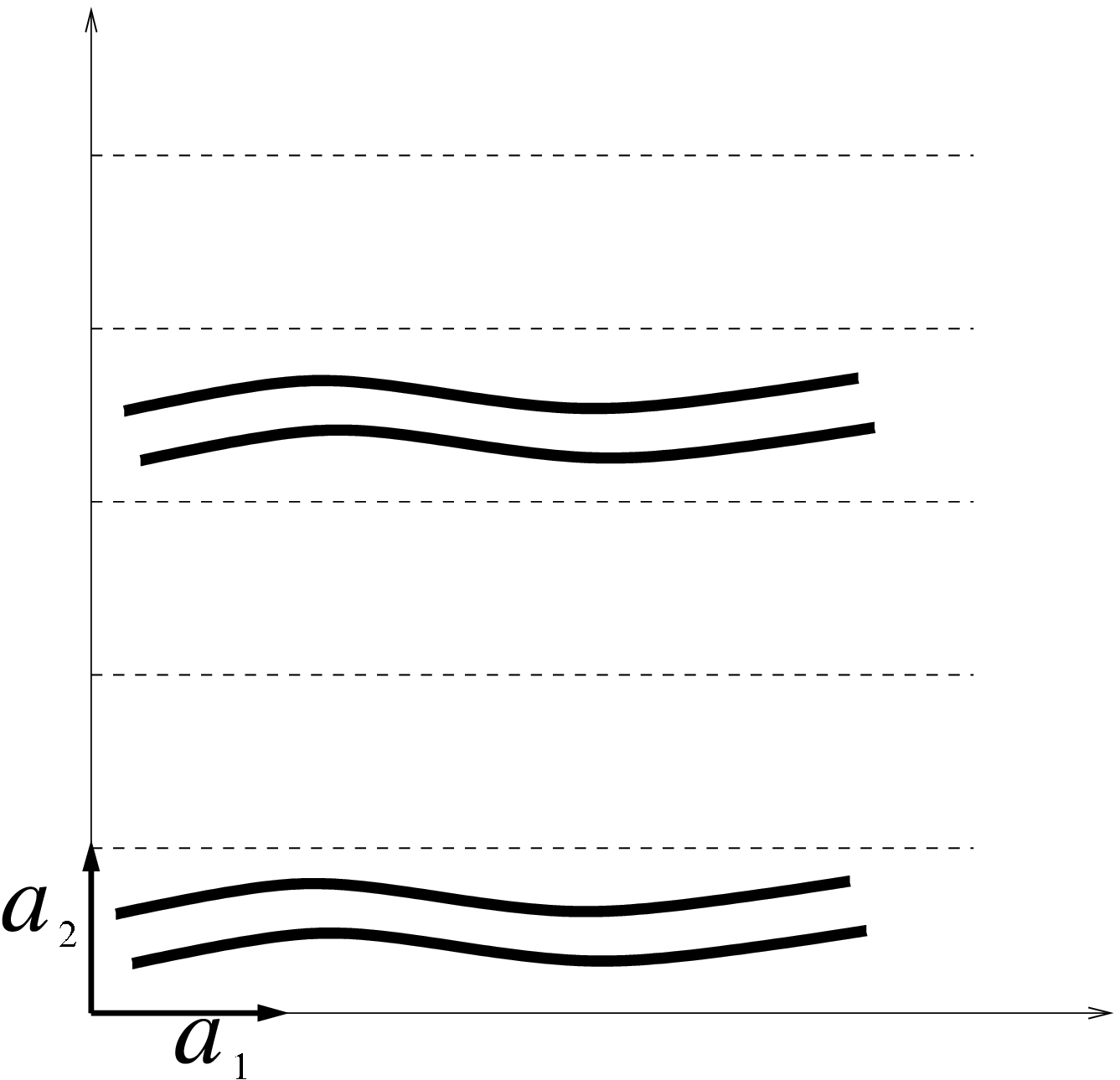}\\(a)
\end{minipage}
\begin{minipage}{60mm}\centering
\includegraphics[width=45mm]{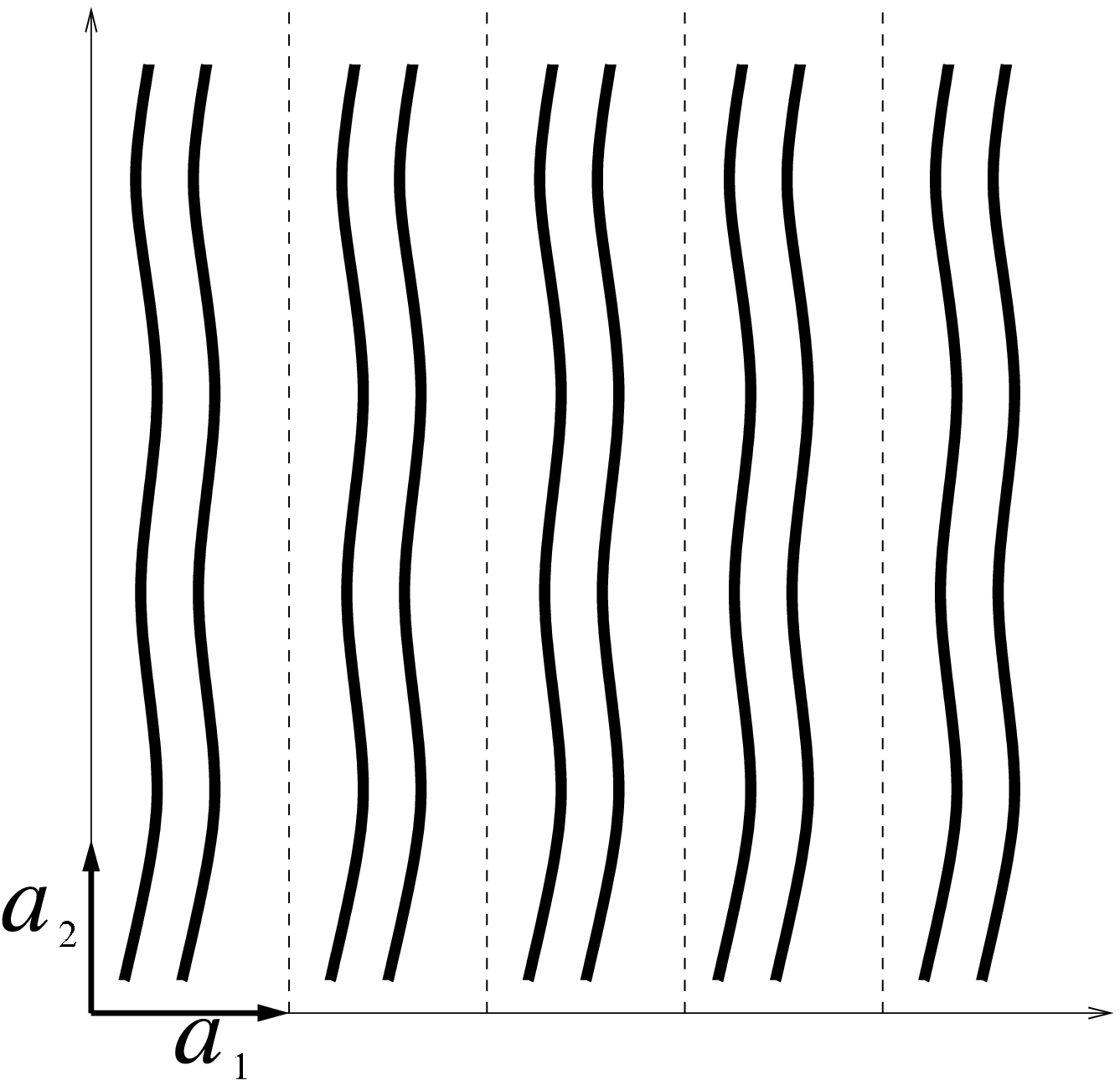}\\(b)
\end{minipage}
\caption{The structure of the asymptotic support
for the magneto-Bloch quasimodes $\Psi^{j}_{\mu}$:
(a) the drift vector is $(1,0)$; (b) the drift vector is $(0,1)$}
\end{figure}

The expressions for the coefficients $C^{j}_k$ for the cases
$d\ne(\pm1,0)$ are rather complicated. At least, it is clear
from the system~\eqref{cjk-d} that all the coefficients
are non-zero in this case. From this point of view,
the case $d=(\pm1,0)$ is the most useful one, because
the asymptotic support of any magneto-Bloch quasimode
is minimal in this case.
From the other side, a situation with an arbitrary drift vector
can be reduced to the case $d=(\pm1,0)$,
if one turns the coordinates, applies a gauge transformation,
and transform the magneto-Bloch conditions accordingly.
But it is important to emphasize, that this reduction
is \emph{not} global, because the drift vector can jump
when passing from regime to regime even in the simplest cases,
and to obtain reasonable formulas
for the magneto-Bloch quasimodes one has to apply different
transformation for different regimes (or, respectively,
these transformations depend on the Landau band),
and \emph{there exist no ``globally good'' coordinates}.

\section{Discussion and heuristic estimate of numbers of subbands}

\subsection{Relationship between the true and the semiclassical spectra.
Dispersion relations}

Let us give a qualitative description of the semiclassical spectrum
in the rational flux case basing on the considerations of the previous
section.

Let us consider a fixed semiclassical Landau band with index~$\mu$.
Its part corresponding to a boundary section is discrete,
and each point $E^{\mu,\nu}_r$ is $M^2$-fold degenerated.
The part of the band corresponding to an interior regime
is continuous, but now each point
$\Tilde{E}^{\mu}_r(\cv{I}_2,h,\varepsilon )$ is only $2M$-fold degenerated
as $\Tilde{E}^{\mu}_r(\cv{I}_2,h,\varepsilon)$ can coincide for different edges
of the~Reeb graph.
(Here we recall that we consider the situation with the simplest
Reeb graphs, see \S4.2; otherwise these estimations
are estimations from below.) From the other side, if the number
$\eta$ is rational,
then the true spectrum of $\Hat{H}$ has band structure
and each point of the true spectrum is $M$-fold degenerated.
Let us try to give an interpretation of the ambiguous degeneracy
of the points $E^{\mu,\nu}_r$.

The existence of the isolated points $(\cv{I}_1^\mu,\cv{I}_2^\nu)$
can be explained then as follows. Our previous considerations
have a non-avoidable error $O(h^\infty+\varepsilon^\infty)$.
Therefore, we can expect that these isolated points really
approximate subbands of width $O(h^\infty+\varepsilon^\infty)$.
The presence of $M^2$-fold degeneracy of these points
probably means that in a neighborhood of each such point
there are $M$ true spectral subbands (minibands) of the operator $\Hat{H}$.
(Additional arguments can be given basing
on the following idea: the formulas~\eqref{Cqq}
realize a representation of the magnetic translation
group on the space of asymptotic eigenfunctions;
as all such representations are $M$-dimensional~[95],
small variation of parameters leads to the splitting
of each energy level into $M$ numbers.)
Enumerate these bands by an index $s=0,\ldots,M-1$,
then the dispersion relations $E^{\mu,\nu}_{r,s}(q_1,q_2,h,\varepsilon)$
in all these minibands look as follows:
\begin{equation*}
E^{\mu,\nu}_{r,s}(q_1,q_2,h,\varepsilon)=E^{\mu,\nu}_{r,K,L}
(h,\varepsilon)+O(h^L+\varepsilon^K)
\end{equation*}
for any positive $K$ and $L$.
Therefore, the approximation used does not give expressions
for the dispersion relations, and the presence of
the separation of bands cannot be found rigorously.
Nevertheless, this splitting is supported by
the analogy with the one-dimensional periodic problem
(subsection 2.3). Assume that there exists a function
$\Breve\psi^{\mu,\nu}_{r,0}$ such that the \emph{true}
magneto-Bloch functions can be represented as
\begin{equation*}
\Psi^{r,s,j}_{\mu,\nu}(x,q,h,\varepsilon)=
\sum_{l=(l_1,l_2)\in\mathbb{Z}^2}
C^{s,j}_l(q,h) \Breve\psi^{\mu,\nu}_{r,0}(x-l\cdot a,h,\varepsilon)
e^{-\frac{i}{h}l_2a_{22}x_1},
\end{equation*}
where the coefficients $C^{s,j}_l$ are defined in Proposition~\ref{p_7_1}.
Similarly to (2.23) we obtain that if $\Psi_1$ and $\Psi_2$
are (generalized) eigenfunctions of $\Hat{H}$ with
eigenvalues $E_1$ and $E_2$, then
\begin{equation}
             \label{E-Psi}
E_1-E_2=\dfrac{
\Re\oint\limits_{\partial D}
\Big[h^2(\overline{\Psi}_1\nabla\Psi_2-\nabla\overline{\Psi}_1\Psi_2)
-ih\overline{\Psi}_1\Psi_2 A\Big] ds
}{\Re\int_D \overline{\Psi}_1\Psi_2 dx_1 dx_2},
\end{equation}
where $A=(-x_2,0)$ is the vector potential of the magnetic field,
$ds=(dx_1,dx_2)$, and $D\subset\mathbb{R}^2$ is any domain
with boundary $\partial D$.
Assume that the asymptotic support of $\Breve\psi^{\mu,\nu}_{r,0}$
belongs to the unit cell generated by the vectors
$a_1$ and $a_2$. Let us choose this unit cell as the domain $D$.
Put $\Psi_{1/2}=\Psi_{\mu,\nu}^{r,s,j}(x,q^{1/2},h,\varepsilon)$
and $E_{1/2}=E^{\mu,\nu}_{r,s}(q^{1/2},h,\varepsilon)$.
Substituting all these expressions into~\eqref{E-Psi},
one obtains (cf.~(2.25)):
\begin{multline}
E^{\mu,\nu}_{r,s}(q^1,h,\varepsilon)-E^{\mu,\nu}_{r,s}(q^2,h,\varepsilon)\\
{}\approx
\sum_{l_1,n=\overline{0,\pm 1}} \rho^{\mu,\nu,r}_{n,l_1}
(e^{2\pi(q^1_1l_1+q^1_2n)}-e^{2\pi(q^2_1l_1+q^2_2n)})
e^{2\pi i \eta l_1 s},
\end{multline}
where
\begin{gather*}
\rho^{\mu,\nu,r}_{n,l_1}=
\dfrac{
\Re\oint_{\partial D}
\Big[h^2({{\Breve\psi}^{\mu,\nu}_{r,l}}\overline{\nabla{\Breve\psi}^{\mu,\nu}_{r,0}}
-\nabla{\Breve\psi}^{\mu,\nu}_{r,l}\overline{{\Breve\psi}^{\mu,\nu}_{r,0}})
-ih\overline{{\Breve\psi}^{\mu,\nu}_{r,0}}{\Breve\psi}^{\mu,\nu}_{r,l}
A \Big]ds
}{\Re\int_D \big|\Breve\psi^{\mu,\nu}_{r,0}\big|^2 dx_1 dx_2}=O(h^\infty),\\
l=(l_1,nM).
\end{gather*}

\begin{figure}\centering
\begin{minipage}{65mm}\centering
\includegraphics[height=60mm]{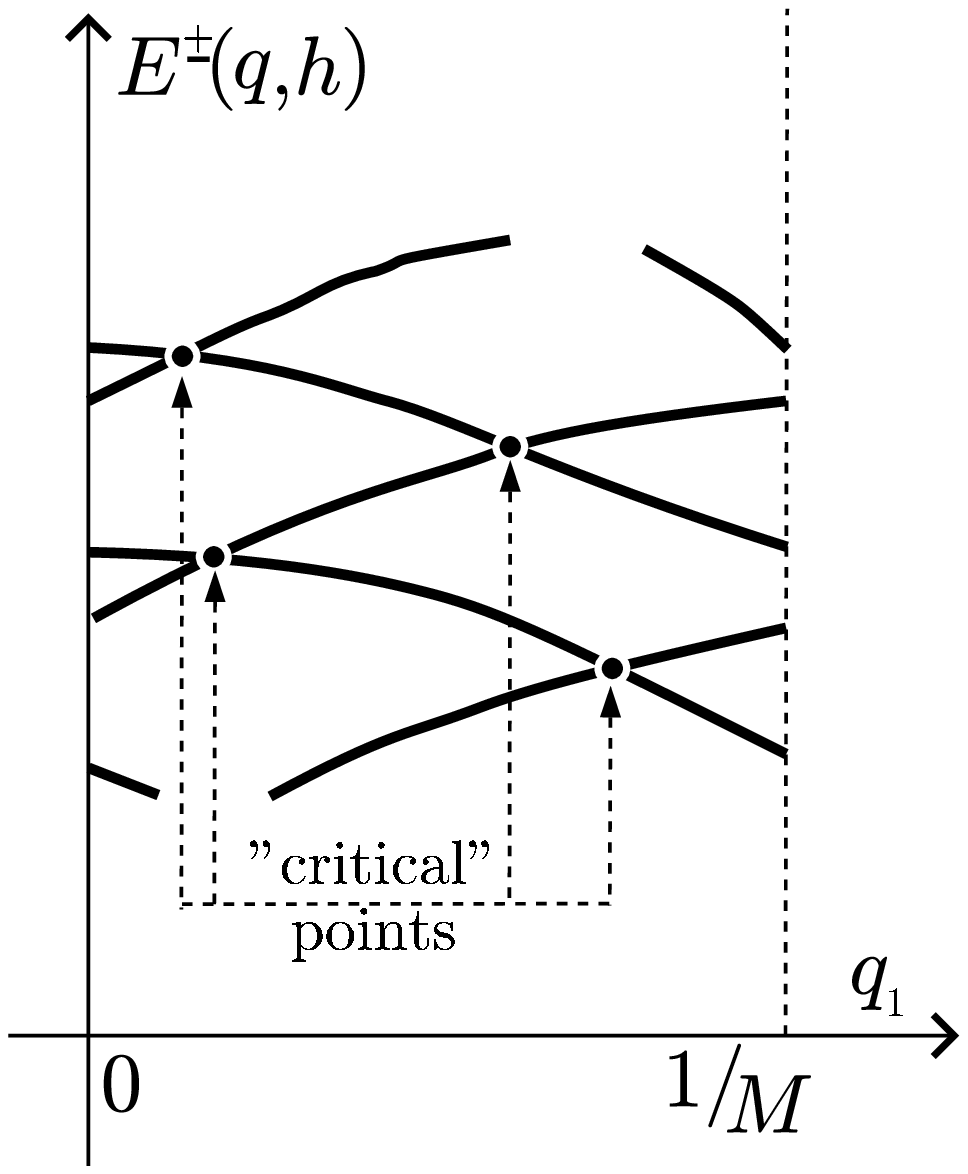}
\caption{Semiclassical dispersion relations}\label{f81}
\end{minipage}~~
\begin{minipage}{65mm}\centering
\includegraphics[height=60mm]{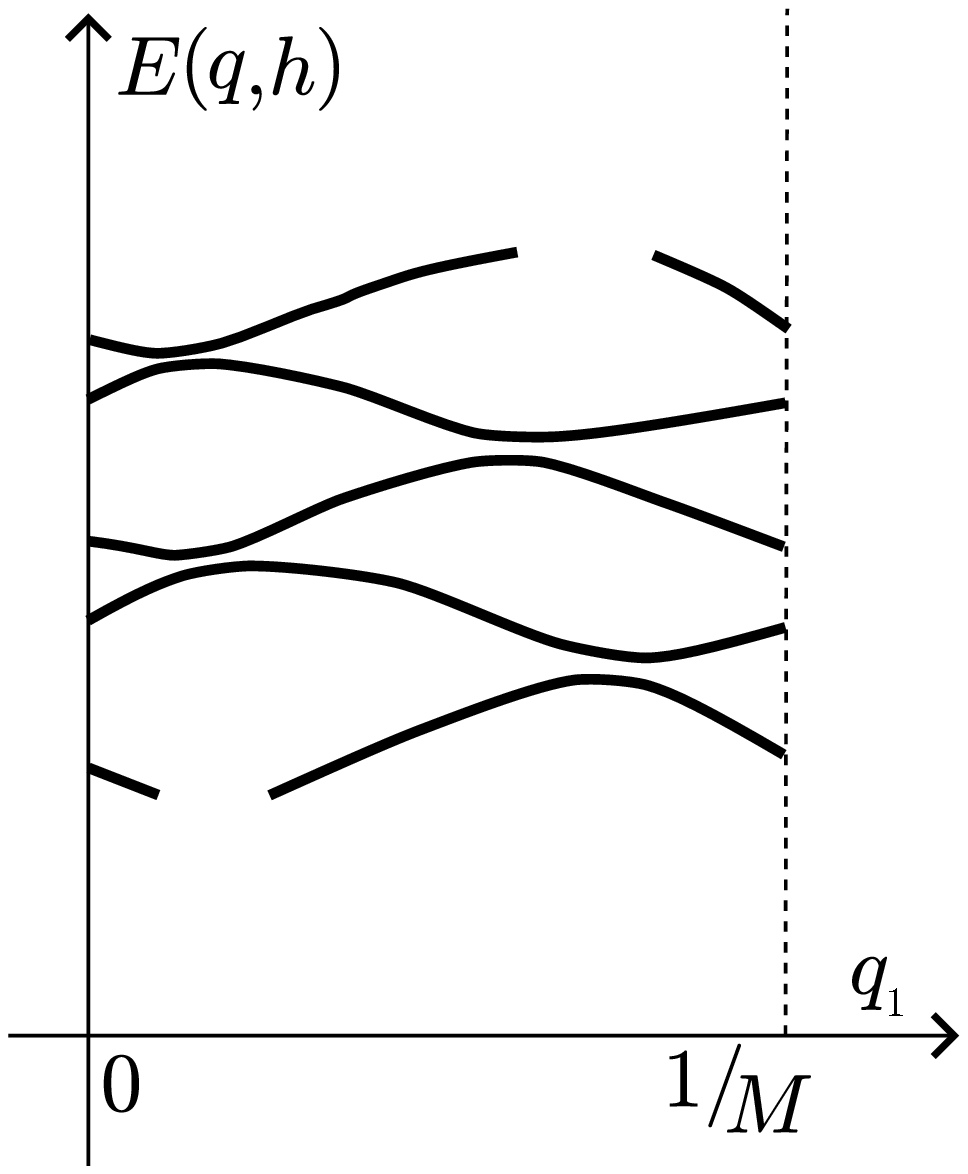}
\caption{True dispersion relations (hypotesis)}\label{f82}
\end{minipage}
\end{figure}

In the interior regimes,
we have the following dependences of the energy
on the quasimomenta (semiclassical dispersion relations):
\begin{equation*}
E^\pm(q,h)=E_{n^\pm}(q,h,\varepsilon)=E^\mu_{r}\big(\cv{I}_2^\pm(q,n^\pm,h),h,\varepsilon\big).
\end{equation*}
Consider the case with drift vector $(\pm1,0)$,
then these functions depends essentially only on 
$q_1$, and the dependence on $q_2$ is absent
up to $O(h^L+\varepsilon^K)$.
As some of these functions increases in $q_1$
and others decreases, for some critical values
of $q_1=q_1^*$ one has
\begin{equation*}
E_{n_1^+}(q_1^*,h,\varepsilon)=E_{n_2^-}(q_1^*,h,\varepsilon)
\end{equation*}
(see Fig.~\ref{f81}).
For the example (1.7), these points correspond
to the values $\cv{I}^{2/3}_2=h(2n+1)/(2M)$.
We can expect that these points together with the ``end'' points
$\cv{I}^{2/3}=h n/M$
are $O(h^\infty+\varepsilon^\infty)$-approximations
of the gaps in the spectrum of $\Hat{H}$, see Fig.~\ref{f82}.
This expectation is based, in particular, on the analogy with
the one-dimensional periodic problem (subsection 2.5),
cf.~[39].
If the interior regime has the drift vector other than
$(\pm1,0)$, then these semiclassical dispersion relations
depend on a certain linear combination of $q_1$ and $q_2$.

Our hypotesis about the structure of the Landau
bands are illustrated in~Fig.~1.5.

Note that in our spectral estimates we have used
the almost invariant Liouville tori and cylinders.
It is known that even an exponentially
small correction, which is kept after applying
the averaging procedure,
can destroy some of these objects, and non-Kolmogorov sets
may appear.
These fact implies the following question: what do non-Kolmogorov sets
mean for the exact spectrum of the operator $\Tilde H$, in particular
in situation when the flux $\eta$ is rational?
It seems that the answer cannot be given using the additive
asymptotics.

\subsection{Heuristic estimate for the numbers of subbands}

If the hypotesys of the previous subsection is true,
then it is possible to count the number of spectral subbands
corresponding to a fixed (semiclassical) Landau band.
Let us consider a certain fixed value $\cv{I}^\mu_1$.

For the edge $i_1$ of the corresponding Reeb graph,
the number of the quantization points $\cv{I}_2^\nu$ is equal approximately
to $-\cv{I}_2^{1+}/h$,
and for the edge $i_4$ this number is equal to $\cv{I}_2^{4-}/h$.
Each of these points subbands, so the end edges
$i_1$ and $i_4$ give us approximately (modulo singular boundaries effects)
\begin{equation*}
\frac{M}{h}\ (\cv{I}_2^{1+}-\cv{I}_2^{4-})=
2\pi N\ \frac{\cv{I}_2^{1+}-\cv{I}_2^{4-}}{a_{11}a_{22}}
\end{equation*}
bands.

The expected numbers of subbands implied by the edges $i_{2/3}$ 
depends may depend on the symmetry properties of the potential $v$.
For the example~(1.7), we obtain approximately
(again modulo singular boundaries effects):
\begin{equation*}
2M\ \frac{\cv{I}_2^{2+}-\cv{I}_2^{2-}}{h}=
4\pi N\ 
\frac{
\cv{I}_2^{2+}-\cv{I}_2^{2-}
}{a_{11}a_{22}}
\end{equation*}
bands.

Therefore, the total number of the subbands
is approximately equal to
\begin{equation*}
2\pi N\ 
\frac{\cv{I}_2^{1+}-\cv{I}_2^{4-}+\cv{I}_2^{2+}-
\cv{I}_2^{2-}+\cv{I}_2^{3+}-\cv{I}_2^{3-}}{a_{11}a_{22}}
\end{equation*}
(we have used the symmetry property $\cv{I}^{2+}_2-\cv{I}^{2-}_2=\cv{I}^{3+}_2-\cv{I}^{3-}_2$)
and the latter number is precisely  equal to $N$
according to the Kirchhoff law (4.5) for the action variables.

\subsection{Correspondence to difference equations}
In conclusion, let us discuss the relationship
between the problem under consideration
and the difference equation (subsection~\ref{s-harper})
in the rational flux case. It is known that the spectrum
of a rational Harper-like operator with flux $\eta=N/M$
consists of $N$ bands [39]; the presence of $M$-grouped
bands corresponding to finite motion (boundary regimes)
was studied numerically for some cases [39].
Therefore, we can expect, that our hypotesis about
the spectrum of $\Hat{H}$, in particular, the estimate
of the number of subbands in each Landau band is not
connected with the simplicity of the potential.

%

\appendix

\section{The canonical operator and spectral estimates}

As it was noted above, the asymptotics of the spectrum
can be found using the canonical operator.
Let us remind some basic properties of the canonical
operator; a more detailed constructions can be found,
for example, in~[60] or~[68].

Let $\Lambda$ be a closed Lagrangian manifolds without
boundary in the space $\mathbb{R}^{2n}_{p,x}$.
The canonical operator $\cv{K}_{\Lambda}$ corresponding
to this manifold maps from $C^\infty(\Lambda)$ to $C^\infty(\mathbb{R}^n)$,
and for any function $f\in C^\infty(\Lambda)$ one has
$Kf=0$ outside certain $\delta$-vicinity of the projection $\pi_x\Lambda$
The canonical operator on $\Lambda$ can be constructed
iff
\begin{equation*}
\frac{1}{2\pi}\oint_\gamma p\,dx =h\Big(n+\frac{{\rm Ind}~\gamma}{4}\Big),
\quad n\in\mathbb{Z},
\end{equation*}
for each basis cycle on $\Lambda$, where ${\rm Ind}$ denotes
the Maslov index.
Applying this consideration to the tori $\Lambda^r_l$
and the cylinders $\Tilde\Lambda^r_k$ we obtain the quantization
condition~\eqref{e_6_1} and~\eqref{e_6_2} for the tori (they
have two basis cycles), and~\eqref{e_6_1} for the cylinders
(only one basis cycle).

The following \emph{commutation formula} is one of the most important
properties of the canonical operator.

\begin{prop}[Commutation formula]
Assume that $\Lambda$ is an invariant Lagrangian manifold
of a Hamiltonian system for a certain Hamiltonian $H$,
and that on $\Lambda$ there exists a volume form
which is also invariant under the Hamiltonian system.
There exists a sequence of differential
operators $\{R^j\}_{j=1}^\infty$,
\begin{equation*}
R^j: C^\infty(\Lambda)\mapsto C^\infty(\Lambda),
\end{equation*}
with smooth coefficients such that
for any function $\varphi\in C^\infty(\Lambda)$
and any number $N\in\mathbb{N}$ one has
\begin{equation}
                  \label{comm-improved}
\Hat{H}\cv{K}_{\Lambda}\varphi=
\cv{K}_{\Lambda}\Big(\sum_{j=0}^N (ih)^j R^j\varphi\Big)+O(h^{N+1}).
\end{equation}
In particular, $R^0$ is the operator of multiplication
by the scalar function $H|_{\Lambda}$,
and
\begin{equation*}
R^1=-\frac{d}{dt}\Leftrightarrow-\frac{\partial H}{\partial p}
\frac{\partial}{\partial x}+
\frac{\partial H}{\partial x}
\frac{\partial}{\partial p}.
\end{equation*}
Here $\Hat{H}$ is the Weyl quantization of the Hamiltonian $H$.
\end{prop}

Now let us try to construct an approximate solutions
of the equation $\Hat{H}\Psi=E\Psi$ up to $O(h^L+\varepsilon^K)$
basing on the commutation formula and almost
Liouville tori $\Lambda^r_l$.
Assume that the conditions~\eqref{e_6_1} and~\eqref{e_6_2}
are satisfied.
Let us find requested solutions in the form
\begin{equation*}
\Psi=\cv{K}_{\Lambda^r_l}u,
\quad
u=\sum_{j=0}^{L-1} (ih)^j u_j,\quad
u_j\in C^\infty(\Lambda^r_l),\quad 
E=\sum_{j=0}^{L-1}(ih)^j\lambda_j.
\end{equation*}
Applying the commutation formula, we obtain
\begin{equation*}
(\Hat{H}-E)\Psi=
\Big\{\sum_{n=0}^{L-1}\sum_{s+j=n}(R^j-\lambda_j)u_s\Big\}
+O(h^L+e^{-C/\varepsilon})
\end{equation*}
(the presence of the term $O(e^{-C/\varepsilon})$ is implied by the fact
that the manifolds $\Lambda^r_l$ are not invariant under $H$,
but only almost invariant).
We request that the expression in the curly brackets vanishes
at least up to $O(\varepsilon^K)$ for some $K>0$.

For $n=0$ we obtain the equation $(H|_{\Lambda^r_l}-\lambda_0)u_0=0$,
and we can put $\lambda_0=\cv{H}|_{\Lambda^r_l}$ and $u_0=1$.

For $n=1$ we obtain $(d/dt+\lambda_1)u_0=0$ and we set $\lambda_1=0$.

The equations for $n\ge2$ have the form
\begin{equation*}
-\frac{d}{dt}u_{n-1}=\sum_{j=0}^{n-2} (R^{n-j}-\lambda_{n-j})u_j
\end{equation*}
(they are called \emph{homological equations}).
Let us show that all these equations can be solved up to $O(\varepsilon^m)$,
where $m$ is arbitrary positive number.
To do this, let us note first, that the operator $d/dt$
on each torus can be written as
\begin{equation*}
\frac{d}{dt}=\omega_1\frac{\partial}{\partial\varphi_1}+
\omega_2\frac{\partial}{\partial\varphi_2},\quad 
\omega_{1/2}=\frac{\partial\cv{H}}{\partial\cv{I}_{1/2}}\Big|_{\Lambda^r_l}.
\end{equation*}
It is important for us that $\omega_1=1+O(\varepsilon)$
and $\omega_2=O(\varepsilon)$; both these numbers do not depend on
$\varphi_1$ and $\varphi_2$.

Rewrite all the homological equations in a common form
\begin{equation}
            \label{homol-eq}
\frac{d}{dt}f=E+g,\quad g\in C^\infty(\Lambda^r_l).
\end{equation}
Let us expand all the functions into their Fourier series:
\begin{equation*}
f=\sum_{(k_1,k_2)\in\mathbb{Z}^2} f_{k_1,k_2} e^{i (k_1\varphi_1+k_2\varphi_2)},\quad
g=\sum_{(k_1,k_2)\in\mathbb{Z}^2} g_{k_1,k_2} e^{i (k_1\varphi_1+k_2\varphi_2)}.
\end{equation*}
Substituting these series into~\eqref{homol-eq}, we obtain
\emph{formally}:
\begin{equation*}
f_{k_1,k_2}=\frac{1}{k_1\omega_1+k_2\omega_2}g_{k_1,k_2},
\quad E=-g_{0,0}.
\end{equation*}
Note that these coefficients $f_{k_1,k_2}$ can be very large
because of the denominator, and the function $f$ is, generally
speaking, not defined.
To avoid this obstacle, let us use additional estimates. 
As $g\in C^\infty$, its Fourier coefficients decay very fast,
and for any $\alpha>0$ there exists positive numbers $C(\alpha)$
and $N(\alpha)$
such that $|g_{k_1,k_2}|\le C(\alpha)/(|k_1|+|k_2|)^\alpha$
as $|k_1|+|k_2|>N(\alpha)$.
Let us put $\alpha=2m$.

Introduce a set $Q(\varepsilon,\alpha)$ as follows:
\begin{multline*}
Q(\varepsilon,\alpha)=\Big\{
(k_1,k_2)\in\mathbb{Z}^2:~
|k_1|+|k_2|\le N(\alpha)
\Big\}\\
{}\cup\Big\{
(k_1,k_2)\in\mathbb{Z}^2:~
|k_2|\le \frac{1}{\sqrt{\varepsilon}}
\Big\}.
\end{multline*}
Clearly, $|k_1\omega_1+k_2\omega_2|\ge 1/2$
as $(k_1,k_2)\in Q(\varepsilon,\alpha)$ and $\varepsilon$
is small enough.

Now set
\begin{equation*}
G=\sum_{(k_1,k_2)\in Q(\varepsilon,\alpha)} g_{k_1,k_2}
e^{i(k_1\varphi_2+k_2\varphi_2)},\quad
\Tilde{g}=g-G.
\end{equation*}
The equation $df/dt=G$ can be solved in Fourier series,
and the function $\Tilde{g}$ gives a discrepancy $O(\varepsilon^K)$,
because
\begin{equation*}
\Tilde{g}_{k_1,k_2}\le C(\alpha)/(|k_1|+|k_2|)^\alpha
\le C(\alpha)\varepsilon^K.
\end{equation*}

Therefore, we can construct a function $u\in L^2(\mathbb{R}^2)$
and a number $E\in \mathbb{R}$ such that $\|u\|_{L^2}\ge c>0$
and $\|(\Hat{h}-E)u\|_{L^2}=O(h^L+\varepsilon^K)$ as $h,\varepsilon\to0$.
Then
\begin{equation}
                \label{spectr-est}
\dist(E,\spec\Hat H)\le\frac{\|(\Hat H-E)u\|}{\|u\|}=O(h^L+\varepsilon^K).
\end{equation}

The same procedure can be applied to each of the
quantized cylinders $\Tilde\Lambda^r_k$,
but as a result we obtain a function $u\in L^2_{\text{loc}}$
and a number $E\in\mathbb{R}$ such that
\begin{gather}
                    \label{U1-U3}
\begin{gathered}
u(x+d\cdot a,h,\varepsilon)=u(x,h,\varepsilon)
e^{\frac{i}{h}\big(2\pi\cv{I}_2
-(d\cdot a)_1 x_1-\frac{1}{2}(d\cdot a)_1(d\cdot a)_2\big)},\\
\|u\|_{L^2(\Pi)}\ge c>0,
\text{ and }\|(\Hat{H}-E)u\|_{L^2(\Pi)}=O(h^L+\varepsilon^K),
\end{gathered}\\
\intertext{where}
\Pi=\big\{
x=\tau_1 (d\cdot a) +\tau_2 J(d\cdot a),
\quad
\tau_1\in [0,1],\quad \tau_2\in\mathbb{R}
\big\}.\notag
\end{gather}
As such function $u$ does not belong to $L^2_{\mathrm{loc}}(\mathbb{R}^2)$,
one cannot apply the inequality~\eqref{spectr-est} directly.

\begin{prop}
Let a function $u$ and a number $E$ satisfy
the conditions~\eqref{U1-U3},
then $\dist(E,\spec\Hat{H})=O(h^L+\varepsilon^K)$.
\end{prop}
\begin{proof}
Introduce new coordinates
\begin{equation*}
y=Ax,\quad A=\begin{pmatrix}
\alpha & \beta\\
-\beta &\alpha
\end{pmatrix},\quad
\begin{pmatrix}
\alpha\\ \beta
\end{pmatrix}=\frac{d\cdot a}{|d\cdot a|}
\end{equation*}
and a function
\begin{equation*}
S(y)=\frac{1}{2}(-\alpha\beta y_1^2+\alpha\beta y_2^2+2\beta^2 y_1 y_2).
\end{equation*}

Define a unitary operator $\Hat{U}$ in $L^2(\mathbb{R}^2)$ by the rule
\begin{equation*}
f(x)\stackrel{\Hat{U}}{\mapsto} g(y)=e^{-\frac{i}{h}S(y)}f(A^{-1}y);
\end{equation*}
it is easy to see that $U$ is well-defined also on $L^2_{\mathrm{loc}}(\mathbb{R}^2)$.
Now we set $\Tilde{H}=\Hat{U}\Hat{H}\Hat{U}{}^{-1}$, i.~e.
\begin{equation*}
\Tilde{H}=
\frac{1}{2}\Big(-ih\frac{\partial}{\partial y_1}+y_2\Big)^2
+\frac{1}{2}\Big(-ih\frac{\partial}{\partial y_2}\Big)^2
+\varepsilon w(y),\quad w(y)=v(A^{-1}y).
\end{equation*}
As the operator $\Hat{U}$ is unitary, the spectra of $\Hat{H}$
and $\Tilde{H}$ coincide.

Put $\varphi=\Hat{U}u$, then
\begin{equation*}
\varphi(y_1+|d\cdot a|,y_2,h,\varepsilon)=
\varphi(y_1,y_2,h,\varepsilon) e^{2\pi\frac{i}{h}\cv{I}_2}.
\end{equation*}
Denote
\begin{equation*}
\Tilde{\Pi}_{s}=\Big\{
(y_1,y_2)\in\mathbb{R}^2:\quad -s|d\cdot a|\le y_1 \le s| d\cdot a|
\Big\},\quad s\in\mathbb{Z},
\end{equation*}
Note that
\begin{equation*}
\|f\|_{L^2(\Tilde{\Pi}_{s})}=\sqrt{s}\|f\|_{L^2(\Tilde{\Pi}_{1})}
\end{equation*}
for any function
satisfying $f(y_1+|d\cdot a|)=e^{i\alpha}f(y_1,y_2)$, $\alpha\in\mathbb{R}$;
in particular, this holds for $f=\varphi$
and for $f=\Phi:=(\Tilde{H}-E)\varphi$.

Choose now a smooth function $e(\xi)$ such that
\begin{align*}
0\le e(\xi)\le 1,\\
e(\xi)=1 &\quad \text{ as } \xi\in (-| d\cdot a|,|d\cdot a|),\\
e(\xi)=0 &\quad \text{ as } \xi\notin (-2| d\cdot a|,2|d\cdot a|),
\end{align*}
and choose a constant $C_0$ such that
\begin{equation*}
|e|+|e'|+|e''|\le  C_0.
\end{equation*}
Put $e_{s}(y_1,y_2):=e(y_1/s)$.
Now we have the following chain of equalities and inequalities:
\begin{gather*}
\sqrt{s}\dist\big(E,\spec \Hat{H}\big) \|\varphi\|
   \le \dist\big(E,\spec \Hat{H}\big)\|e_s\varphi\|_{L^2(\Tilde{\Pi}_s)}\\
        \le \Big\|\big(\Tilde{H}-E\big)(e_s\varphi)\Big\|
\le \Big\|e_s\Phi-\frac{1}{2}h^2\Delta e_s \varphi -h^2\langle\nabla e|\nabla\varphi\rangle
    -ihx_2\frac{\partial e_s}{\partial x_1}\Big\|\\
 \le\|e_s\Phi\|+\frac{1}{2}h^2\|\Delta e_s \varphi\|
       +h^2\big\|\langle\nabla e_s|\nabla\varphi\rangle\big\|+
        h\Big\|x_2\frac{\partial e_s}{\partial y_1}\varphi\Big\|\\
 \le C_0 \sqrt{2s}\|\Phi\|_{L^2(\Tilde{\Pi}_1)}+
  \frac{h^2 C_0\sqrt{s}}{2s^2}\big\|\varphi\big\|_{L^2(\Tilde{\Pi}_1)}
  +\frac{h^2 C_0\sqrt{s}}{s}\big\| \frac{\partial\varphi}{\partial y_1}\big\|_{L^2(\Tilde{\Pi}_1)}\\
  +\frac{h^2 C_0\sqrt{s}}{s}\big\| \frac{\partial\varphi}{\partial y_2}\big\|_{L^2(\Tilde{\Pi}_1)}
  +\frac{h C_0\sqrt{s}}{s} \Big\|x_2\varphi\Big\|_{L^2(\Tilde{\Pi}_1)}.
\end{gather*}
Tending $s$ to $+\infty$, we obtain the inequality
\begin{equation*}
\dist(E,\spec\Hat{H})
\big\|\varphi\big\|_{L^2(\Tilde{\Pi}_1)}\le
C_0\sqrt{2}\big\|\big(\Tilde{H}-E\big) \varphi\big\|_{L^2(\Tilde{\Pi}_1)}.
\end{equation*}
Now one only has to use the conditions~\eqref{U1-U3}.
\end{proof}

Note that in the problem under consideration there
exist non-Lagrangian (but isotropic) invariant manifolds
of the averaged Hamiltonian. The construction
of spectral series corresponding to such manifolds
is studied in~\cite{bdp-tmf}.

\end{document}